\documentclass[12pt,a4paper]{article}

\usepackage{amssymb,amscd,amsmath,graphicx,cite}
\usepackage{epsfig,float}

\title{Quantum dynamics in strong fluctuating fields}
\author{Igor Goychuk\thanks{Corresponding author, 
e-mail: goychuk@physik.uni-augsburg.de}$\;$  and Peter H\"anggi,\\ 
Universit\"at Augsburg, Institut f\"ur Physik, Universit\"atsstr. 1,
\\ D-86135 Augsburg, Germany
}

\begin{document}

\maketitle

\vspace{-1cm}

\begin{abstract}

A large number of multifaceted quantum transport processes in
molecular systems and physical nanosystems, such as e.g.
nonadiabatic electron transfer in proteins, can be treated in terms
of quantum relaxation processes which couple to one or several
fluctuating environments. A thermal equilibrium environment can
conveniently be modelled by a thermal bath of harmonic oscillators.
An archetype situation provides a two-state dissipative quantum
dynamics, commonly known under the label of a spin-boson dynamics.
An interesting and nontrivial  physical situation emerges, however,
when  the quantum dynamics evolves far away from thermal
equilibrium. This occurs, for example,  when a  charge transferring
medium possesses  nonequilibrium degrees of freedom, or when a
strong time-dependent control field is  applied externally.
Accordingly, certain parameters of underlying quantum subsystem
acquire stochastic character. This may occur for example for the
tunnelling coupling between the donor and acceptor states of
transferring electron, or for the corresponding energy difference
between electronic states which assume via the coupling to the
fluctuating environment an explicit stochastic or deterministic
time-dependence. Herein, we review the general theoretical framework
which is based on  the method of projector operators, yielding the
quantum master equations for systems that are exposed to strong
external fields. This allows one to investigate on a common basis
the influence of nonequilibrium fluctuations and periodic electrical
fields on those already mentioned dynamics and related quantum
transport processes. Most importantly, such strong fluctuating
fields induce a whole variety of nonlinear and nonequilibrium
phenomena. A characteristic feature of such dynamics is the absence
 of thermal (quantum) detailed balance.

\end{abstract}

\tableofcontents

\section{\label{Intro}Introduction}

The description and analysis of the dynamics of open quantum
systems, i.e. quantum systems interacting with a dissipative
environment, presents a key challenge for nonequilibrium statistical
physics. Moreover, this theme is also of prominent importance for
many applications in physics, physical chemistry and physical
biology. This can be exemplified by the relaxation dynamics
occurring in a two-level quantum system that is coupled to the
vibrational degrees of freedom of an environment. This latter theme
gained great popularity and is known under the label of  "spin-boson
dynamics'' \cite{LeggettReview,WeissBook,GrifHangRev,KohlerRev}.
Several apparently different physical problems can  formally  be
unified within such a common mathematical description. For example,  the
relaxation dynamics of a nuclear spin $1/2$ in solids,  the
tunnelling of defects in metals, the relaxation dynamics of atoms in
optical cavities can all be modelled by a (pseudo)-spin-boson
dynamics. Another important and relevant situation
refers to donor-acceptor electron transfer reactions in various
molecular structures
\cite{Garg,RipsJortner,Schulten,advance,MayBook}. For spatially
extended  quasi-periodic molecular structures like those formed by
protein $\alpha$-helices
\cite{Davydov1,PetrovBook,KPU78,SkourtisBeratan}, or DNA's
\cite{Jortner98,Bruinsma,Giese,BraunKeren,Conwell} many quantum
states are generally required to describe
 charge  transfer processes.  Here, a
multi-state tunnelling problem naturally emerges with
the tight-binding model serving  as one of simplest theoretical
frameworks.

The primary problem is to describe the influence of the
environmental degrees of freedom on the quantum dynamics of interest. Many
different approaches have been developed  to tackle this
challenge. The fundamental methodology consists in
separating the total  system into
the two (or more) mutually
interacting parts: the dynamical subsystem with a small number of relevant
degrees of freedom and a thermal bath represented by a huge number of
microscopic degrees of freedom which are at thermal equilibrium.   A most
general quantum-mechanical description is provided by the density
operator of the whole, combined system which depends both on the variables of
the considered dynamical subsystem (relevant variables) and the
variables of the thermal bath (irrelevant variables). The dynamical
behaviour of a small quantum subsystem presents the focus of interest with
the thermally equilibrated bath degrees of freedom serving as a source of
randomness for the relevant dynamics. This randomness
can effectively be eliminated via a course-grained  description
of the system of interest. A corresponding averaging procedure
results in a contracted, reduced
dynamics  which generally entails  memory effects,
decoherence and dissipation.

Different approaches have been developed over the years
within this general line of reasoning.
Within a variety of different approaches, the method of
path integrals in real time
\cite{Feynman,CaldeiraLeggett,LeggettReview,WeissBook,GIS}
and the projection  operator
method \cite{Nakajima,Zwanzig,ArgyresKelley,Hynes,Haake,GrabertBook,
PA1,PA2,PA3,ZwanzigBook}
provide some of the most frequently used methods.
The path-integral approach can, however, be technically
cumbersome in practical applications of interest.
The projection operator method appeals to its generality and
technical elegance. It allows one to obtain
formally  exact
generalised master equations (GMEs) for the reduced density matrix
 in a straightforward way.
By and large, however, such exact GMEs cannot be analytically elaborated
further without invoking some sort of a perturbation technique with
corresponding approximations. For example, already the seemingly
simple  spin-boson dynamics cannot be solved analytically exactly.  The
weak-coupling approximation of the system-bath coupling
is one of the most useful and commonly employed
scheme. Moreover, a strong-coupling problem can often be
mapped onto a (different) weak-coupling problem
within a canonically transformed
basis of the {\it total} system.
 The projection operator method,  combined with appropriate canonical
transformations, further improved by use of variational approaches,
presents a powerful
and  general method of wide acceptance. This well-established methodology is,
however, also rather demanding.

Yet another popular methodology consists in modelling the thermal bath
influence through a classical stochastic field which acts upon the
considered dynamical system. Formally, this methodology corresponds to
introduction of randomly fluctuating time-dependent forces in the
Hamiltonian of considered quantum system \cite{Anderson53,Kubo54,Burstein,Haken72}
and finding subsequently the
stochastically averaged evolution of the considered system which is
governed by a stochastic Liouville-von-Neumann equation. This methodology
is known under the label of
 stochastic Liouville equation (SLE) approach
\cite{Kubo63,Burstein,Haken72,Fox,BlumenSilbey,MukamelOppenRoss,ReinekerBook,Lindenberg}.
Due to a reasoning that involves the central limit theorem classical
random forces with Gaussian statistics are most frequently used in
this kind of approximate modelling. The Gaussian  white noise serves
here as a simplest implementation for the corresponding classical
stochastic bath. It corresponds to a bath with an infinite spectrum
of excitations. Such models can be solved exactly in a number of
specific cases
\cite{Haken72,OvchinErikh,Fox,ReinekerBook,Lindenberg,ShaoZerbeHan98,Shao97}.
All the thermal baths possess, however, finite energy spectra. This
circumstance gives rise to temporal autocorrelations in the
bath-induced classical stochastic fields. Gaussian Markov noise with
the exponentially decaying temporal autocorrelations presents one of
the simplest models of such more realistic, coloured noise
\cite{HanggiJung}. Yet, even in the simplest case of a two-state
tunnelling system this model  cannot be solved exactly except for
some limiting cases (see, e.g., in \cite{Kayanuma1} for the
Landau-Zener model with a stochastic modulation). Generically one
must invoke some approximations; e.g., in the case of a weakly
coloured Gaussian noise some kind of cumulant expansion technique
\cite{MukamelOppenRoss,Fox,KampenBook} can be used.

There exists a different
possibility.
Continuous state noises can be approximated by noise sources with  a large
number of discrete states (e.g., by a discretisation procedure of a continuous diffusion
process in a potential). Certain Markovian
discrete state noises provide then a rather general
framework for a formally
exact stochastic averaging \cite{Kubo62,Burstein,FrischBrissaud}.
Moreover, the two-state Markovian noise (also known as {\it dichotomous noise})
presents such a
simple discrete noise source which allows for an exact study of
noise driven two-level quantum systems
\cite{AverPouq,ShapLog1,HanggiJung,Reineker1,PRE95_3,PLA96,Iwan,AnkerPech}.
In addition, the multistate case of exciton transfer
in molecular
aggregates with many quasi-independent noise sources modelled by
independent two-state Markovian noises presents
another analytically tractable case, in the sense that it can be reduced
to the solution
of a system of linear differential equations with constant coefficients
for averaged dynamics \cite{Reineker2}. The
discussed dichotomous noise can  model a
quasi-spin 1/2 stochastic bath variable.
In the case of electron transfer in molecular systems such a quasi-spin
stochastic variable can simulate, for example, the bistable
fluctuations of a charged
molecular group nearby the donor, or acceptor site, or the conformational
fluctuations of a bistable molecular bridge.

A well-known drawback
of the SLE approach consists, however, in
the asymptotic equal-population of the energy
levels of quantum system  which occurs
for arbitrary energy differences  \cite{Lindenberg,Fox,ReinekerBook}.
This means that the SLE approach corresponds formally to an infinite bath
temperature. At least, the thermal energy $k_BT$ should thus be larger
than the characteristic energy scale of the quantum system, e.g., larger than
the energy width of the corresponding excitonic band.
This corresponds to
a high-temperature approximation \cite{ReinekerBook,Lindenberg,Reineker1,Reineker2}.
The reason for this intrinsic restriction
is that the stochastic
field unidirectionally drives the  quantum system  without being
modified by the
system's feedback (no back reaction). This drawback within
the SLE approach requires some {\it ad hoc} corrections to enforce
the correct thermal equilibrium \cite{OvchinErikh,Zusman,FaidFox87}.
Nevertheless, the SLE approach can yield a very useful tool,
notably in the nuclear magnetic resonance (NMR) theory
\cite{ReinekerBook,Slichter}, the theory of exciton transfer in
molecular aggregates \cite{ReinekerBook} and within the theory of
single-molecular spectroscopy \cite{JungSilbey}.

Combined scenarios have been used in several
works \cite{FaidFox87,PetTes90,Vill91}.
Initially, those were aimed to model the influence of relaxation processes
in the thermal bath \cite{FaidFox87}, or to account
for non-Gaussian large-amplitude fluctuations of molecular charged groups
\cite{PetTes90}.
However, it was soon recognised that the addition of
classical noise to a dissipative quantum dynamics generally violates the
detailed balance symmetry at the environmental temperature
\cite{PRE94}. Therefore, the stochastic field in these approaches
 correspond physically to a nonequilibrium noise influence.
It has been shown theoretically that such
a nonequilibrium non-Gaussian (e.g., two-state) noise can regulate the quantum
transition rates by several
orders of magnitude \cite{PetTes90,UFZ93,PRE94,PRE96}. Moreover,
it may pump energy into the quantum
system. This in turn  gives rise to various interesting
nonlinear nonequilibrium  phenomena such as
a noise-induced enhancement of thermally assisted
quantum tunnelling \cite{JCP95},
an inversion of population in discrete quantum dissipative systems
\cite{PRE97}, a
noise-induced absolute negative mobility for
quantum transport \cite{PLA98}, or also
a fluctuation-induced transport of quantum particles within a tight-binding
description \cite{PRL98}, to name  the most prominent ones.
From a thermodynamical perspective these nonequilibrium effects are due to a
virtual presence of
two heat baths of different nature:
a first one assuming the temperature of the environment $T$
(modelled by a thermal bath of harmonic oscillators that are
bi-linearly coupled to the
relevant system), another one possessing a
virtually infinite temperature $T_{\sigma}=\infty$ (stochastic bath).
In this intuitive  picture,
a nonequilibrium stochastic field
is expected to heat
the quantum-mechanical degrees of freedom, causing various, surprising
 nonequilibrium phenomena.

The study of the dynamics of such quantum dissipative
systems which are driven far from thermal equilibrium by nonequilibrium
fluctuations is  the focus of
present work. The situation here is similar in spirit to one in the 
recently emerged field of
(classical) Brownian motors
\cite{Magnasco93,Prost94,BartHangKis94,AstumBier94,Doering94,Bartussek1996},
see, e.g., in \cite{HanggiBartussek,AstumianHanggi,HanggiReimann,ReimannRev,HanggiMarchesoniNori}
 for surveys and further references.

In this review, we present
a general outline with many important
examples given of the following methodology:
The nonequilibrium stochastic field is represented
by an external time-varying
classical field in the Hamiltonian of the quantum system. This field is
treated without invoking any  further approximation, until it becomes necessary to do so.
In doing so, a formally exact generalised master equation is obtained
which includes the external field both in the dynamical part and
in the dissipative kernel of the GME exactly. Subsequently,
the dissipative kernel
is expanded to the lowest order, i.e. the second order
in the system-bath coupling.
[In a properly canonically transformed basis this scheme
allows one to study the opposite limit of
strong dissipation/weak tunnelling as well.]
The overall procedure results  in an
approximate, generalised master equation for the
reduced density matrix of the considered, relevant quantum system.
We recall that within this methodology  the external
field is not only included
exactly in the dynamical part, but it  modifies as well
the dissipative kernel in a very profound manner.
In particular, the dissipative kernel becomes a retarded functional of
the driving field. Thereby, the field influence on quantum dynamics
 is taken rigorously into account within the given order of the
system-bath coupling. Such a corresponding modification of the dissipative kernel
becomes crucial for  strong driving fields.

This so tailored approach allows one to describe  stochastic  and
time-periodic fields on  equal grounds. The influence of a
time-periodic driving on the dissipative quantum dynamics has been
investigated in Refs.
\cite{Petrov86,CPL96,PRE96R,JCP97,PRE97,PRE2000}. Other related work
has been done at the same time and in parallel by several other
research groups elaborating similar
\cite{Dakh94,DakhCoal95,Morillo96}, or different
\cite{Dittrich93,Grifoni93,GSHW95,Morillo93,GWW,Kohler,Kayanuma,JCP2000}
approaches which reconcile within some approximations
\cite{GrifSasWeis,GrifHartHang97,GrifHangRev,PRE2000,KohlerRev}.

The difference between the stochastic fields and the periodic fields
enters within our approach on the level of averaging of the
corresponding field-driven generalised master equations. In this
context,  one must refer to some further approximations, which
generally are based on the separation of time scales involving the
external driving, the contracted quantum dynamics, and the decay of
dissipative kernels in the generalised master equations. Remarkably,
in the case of dichotomous fluctuations this averaging can be done
exactly without further approximations \cite{Goychuk95,JCP95,PRE95}.
The spin-boson model driven by such dichotomous Markovian
fluctuations presents one instance of general interest which has
been studied in detail in Refs.
\cite{JCP95,PRE95,PRE96R,JCP97,PRE97}. Two other important
situations, where an exact averaging is feasible for a broad class
of stochastic and periodic processes, are given by an infinitely
extended tight-binding model. These are: (i) the case of coherent 
tunnelling in the absence of dissipation and (ii) the regime of
incoherent tunnelling (strong dissipation) when
the tunnelling is weak (high tunnelling barriers).
Some explicit pertinent examples are
discussed in Refs. \cite{PRL98,EPL98,LNP2000,JPCB2001}.

\begin{figure}
\centerline{\epsfxsize=13cm \epsfbox{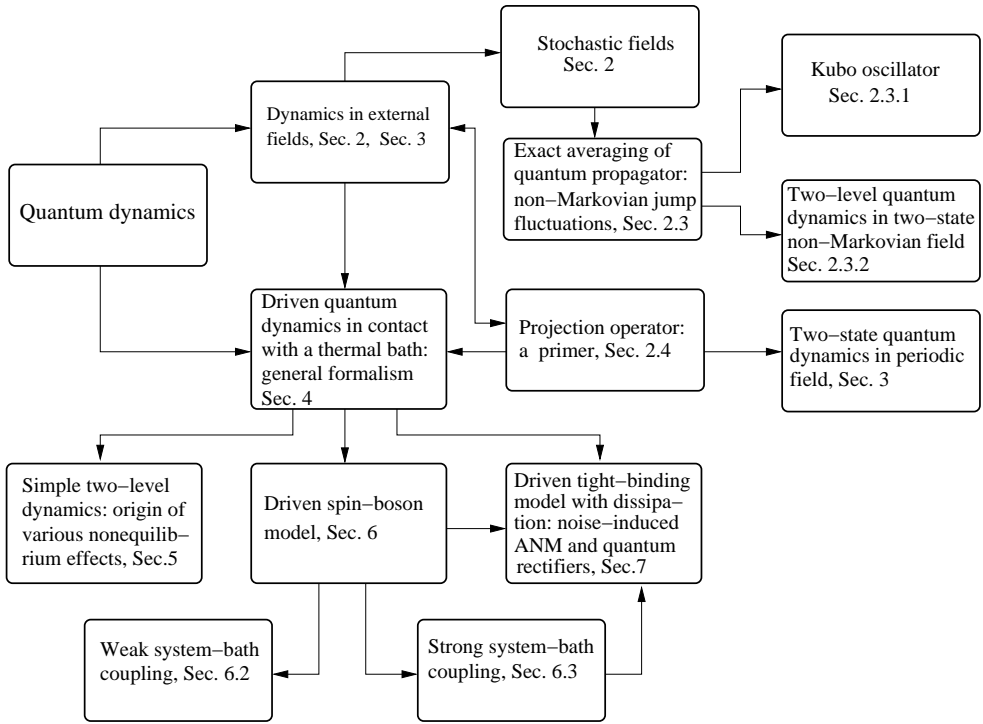}} \caption{A flow
diagram depicting the various topics and their mutual interrelations
which are covered by this review.} \label{chart}
\end{figure}

This review is organised as follows, see in Fig. \ref{chart}.
The study of a quantum dynamics subjected to non-Markovian stochastic fields
that are modelled by discrete state processes of the
renewal type is presented in Section 2. Therein, a formally exact
averaging of the quantum evolution over the stationary realizations
of stochastic fields is given. The general results are illustrated
by a new Laplace-transformed exact solution of averaged two-level
quantum dynamics driven by a symmetric non-Markovian two-state
field. The prior results for a quantum two-level dynamics driven by
a dichotomous Markovian field are reproduced as a particular
limiting case. This section contains also the results for the
fluctuating Kubo oscillator. These are used in the following. This
section also contains  a short primer into the projection operator
formalism. Two-state quantum dynamics in strong periodic fields is
considered in Sec. 3. The Section 4 outlines the general formalism
of dissipative quantum dynamics in strong, time-varying fields
within the reduced density matrix approach. The corresponding
weak-coupling generalised master equations and the generalised
Redfield equations are presented there. These equations serve as a
basis for subsequent applications and analysis of stochastic
field-induced phenomena. Section 5 contains a simple implementation
of our general approach which manifests the origin and basic
features of strongly nonequilibrium phenomena described in the
subsequent Sections for more realistic models. The stochastically
and periodically driven spin-boson model is discussed in Section 6,
including quantum stochastic resonance features. Section 7 is
devoted to the phenomenon of noise-induced, absolute negative
mobility in quantum transport and to the analysis of dissipative
quantum rectifiers. Concluding remarks are drown in the Section 8.

\section{\label{rand}Quantum dynamics in stochastic fields}
\subsection{Stochastic Liouville equation}

To begin, let us consider an arbitrary quantum system with
a Hamilton operator $\hat H[\xi(t)]$ which depends on a classical,
 noisy parameter
 $\xi(t)$. This stochastic process $\xi(t)$ can take on either
continuous or discrete values.  Accordingly, the Hamiltonian
$\hat H$ acquires randomly in time  different operator values
 $\hat H[\xi(t)]$ which generally do not commute, i.e.,
$[\hat H[\xi(t)],\hat H[\xi(t')]]\neq 0$.

The posed problem is to average the corresponding quantum dynamics in
the Liouville space,
which is characterised by the Liouville-von-Neumann equation
\begin{equation}\label{Liouville}
\frac{d}{dt} \rho (t)= -i \mathcal{ L}[\xi(t)]\rho(t) \;,
\end{equation}
for the density operator $\rho(t)$ over the realizations
of noise $\xi(t)$. $\mathcal{ L}[\xi(t)]$
in Eq. (\ref{Liouville})
stands for the quantum Liouville
superoperator, $\mathcal{ L}[\xi(t)](\cdot)=\frac{1}{\hbar}
[\hat H[\xi(t)], (\cdot)]$. In other words, the objective is to evaluate
the noise-averaged propagator
\begin{equation}\label{propagator}
\langle S (t_0+t,t_0) \rangle =
\langle \mathcal{ T} \exp[-i\int_{t_0}^{t_0+t}
\mathcal{ L}[\xi(\tau)] d\tau] \rangle,
\end{equation}
where $\mathcal{ T}$ denotes the time-ordering operator.

\subsection{Non-Markovian {\it vs.} Markovian discrete state
fluctuations}

\begin{figure}
\centerline{\epsfxsize=8cm
\epsfbox{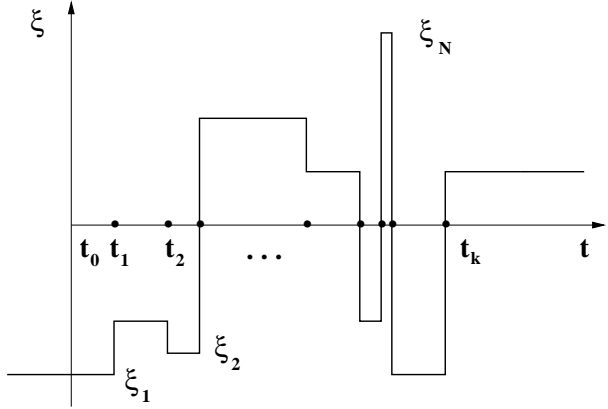}}
\caption{Typical trajectory of the considered process \cite{JCP05}.}
\label{Fig1}
\end{figure}

We specify this task for a discrete state
noise with $N$  states $\xi_i$ (cf. Fig. \ref{Fig1}).
The noise is generally assumed to be a non-Markovian renewal process which is
fully characterised by
the set of  transition probability densities $\psi_{ij}(\tau)$
for making random transitions within the time
interval $[\tau,\tau+d\tau]$ from
the  state $j$ to the state
$i$. These probability densities are obviously  positive and do obey the
normalisation conditions
\begin{equation}\label{norm}
\sum_{i=1}^{N}\int_{0}^{\infty}
\psi_{ij}(\tau)d\tau =1 \;,
\end{equation}
for all $j=1,2,...,N$.

The subsequent residence time-intervals between jumps
are assumed to be mutually
uncorrelated.
The residence time distribution (RTD) $\psi_j(\tau)$ in the state $j$
reads
\begin{equation}\label{RTD}
\psi_j(\tau)=\sum_i \psi_{ij}(\tau)=-\frac{d\Phi_j(\tau)}{d\tau}.
\end{equation}
 The survival probability $\Phi_j(\tau)$
of the state $j$ follows then as
\begin{equation}\label{survival}
\Phi_j(\tau)=\int_{\tau}^{\infty}
\psi_j(\tau)d\tau.
\end{equation}
This constitutes the  general description for continuous time
random walk
(CTRW) theory \cite{MontrolWeiss,LaxSher,Shlesinger,Hughes}.

Several descriptions
used for such simplest non-Markovian processes of the renewal type
are worth mentioning. The approach in Ref. \cite{Kampen79}
with the time-dependent aging rates $k_{ij}(t)$ for the transitions
from state $j$ to state $i$ corresponds to a particular choice
\begin{equation}\label{kamp}
\psi_{ij}(\tau):=k_{ij}(\tau)\exp[-\sum_i\int_{0}^{\tau}k_{ij}(t)dt].
\end{equation}
The Markovian case corresponds to time-independent transition rates
$k_{ij}(\tau)=const$.
Any deviation of $\psi_{ij}(\tau)$ from the corresponding
strictly exponential
form which can be accounted for by introducing a
time-dependence of the transition rates $k_{ij}(\tau)$
amounts to a non-Markovian behaviour\footnote{This observation which can
 be traced back to Ref. \cite{Feller} can be rationalised as follows.
Let us consider a sojourn in the state $j$
characterised by the survival
probability $\Phi_j(\tau)$. The
corresponding
residence time interval $[0,\tau]$ can be arbitrarily
divided into two pieces
$[0,\tau_1]$
and $[\tau_1,\tau]$.
If no memory effects are present, then
$\Phi_j(\tau)=\Phi_j(\tau-\tau_1)\Phi_j(\tau_1)$.
The only nontrivial solution of this latter functional
equation which decays
in time reads $\Phi_j(\tau)=\exp(-\gamma_j\tau)$, with $\gamma_j>0$.}. 
Furthermore,
the survival probability $\Phi_j(\tau)$ in the
state $j$  is determined by
\begin{equation}\label{kamp2}
\Phi_j(\tau)=\exp[-\sum_{i=1}^{N}
\int_{0}^{\tau}k_{ij}(t)dt]
\end{equation}
and Eq. (\ref{kamp})  can be recast as
\begin{equation}\label{kamp3}
\psi_{ij}(\tau):=k_{ij}(\tau)\Phi_j(\tau).
\end{equation}
The introduction of  time-dependent ``aging''
rates presents one possibility to describe the non-Markovian effects.
It is not unique though.
A different
scheme  follows by defining  \cite{Chvosta}:
\begin{equation}\label{alter}
\psi_{ij}(\tau):=p_{ij}(\tau)\psi_j(\tau)
\end{equation}
 with $\sum_i p_{ij}(\tau)=1$.
The physical interpretation is as follows: The process stays
in a state $j$ for a random time interval characterised by the
probability density $\psi_j(\tau)$. At the end of this time interval
it jumps into another state $i$ with a generally
time-dependent conditional probability $p_{ij}(\tau)$. Evidently,
any process of the considered type can be interpreted in this way.
By equating Eq. (\ref{kamp3}) and Eq. (\ref{alter}) and taking into
account $\psi_j(\tau):=-d\Phi_j(\tau)/d\tau$ one can deduce
that the approach in Ref. \cite{Kampen79} can be reduced to that
in Ref. \cite{Chvosta}
with the time-dependent transition probabilities
\begin{equation}\label{corres}
p_{ij}(\tau)=\frac{k_{ij}(\tau)}{\sum_i k_{ij}(\tau)}
\end{equation}
and with the non-exponential probability densities $\psi_j(\tau)$ which
follow as  $\psi_j(\tau)=
\gamma_j(\tau)\exp[-\int_{0}^{\tau}\gamma_j(t)dt]$ with
$\gamma_j(\tau):=\sum_ik_{ij}(\tau)$.

The description of non-Markovian effects with the
time-dependent transition probabilities $p_{ij}(\tau)$,
is rather difficult
to deduce immediately from the sample trajectories of
an experimentally {\it observed} random process $\xi(t)$.
The same holds true for
the concept of time-dependent rates.
These rates cannot be measured directly from
the set of stochastic sample trajectories.
On the contrary, the RTD $\psi_j(\tau)$ and the
{\it time-independent} $p_{ij}$ (with $p_{ii}:=0$)
can routinely be deduced from
sample trajectories measured, say, in a {\it single-molecular} 
experiment \cite{Flomenbom}.
Fig. \ref{Fig1} renders these assertions more obvious. The study of the
statistics of the residence time-intervals allows one to obtain
the corresponding probability densities $\psi_j(\tau)$
and, hence, the survival probabilities $\Phi_j(\tau)$. Furthermore,
the statistics of
the transitions from one state into all other states allows one to derive
the corresponding conditional probabilities $p_{ij}$. From this primary
information
a complementary interpretation of experimental data in terms of
time-dependent rates
$k_{ij}(\tau)$ can readily be given as
\begin{equation}\label{reduce}
k_{ij}(\tau)=-p_{ij}\frac{d\ln[\Phi_j(\tau)]}{d\tau},
\end{equation}
if one prefers to use this language to describe the
non-Markovian effects.
Moreover, the description with a constant set
$p_{ij}$ provides a consistent approach to construct the stationary
realizations of $\xi(t)$, and thus to find the corresponding  averaged quantum evolution
 \cite{Goychuk04}.

\subsection{Averaging the quantum propagator}

The task of performing the noise-averaging of the quantum dynamics
in Eq. (\ref{propagator}) can be solved exactly due the piecewise
constant character of the noise $\xi(t)$
\cite{Burstein,FrischBrissaud}. Let us consider the time-interval
$[t_0,t]$ and take a frozen realization of $\xi(t)$ assuming $k$
switching events within this time-interval at the time-instants
$t_i$,
\begin{equation}\label{time}
t_0<t_1<t_2<...<t_k<t.
\end{equation}
Correspondingly, the noise takes on the values
$\xi_{j_0},\xi_{j_1},...,\xi_{j_k}$ in the time sequel.
Then, the propagator $S(t,t_0)$ reads
\begin{equation}\label{pathU}
S(t,t_0)=e^{-i \mathcal{ L}[\xi_{j_k}](t-t_{k})}
e^{-i \mathcal{ L}[\xi_{j_{k-1}}](t_k-t_{k-1})}...e^{-i \mathcal{ L}
[\xi_{j_0}](t_1-t_{0})} \;.
\end{equation}
Let us assume further that the process $\xi(t)$ has been {\it prepared}
in the state $j_0$ at $t_0$. Then,
the corresponding $k-$times probability density
for such noise realization is
\begin{equation}\label{prob}
P_k(\xi_{j_k},t_{k};\xi_{j_{k-1}},t_{k-1};...;\xi_{j_1},t_{1}|\xi_{j_0},t_{0})=
\Phi_{j_k}(t-t_k)\psi_{j_kj_{k-1}}(t_k-t_{k-1})...
\psi_{j_1j_0}(t_1-t_0)
\end{equation}
for $k\neq 0$ and $P_0(\xi_{j_0},t_0)=\Phi_{j_0}(t-t_0)$ for $k=0$.
In order to obtain the noise-averaged propagator
$\langle S(t|t_0,j_0)\rangle$
conditioned on such {\it nonstationary} initial noise preparation
in the state $j_0$
one has to
average (\ref{pathU}) with the probability measure in (\ref{prob})
(for $k=\overline{0,\infty}$).
This task can be easily done formally by use of the
Laplace-transform [denoted in the following as $\tilde A(s):=
\int_{0}^{\infty}\exp(-s\tau) A(\tau) d\tau$ for any time-dependent
quantity $A(\tau)$]. The result for
$\langle \tilde S(s|t_0,j_0)\rangle=
\int_{0}^{\infty}\exp(-s\tau) \langle S(t_0+\tau|t_0,j_0)
\rangle d\tau$ reads \cite{Kampen79,Goychuk04}
\begin{equation}\label{result1}
\langle \tilde S(s|t_0,j_0)\rangle=\sum_i \Big
(\tilde A(s)[I-\tilde B(s)]^{-1} \Big )_{i j_{0}},
\end{equation}
where the matrix operators  $\tilde A(s)$ and
$\tilde B(s)$ reads in components
\begin{equation}\label{aux1}
\tilde A_{kl}(s):=\delta_{kl}
\int_{0}^{\infty}\Phi_l(\tau)e^{-(s+i \mathcal{ L}[\xi_{l}])\tau}
d\tau,
\end{equation}
and
\begin{equation}\label{aux2}
\tilde B_{kl}(s):=
\int_{0}^{\infty}\psi_{kl}(\tau)e^{-(s+i \mathcal{ L}[\xi_{l}])\tau}
d\tau\;,
\end{equation}
correspondingly, and $I$ is the unity matrix.

To obtain the stationary noise averaging it necessary to average
(\ref{result1}) in addition
over the stationary initial probabilities $p_{j_0}^{st}$.
The averaging over the initial distribution alone
is, however, not sufficient to arrive at the stationary noise-averaging
in the case of non-Markovian processes since the noise realizations
constructed in the way just
described  still remain  non-stationary. This principal
problem is rooted in the
following observation.
By preparing the quantum system at $t_0=0$ in a nonequilibrium
state $\rho(0)$, the noise will be
picked up at random in some initial state
 $\xi_{j_0}$ with the probability $p_{j_0}^{st}$ (stationary noise).
 However, every time when we  repeat the preparation of the quantum
 system in its initial state, the noise will
 already occupy a (random) state $\xi_{j_0}$
 for some unknown, random time interval $\tau_{j_0}^{*}$ (setting a clock at
 $t_0=0$ sets the initial time for the quantum
 system, but not for the noise, which is assumed to start in the infinite past,
 cf. Fig. \ref{Fig1}, where
 $\xi_{j_0}=\xi_1$ at $t_0=0$).
Therefore, in a stationary
setting a proper averaging over this unknown
time $\tau_j^{*}$
is necessary. The corresponding procedure
 implies that the mean residence time $\langle\tau_j\rangle$ is
finite, $\langle\tau_j\rangle\neq \infty$, and
yields a different residence time
distribution for the initial noise state, $\psi_j^{(0)}(\tau)$; namely, it is obtained as
$\psi_j^{(0)}(\tau)=\Phi_j(\tau)/\langle \tau_j\rangle$ \cite{cox}.
Only for Markovian processes where  $\Phi_j(\tau)$ is strictly
exponential, 
does $\psi_j^{(0)}(\tau)$ coincides
with $\psi_j(\tau)$.
Using this $\psi_j^{(0)}(\tau)$ instead of $\psi_j(\tau)$
for the first sojourn in the corresponding state
and for the {\it time-independent} set $p_{ij}$,
the noise realizations
become stationary \cite{cox,Goychuk04,Bursh86}. The corresponding expression
for the quantum propagator averaged over such stationary noise realizations
has been obtained in Ref. \cite{Goychuk04}, cf. Eqs. (25), (29) therein.
In a slightly modified form it reads
\begin{equation}\label{final}
\langle \tilde S(s)\rangle=\langle \tilde S(s)\rangle_{static}
-\sum_{ij} \Big
(\tilde C(s)-\tilde A(s)[I-P\tilde D(s)]^{-1} P\tilde A(s)
\Big )_{i j}\frac{p_j^{st}}{\langle \tau_j\rangle},
\end{equation}
where $\langle \tilde S(s)\rangle_{static}$ is the Laplace-transform of
the statically averaged Liouville propagator
\begin{equation}\label{static}
\langle S(\tau)\rangle_{static}:=\sum_k e^{-i \mathcal{ L}[\xi_{k}]\tau}p_k^{st},
\end{equation}
$p_j^{st}=\lim_{t\to\infty}p_j(t)$ are the stationary probabilities
which are determined by a
system of linear algebraic equations \cite{Goychuk04,Bursh86},
\begin{eqnarray}\label{eqpop}
\frac{p_j^{st}}{\langle \tau_j\rangle}=\sum_n p_{jn}\frac{p_n^{st}}
{\langle \tau_n\rangle}\;,
\end{eqnarray}
and $P$ is the matrix of transition
probabilities $p_{ij}$ (``scattering matrix'' of the random process
$\xi(t)$). Furthermore, the auxiliary matrix operators $\tilde { C}(s)$
and $\tilde { D}(s)$ in (\ref{final}) read in components:
\begin{equation}\label{aux1new}
\tilde C_{kl}(s):=\delta_{kl}
\int_{0}^{\infty}e^{-(s+i \mathcal{ L}[\xi_{l}])\tau}
 \int_{0}^{\tau}\Phi_l(\tau')d\tau'd\tau
\end{equation}
and
\begin{equation}\label{aux2new}
\tilde D_{kl}(s):=\delta_{kl}
\int_{0}^{\infty}\psi_l(\tau)e^{-(s+i \mathcal{ L}[\xi_{l}])\tau}
 d\tau \;.
\end{equation}

This very same averaging procedure can be applied to any system of
linear stochastic differential equations.

\subsubsection{\label{KuboOsc}Kubo oscillator}

A prominent application of this general procedure is the noise-averaging procedure
for the Kubo phase oscillator \cite{Kubo62,Shlesinger}; reading
\begin{equation}\label{kubo-osc}
\dot X(t)=i\epsilon[\xi(t)] X(t)\;.
\end{equation}
This particular equation emerges in the theory of optical line shapes,
in the nuclear magnetic resonance
\cite{Kubo62,Anderson53}, and for
single molecule
spectroscopy \cite{JungSilbey}. It appears also naturally
within our approach, see below, where $X(t)$ corresponds to a diagonal
matrix element of the evolution operator of a quantum system with
fluctuating eigenenergies. In the context of the stochastic theory
of spectral line shapes \cite{Kubo62,Anderson53,JungSilbey}, $\epsilon[\xi(t)]$
in Eq. (\ref{kubo-osc}) corresponds to a stochastically modulated
frequency of quantum
transitions between the levels of a ``two-state atom'', or transitions between the
eigenstates of a spin 1/2 system.

The spectral line shape is determined via
the corresponding stochastically averaged propagator of the Kubo oscillator
as \cite{Kubo62}
\begin{equation}\label{shape}
I (\omega)=\frac{1}{\pi}\lim_{\eta \to +0} {\rm Re}
[ \tilde S(-i\omega+\eta)]\;.
\end{equation}
Note that the limit $\eta\to +0$ in Eq. (\ref{shape}) is necessary
for the regularisation of the corresponding integral in the quasi-static
limit $\langle \tau_j\rangle \to \infty$.
Upon identifying $\mathcal{ L}[\xi_k]$ with $-\epsilon_{k}$ in Eq. (\ref{final})
we end up with
\begin{eqnarray}\label{propN}
\langle \tilde S(s) \rangle & = & \sum_{k}\frac{p_k^{st}}{s-i\epsilon_k}
-\sum_{k}\frac{1-\tilde \psi_k(s-i\epsilon_k)}{(s-i\epsilon_k)^2}
\frac{p_k^{st}}{\langle \tau_k\rangle }\\
\nonumber
& + &
\sum_{n,l,m}\frac{1-\tilde \psi_l(s-i\epsilon_l)}{s-i\epsilon_l}
\Big (\frac{1}{I-P\tilde D(s)}\Big )_{lm}p_{mn}
\frac{1-\tilde \psi_n(s-i\epsilon_n)}
{s-i\epsilon_n}\frac{p_n^{st}}{\langle \tau_n\rangle }\,,
\end{eqnarray}
where $\tilde D_{nm}(s)=\delta_{nm}\tilde \psi_m(s-i\epsilon_m)$.
\footnote{Note that the formal solution of another prominent problem of
the first order relaxation kinetics with a
fluctuating rate, $\dot p(t)=-\Gamma[\xi(t)] p(t)$
follows immediately from (\ref{propN}) upon substitution
$\epsilon_j\to i\Gamma_j$, see in Ref.  \cite{JCP05} for some nontrivial
non-Markovian examples and the corresponding discussion.}
The corresponding
line shape follows immediately from Eq. (\ref{propN}) by virtue of
Eq. (\ref{shape}). This result presents a non-Markovian
generalisation of the pioneering result by Kubo \cite{Kubo62} for
arbitrary $N$-state
discrete Markovian processes. The generalisation
consists in allowing for
arbitrary non-exponential RTDs $\psi_k(\tau)$, or, equivalently, in
accordance with Eq. (\ref{reduce}) also for time-dependent transition
rates $k_{ij}(\tau)$. This generalisation was put forward originally  in Ref.
\cite{Goychuk04} for a
particular case, $p_j^{st}=\langle \tau_j \rangle/
\sum_k\langle \tau_k \rangle$, which corresponds to an ergodic
process with uniform mixing (meaning that in a long time run each state $j$
is visited equally often).

Let us next
apply this result to the case of two-state
non-Markovian noise with $p_{12}=p_{21}=1$ and
$p_{1,2}^{st}=\langle \tau_{1,2}\rangle/[\langle \tau_{1}\rangle
+\langle \tau_{2}\rangle]$. Then, Eq. (\ref{propN})
yields after some simplifications:
\begin{eqnarray}\label{prop2}
\langle \tilde S(s)\rangle & = & \sum_{k=1,2}\frac{1}{s-i\epsilon_k}
\frac{\langle \tau_k \rangle}{\langle \tau_1 \rangle+
\langle \tau_2 \rangle}
 + \frac{(\epsilon_1-\epsilon_2)^2}{(\langle \tau_1 \rangle+
\langle \tau_2 \rangle)(s-i\epsilon_1)^2(s-i\epsilon_2)^2}\nonumber \\
&\times &
\frac{[1-\tilde \psi_1(s-i\epsilon_1)][1-\tilde \psi_2(s-i\epsilon_2)]}
{1-\tilde \psi_1(s-i\epsilon_1)\tilde \psi_2(s-i\epsilon_2)}\; .
\end{eqnarray}
 With (\ref{prop2}) in (\ref{shape}) one obtains the result
 for the corresponding
spectral line shape which is equivalent to one presented
recently in Ref. \cite{JungBarkSilb}
by use of a  different method. It is reproduced within our treatment as a
particular two-state limiting case. Moreover,
in the simplest case of Markovian two-state fluctuations
with $\tilde \psi_{1,2}(s)=1/(1+\langle \tau_{1,2}\rangle s)$
and with zero mean, $\langle \xi(t)\rangle =\langle \tau_1\rangle
\epsilon_1 +\langle \tau_2\rangle
\epsilon_2=0$, this
result simplifies further to  read
\begin{eqnarray}\label{Markov}
\langle \tilde S(s)\rangle =\frac{s+2\chi}{s^2+2\chi s +\sigma^2}\;.
\end{eqnarray}
In (\ref{Markov}),
$\sigma=\sqrt{\langle \xi^2(t)\rangle}=|\epsilon_2-\epsilon_1|
\sqrt{\langle \tau_1\rangle \langle \tau_2\rangle}/
(\langle \tau_1\rangle + \langle \tau_2\rangle)$ denotes the root mean
squared (rms) amplitude of fluctuations. Moreover,
$\chi=\nu/2+i\sigma\sinh(b/2)$
is a complex frequency parameter, where $\nu=1/\langle \tau_1\rangle+
1/\langle \tau_2\rangle$ is the inverse of the autocorrelation time
of the considered process
\footnote{Note that throughout this work $\nu$ is the inverse
of the autocorrelation time. It is equal to the sum of two rates.}
 which possesses the autocorrelation
function $\langle \xi(t)\xi(t')\rangle=\sigma^2\exp(-\nu|t-t'|)$.
Furthermore, $b=\ln(\langle \tau_1\rangle/
\langle \tau_2\rangle)=\ln|\epsilon_2/\epsilon_1|$
is an asymmetry parameter.
The spectral line shape corresponding to (\ref{Markov}) has been
first obtained by Kubo \cite{Kubo62}.
It reads \cite{Kubo62,PRE94},
\begin{eqnarray}\label{kuboline}
I(\omega)=\frac{1}{\pi}\frac{\sigma^2\nu}
{(\omega+\epsilon_1)^2(\omega+\epsilon_2)^2+\omega^2\nu^2}.
\end{eqnarray}
Moreover, the expression (\ref{Markov}) can be readily inverted into
the time domain.
It is crucial that the corresponding averaged propagator
$\langle S(t)\rangle $ of
Kubo oscillator \cite{PRL98}, i.e.,
\begin{eqnarray}\label{propKubo}
\langle S(t)\rangle=e^{-\chi t}\Big [\cos(\sqrt{\sigma^2-\chi^2}t)+
\frac{\chi}{\sqrt{\sigma^2-\chi^2}}\sin(\sqrt{\sigma^2-\chi^2}t)\Big],
\end{eqnarray}
is complex  when the process $\xi(t)$ is asymmetric, $b\neq 0$. This
correlates with the asymmetry of the corresponding spectral line shape,
$I(-\omega)\neq I(\omega)$.
Derived in a different form \cite{ReilSkin} (for a two-state Markovian
process with
a nonzero mean and in quite different notation)
an expression equivalent to (\ref{propKubo}) is
used in
the theory of single-molecule spectroscopy
\cite{ReilSkin,GevaSkinner,Barkai}.
For a  symmetric dichotomous
process (with $b=0$) Eq. (\ref{propKubo}) reduces to the expression
(6.10) (with $\omega_0=0$) in Ref.
\cite{KampenBook}.

\subsubsection[Averaged 
dynamics of two-level quantum systems]
{Averaged 
dynamics of two-level quantum systems \\ 
exposed to  two-state
stochastic fields \label{TLSclassic}}

The  outlined non-Markovian stochastic theory of quantum relaxation
can be exemplified for the instructive and relevant case of a two-level 
quantum system, reading
\begin{eqnarray}
\label{eq:TLS0}
H(t)=E_1|1\rangle\langle 1|+
E_2|2\rangle\langle 2|+\frac{1}{2}\hbar\xi(t)
(|1\rangle \langle 2|+|2\rangle \langle 1|),
\end{eqnarray}
which is driven by a two-state non-Markovian stochastic
field $\xi(t)=\pm \Delta$ with identical RTDs,
$ \psi_1(\tau)=\psi_2(\tau)=\psi(\tau)$. This stochastic field
causes (dipole) transitions between two states,
$|1\rangle$ and $|2\rangle$, and is zero on average.

This archetype  model exhibits a  very rich behaviour.
In particular, it allows one to
study the problem of quantum decoherence of a two-state atom
under the influence of
two-state ``$1/f^{\alpha}$''
noises exhibiting long range time-correlations
with a power law decay (for $\psi(\tau)$ possessing a long-time
algebraic tail, $\psi(\tau)\propto 1/\tau^{3-\alpha}$,
$0<\alpha<1$)\cite{LowenTeich,PRL93}.
It thus presents a prominent problem of general interest. Moreover, it relates
to activities for solid state quantum computing \cite{Wilhelm}.
It is convenient
to express the Hamiltonian (\ref{eq:TLS0}) in terms of Pauli
matrices, $\hat \sigma_z:=|1\rangle \langle 1|-|2\rangle \langle 2|$,
$\hat \sigma_x:=|1\rangle \langle 2|+|2\rangle \langle 1|$,
$\hat \sigma_y:=i(|2\rangle \langle 1|-|1\rangle \langle 2|)$ and
the unity matrix $\hat I$,
\begin{eqnarray}
\label{eq:TLS1}
H(t)=\frac{1}{2}\hbar\epsilon_0 \hat \sigma_z+\frac{1}{2}\hbar\xi(t)
\hat \sigma_x + \frac{1}{2}(E_1+E_2)\hat I,
\end{eqnarray}
where $\epsilon_0=(E_1-E_2)/\hbar$. Then,
the dynamics of the density matrix of the quantum
two-state quantum system can be given
as $\rho(t)=\frac{1}{2}[\hat I +\sum_{i=x,y,z}\sigma_i(t)\hat \sigma_i]$
in terms of a classical spin dynamics (with components  $\sigma_i(t)=
{\rm Tr}(\rho(t)\hat\sigma_i)$) in a magnetic field.
This latter dynamics evolves on a Bloch sphere of unit radius (i.e.,
the (scaled) magnetic moment is conserved,
\footnote{This means that each and every
stochastic trajectory runs on the Bloch
sphere. The {\it averaged} Bloch vector $\langle\vec \sigma (t)\rangle$
becomes, however, contracted $|\langle\vec \sigma (t)\rangle|\leq 1$,
because $\langle \sigma_i(t)\rangle^2\leq \langle \sigma_i^2(t)\rangle$.
Thus, the averaged density matrix $\langle \rho(t)\rangle$
is always positive  in the considered model, cf.
\cite{KampenBook}, independent
of the particular model used for the stochastic driving $\xi(t)$.}
$|\vec \sigma (t)|=1$). It reads,
\begin{eqnarray}
\label{eq:Bloch}
 \dot \sigma_x(t) &=& -\epsilon_0 \sigma_y(t), \nonumber \\
 \dot \sigma_y(t) &=& \epsilon_0 \sigma_x(t)-\xi(t) \sigma_z(t), \\
 \dot \sigma_z(t) &=& \xi(t) \sigma_y(t) \;.\nonumber
\end{eqnarray}

The above  theory  can  readily be applied to a noise
 averaging of 3-dimensional system of linear differential equations
(\ref{eq:Bloch}) over arbitrary {\it stationary} realizations of $\xi(t)$.
After some algebra, the following result is obtained \cite{ChemPhysHush}
for the Laplace-transformed averaged difference of populations
$\langle \sigma_z(t)\rangle=\langle \rho_{11}(t)\rangle-
\langle \rho_{22}(t)\rangle$ with the initial condition
$\sigma_z(0)=1,\sigma_{x,y}(0)=0$
, i.e., the state ``1'' is populated initially with
the probability one:
\begin{eqnarray}
\label{eq:solution}
\langle \tilde \sigma_z(s)\rangle=\frac{s^2+\epsilon_0^2}{s(s^2+\Omega^2)}-
\frac{2\Delta^2}{\tau s^2(s^2+\Omega^2)^2}\frac{\tilde A_{zz}(s)}
{\tilde B_{zz}(s)},
\end{eqnarray}
where
\begin{eqnarray}
  \label{eq:Azz}
  \tilde A_{zz}(s) & = &\epsilon_0^2 [1-\tilde\psi(s)]
\{ (\Omega^2-s^2)(1-\tilde\psi(s+i\Omega)
\tilde\psi(s-i\Omega))  \nonumber \\
& - &
2 i \Omega\, s\, [\tilde \psi(s+i\Omega)-\tilde\psi(s-i\Omega)]\}
\nonumber \\ & - & \Delta^2 s^2[1+\tilde\psi(s)]
[1-\tilde\psi(s+i\Omega)][1-\tilde\psi(s-i\Omega)],  \\
\tilde B_{zz}(s) &= &\epsilon_0^2[1-\tilde\psi(s)]
[1+\tilde\psi(s+i\Omega)]
[1+\tilde\psi(s-i\Omega)] \nonumber \\ & + &
\Delta^2  [1+\tilde\psi(s)] (1-\tilde\psi(s+i\Omega)
\tilde\psi(s-i\Omega)), \nonumber
\end{eqnarray}
and $\Omega:=\sqrt{\epsilon_0^2+\Delta^2}$.
Furthermore, $\tau$ is the mean
residence time between the field's alternations.
Note that for the considered initial condition,
$\langle \sigma_x(t)\rangle=
\langle  \sigma_y(t)\rangle=0$ for all times.
For $\epsilon_0=0$ the result in (\ref{eq:solution})-(\ref{eq:Azz})
reduces to one for
Kubo oscillator (\ref{prop2}) with identical $\psi_{1,2}(\tau)$. Moreover,
for the Markovian case, $\tilde\psi(s)=1/(1+\tau s)$, Eq.
(\ref{eq:solution}) reduces to
\begin{eqnarray}
\label{sigmaZ}
\langle \tilde \sigma_z(s)\rangle=\frac{s^2+2\nu s+ \nu^2+\epsilon_0^2}
{s^3+2\nu
  s^2+ (\Delta^2 + \epsilon_0^2 +\nu^2)s +\Delta^2 \nu},
\end{eqnarray}
where $\nu=2/\tau$ is the inverse autocorrelation time.
 This latter result reproduces
the result for the averaged populations $\langle \tilde
\rho_{11}(s)\rangle=(1/s+\langle \tilde \sigma_z(s)\rangle)/2$
and $\langle \tilde
\rho_{22}(s)\rangle=(1/s-\langle \tilde \sigma_z(s)\rangle)/2$
in \cite{AverPouq,ShapLog1}.
The same result (\ref{sigmaZ}) can also be reduced from a more
general solution
for the Markovian case with an asymmetric field of non-zero mean
 \cite{PLA96}. It possesses several remarkable
features. First, the asymptotic difference between populations is zero,
$\langle \sigma_z(\infty)\rangle=\lim_{s\to 0}(s
\langle \tilde \sigma_z(s)\rangle)=0$. In other words, the steady
state populations of both energy levels  equal  $1/2$, independently
of the energy difference $\hbar\epsilon_0$.
One can interpret this result in terms of a ``temperature''
$T_{\sigma}$ of the (quasi-)spin system. This spin-temperature is formally introduced
by using for the  asymptotic distribution an Ansatz of the Boltzmann-Gibbs
form,
$\langle\rho_{nn}(\infty)\rangle = \exp[-E_n/k_BT_{\sigma}]/
\sum_{n}\exp[-E_n/k_BT_{\sigma}]$.
Then, \footnote{This is a standard definition
of the temperature of a spin subsystem in nuclear magnetic
resonance and similar areas  \cite{Slichter}. It
is used also to introduce the parlance of formally
{\it negative} temperatures.}
\begin{eqnarray}\label{Tsigma}
T_{\sigma}:=\frac{\hbar\epsilon_0}
{k_B\ln\left(\frac{\langle\rho_{22}(\infty)\rangle}
{\langle\rho_{11}(\infty)\rangle} \right)}\;
\end{eqnarray}
for two-level systems.
In accord with this definition, the result of equal asymptotic populations,
$\langle\rho_{22}(\infty)\rangle=\langle\rho_{11}(\infty)\rangle=1/2$
can be interpreted in terms of an infinite temperature $T_{\sigma}=\infty$.
This constitutes a general finding: a purely stochastic bath corresponds to an apparent infinite
temperature \cite{ReinekerBook,Lindenberg}. For this reason, such stochastic
approaches to describe the relaxation
process in open quantum systems is suitable only
for sufficiently
high temperatures $k_BT\gg \hbar|\epsilon_0|$ \cite{ReinekerBook,Lindenberg}.
An asymmetry of unbiased stochastic perturbations
does not change this  conclusion, see in Ref. \cite{PLA96}.
Moreover, the relaxation to the steady state can be
either coherent, or incoherent,
depending  on the noise strength and the autocorrelation time. In particular,
an approximate analytical expression for the rate $k$ of incoherent
relaxation, $\langle\rho_{11}(t)\rangle=[1+\exp(-k t)]/2$,
has been obtained in a limit of small Kubo numbers, $K:=\Delta/\nu\ll 1$,
which corresponds to a weakly coloured noise
\cite{KampenBook,FrischBrissaud}.
This analytical result reads \cite{AverPouq,ShapLog1,PLA96}
\begin{equation}\label{resonance}
k=\frac{\Delta^2\nu}{\nu^2+\epsilon_0^2}
\end{equation}
and exhibits a resonance feature versus $\nu$ at $\nu=\epsilon_0$.
A similar such resonance feature
occurs also in the theory of nuclear magnetic resonance for a weakly coloured
Gaussian noise \cite{Slichter}. Note that in \cite{PLA96}
this notable
result has been obtained for asymmetric fluctuations of the tunnelling
coupling possessing a non-vanishing mean
value $\langle \xi(t)\rangle \neq 0$.
This corresponds to  a quantum particle transfer between
two sites of localisation which are separated by
a fluctuating tunnelling barrier. A related  problem with the inclusion of
dissipation has been  elaborated in \cite{JCP95} within a stochastically
driven spin-boson model.

Yet another interesting solution can be  obtained for $\langle \tilde
\sigma_x(s)\rangle$ with the initial condition  reading $\sigma_x(0)=1$.
The Laplace transform of the solution is obtained as
\begin{eqnarray}
\label{eq:solution2}
\langle \tilde \sigma_x(s)\rangle=\frac{s^2+\Delta^2}{s(s^2+\Omega^2)}-
\frac{2\Delta^2\epsilon_0^2\Omega^2}{\tau s^2(s^2+\Omega^2)^2}
\frac{\tilde A_{xx}(s)}
{\tilde B_{xx}(s)},
\end{eqnarray}
where
\begin{eqnarray}
  \label{eq:Axx}
  \tilde A_{xx}(s) & = &[1-\tilde\psi(s)]
[1-\tilde\psi(s+i\Omega)][1-\tilde\psi(s-i\Omega)],  \\
\tilde B_{xx}(s) &= &\epsilon_0^2[1+\tilde\psi(s)][1-
\tilde\psi(s+i\Omega)]
[1-\tilde\psi(s-i\Omega)] \nonumber \\ & + &
\Delta^2  [1-\tilde\psi(s)] (1-\tilde\psi(s+i\Omega)
\tilde\psi(s-i\Omega)). \nonumber
\end{eqnarray}
The physical relevance of this solution (\ref{eq:solution2}) is as follows.
In a rotated quasi-spin basis, i.e.,
$\hat \sigma_x\to \hat \sigma_{z'}$,
$\hat \sigma_z\to \hat \sigma_{x'}$, $\hat \sigma_y\to \hat
\sigma_{y'}$, the considered problem becomes mathematically
equivalent to the problem
of the delocalisation of a quantum particle in a symmetric dimer with
the tunnelling coupling $\epsilon_0$ under the influence of a
dichotomously fluctuating energy bias $\xi(t)$.
Therefore, it
describes the corresponding delocalisation dynamics and, in
particular, allows one to determine whether this dynamics is
coherent or incoherent,
depending on the noise features.

For the Markovian case Eq. (\ref{eq:solution2}) reduces to
\footnote{The corresponding
dynamics also exhibits a resonance feature versus $\nu$ in a certain
limit \cite{AnkerPech}.}

\begin{eqnarray}
\label{sigmaX}
\langle \tilde \sigma_x(s)\rangle=\frac{s^2+\nu s +\Delta^2}
{s^3+\nu
  s^2+ (\Delta^2 + \epsilon_0^2)s +\epsilon_0^2 \nu}.
\end{eqnarray}
Note that the denominators in Eq. (\ref{sigmaZ}) and Eq. (\ref{sigmaX})
are different. \footnote{A remarkable feature is, however, that
both corresponding secular cubic equations have the
{\it same} discriminant,
$D(\Delta,\nu,\epsilon_0)=0$,
separating the domains of complex and real roots. Hence,
the transition from a coherent relaxation
(complex roots are present) to an incoherent relaxation
(real roots only) occurs
at the same values of noise parameters,
independently of the initial conditions.
The corresponding phase diagram separating regimes of coherent
and incoherent relaxation (judging from the above criterion)
has been found in \cite{AnkerPech}.
It must be  kept in mind, however, that the weights of the corresponding
exponentials are also of importance for the character of relaxation process.
These weights naturally do depend  on the initial conditions.}
In a more general case of asymmetric Markovian noise,
the corresponding denominator is
 a polynomial of 6th order in $s$, see in \cite{PLA96}.
In the considered case of symmetric noise it factorises
into the product of two  polynomials of 3d order,
those in the denominators of Eq. (\ref{sigmaZ}) and Eq. (\ref{sigmaX}).
Thus, for a general initial condition the relaxation of a two-level
quantum system exposed to a two-state Markovian field
involves generally 6th-exponential terms. As a matter of fact,
this seemingly simple, {\it exactly} solvable model can exhibit
an unexpectedly complex behaviour even in the simplest Markovian case
of a coloured noise driving.
However, for certain initial
conditions, as exemplified above, the general solution being a fraction of two polynomials
of $s$  simplifies to the results in Eq. (\ref{sigmaZ}) and Eq. (\ref{sigmaX}).

In a general case of non-Markovian noise, the analytical solutions
in Eqs. (\ref{eq:solution}) and (\ref{eq:solution2}) can be inverted
to the time domain numerically by use  of a  numerical Laplace inversion procedure
such as the one detailed in Ref. \cite{Stehfest}.

\subsection{\label{primer}Projection operator method: a primer}

Next we shall
introduce the readers, following Ref. \cite{Kenkre}, into  the projection operator
technique. We elucidate this scheme by addressing an  example that is
of physical interest in its own right.

Let us consider the somewhat more general dynamics,
\begin{eqnarray}
\label{Bloch2}
\frac{d\vec \sigma(t)}{dt}:=\left ( \begin{array}{c}
\dot \sigma_x(t)\\
\dot \sigma_y(t)\\
\dot \sigma_z(t)
\end{array} \right ) =
\left ( \begin{array}{ccc}
0            & -\epsilon(t) & 0 \\
\epsilon(t) & 0           &  -\Delta(t) \\
0            & \Delta(t)  & 0
\end{array} \right )
\left ( \begin{array}{c}
\sigma_x(t)\\
\sigma_y(t)\\
\sigma_z(t)
\end{array} \right ) :=
\hat B(t)\vec \sigma(t) ,
\end{eqnarray}
and let us pose the question: How can we extract a single, closed equation for the
evolution of $\sigma_z(t)$
(without any approximation) for an arbitrary time-dependence of the parameters
governing the driven quantum dynamics?
The use of a projection operator method provides an elegant way
to solve this problem \cite{Kenkre}. The key idea is to project the
whole dynamics onto a corresponding subspace of  reduced dimensionality by
using a projection operator
$\mathcal{ P}$ with the idempotent property $\mathcal{ P}^2=\mathcal{ P}$.
In the present case, the choice of this projection operator follows naturally as:
\begin{eqnarray}
\mathcal{ P} \left ( \begin{array}{c}
\sigma_x(t)\\
\sigma_y(t)\\
\sigma_z(t)
\end{array} \right ) =
\left ( \begin{array}{c}
0 \\
0 \\
\sigma_z(t)
\end{array} \right ):=\vec\sigma_0(t)
\;\;.
\end{eqnarray}
 The use of this projection operator allows one to
split  the whole dynamics into the ``relevant'' one
$\vec \sigma_0(t)$, and a remaining,
``irrelevant'', $\vec \mu(t)$, part,  respectively; i.e.,
$\vec \sigma(t)\equiv  \mathcal{ P}
\vec \sigma(t)+ (1-\mathcal{ P}) \vec \sigma(t):=
\vec \sigma_0(t)+ \vec \mu(t)$ by applying $\mathcal{ P}$
and the complementary projection operator $1-\mathcal{ P}$
 to Eq. (\ref{Bloch2}). From the resulting system of two
coupled linear equations for $\vec \sigma_0(t)$ and $\vec \mu(t)$, i.e.,
\begin{eqnarray}
 \frac{d\vec\sigma_0(t)}{dt} & = &
 \mathcal{ P}\hat B(t)\vec \sigma_0(t) +\mathcal{ P}\hat B(t)\vec \mu(t),
 \nonumber \\
\frac{d\vec \mu(t)}{dt} & = &
 (1-\mathcal{ P})\hat B(t)\vec \sigma_0(t) +(1-\mathcal{ P})\hat B(t)\vec \mu(t),
\end{eqnarray}
a single integro-differential equation for $\vec \sigma_0(t)$ follows, reading:
\begin{eqnarray}
\label{eq:exact1}
 \frac{d\vec\sigma_0(t)}{dt} & = & \mathcal{ P}\hat B(t)\vec\sigma_0(t)
\nonumber \\
 & + & \int_{0}^{t} \mathcal{ P}\hat B(t)\mathcal{ T}\exp\left(
\int_{t'}^{t}d\tau
(1-\mathcal{ P})
\hat B(\tau)\right)(1-\mathcal{ P})\hat B(t')
\vec\sigma_0(t') d t'\nonumber \\
& + &
\mathcal{ P}\hat B(t)\mathcal{ T}\exp\left(
\int_{0}^{t}d\tau (1-\mathcal{ P})\hat B(\tau)\right)
\vec\mu(0)\;.
\end{eqnarray}
The exponential matrix operations in (\ref{eq:exact1})
can be done explicitly
without any approximation, yielding the
{\it exact} closed
equation for $\sigma_z(t)$  \cite{Kenkre}, reading
\footnote{Within the path-integral approach,  the same
equation can be derived  from a non-interacting
blip approximation (NIBA) result of the dissipative spin-boson model
\cite{GrifPaladWeiss}
by putting formally therein the strength of the system-bath coupling
to zero.  Astonishingly enough, the NIBA
turns out to provide the exact result for this
 singular limit of zero-dissipation.}
\begin{eqnarray}
\label{eq:exact2}
\dot \sigma_z(t)& = &-\int_{0}^{t}\Delta(t)\Delta(t')\cos[\zeta(t,t')]
\sigma_z(t')dt' \nonumber \\
& + &\Delta(t)\sin[\zeta(t,0)]\sigma_x(0)+
\Delta(t)\cos[\zeta(t,0)]\sigma_y(0)\;.
\end{eqnarray}
In Eq. (\ref{eq:exact2}), the time-dependent phase
\begin{eqnarray}
\label{phase}
\zeta(t,t')& = &\int_{t'}^{t}\epsilon(\tau)d\tau\;\;
\end{eqnarray}
is introduced which is a functional of the
time-varying parameter $\epsilon(t)$.
The projection of the entire dynamics onto the
some subspace typically entails memory effects. Put differently,
a nonlocality
in time emerges for the reduced space dynamics.
Moreover, an explicit dependence on the initial conditions in the
"irrelevant" subspace is necessarily present.

\section{\label{period}Two-state quantum dynamics in periodic fields}

Let us illustrate the practical usefulness of the {\it exact} equation
(\ref{eq:exact2}) by its application to a quantum dynamics
occurring in strong time-periodic fields. Towards this goal, we consider a
quantum two-state tunnelling system,
where the two states $|1\rangle$ and $|2\rangle$ correspond
to the two sites of charge localisation (i.e. we  work in the "tunnelling" representation)
and $\Delta=const$ corresponds
to the tunnelling matrix element. This charge
dynamics is driven by a periodic electric field of frequency $\Omega$
which results
in a periodic modulation of the energy bias between two localised
states $\epsilon(t)=\epsilon_0+A\cos(\Omega t)$ \cite{GrifHangRev}. We assume that
the particle is prepared initially on the site ``1'' at $t=0$, i.e.
$\sigma_z(0)=1$
and $\sigma_x(0)=\sigma_y(0)=0$. Further, one assumes that the
frequency of external field is rather high, $\Omega\gg \Delta$ and
we consider the correspondingly averaged dynamics
$\langle \sigma_z(t)\rangle_{\Omega}$ using the high-frequency
decoupling approximation $\langle\cos[\zeta(t,t')]
\sigma_z(t')\rangle_{\Omega}\approx \langle\cos[\zeta(t,t')]
\rangle_{\Omega}\langle
\sigma_z(t')\rangle_{\Omega}$ \cite{CPL96}. Using Eq. (\ref{phase}) and
the well-known identity
$\exp(iz\sin\theta)=\sum_{n=-\infty}^{\infty}J_n(z)\exp(in\theta)$
($J_n(z)$ denotes  the Bessel function of the first kind)
the high-frequency approximation in Eq. (\ref{eq:exact2}) yields
\begin{eqnarray}
\label{eq:high-freq}
\langle \dot \sigma_z(t)\rangle_{\Omega} = -\Delta^2
\int_{0}^{t}\Gamma(t-t') \langle \sigma_z(t')\rangle_{\Omega}dt'
\end{eqnarray}
with the kernel $\Gamma(t)=\sum_{n=-\infty}^{\infty} J^2_n(A/\Omega)
 \cos[(\epsilon_0+n\Omega)t]$. This latter equation can be solved by
the use of the Laplace-transform. For
$\tilde\sigma_z(s):=
\int_0^t\exp(-st) \\
\langle \sigma_z(t)\rangle_{\Omega} dt$
one obtains
\begin{eqnarray}
\label{eq:high-freq-Laplace}
\tilde\sigma_z(s) = \frac{1}{s}\;\;\frac{1}{1+\Delta^2
\sum_{n=-\infty}^{\infty  }\frac{J_n^2(A/\Omega)}
{s^2+(\epsilon_0-n\Omega)^2}}\;\;.
\end{eqnarray}
From this relation now follow  some  key-results.

\subsection{Coherent destruction of tunnelling (CDT)}

The formal inversion of the result in Eq. (\ref{eq:high-freq-Laplace})
into the time domain reads
\begin{eqnarray}
\langle \sigma_z(t)\rangle_{\Omega}=\sum_{j=0,\pm 1,\pm 2,...}
c_j\exp(i\omega_jt),
\end{eqnarray}
where $\omega_j=is_j>0$ are  the poles $s_j$
of Eq. (\ref{eq:high-freq-Laplace}). From the quasi-periodic character
of the driven dynamics it follows that all these poles
lie on the
imaginary axis in complex conjugated pairs.
Therefore, $\omega_{-j}=-
\omega_{j}$ and
$c_{-j}=c_j^*$. Although there appears an infinite number
of poles, only few of them contribute significantly in the regimes of
interest.

Let us consider  the case of symmetric two-level system (TLS), 
$\epsilon_0=0$.  Then,
the approximate solution reads
(only the term
with $n=0$ in the sum in Eq. (\ref{eq:high-freq-Laplace}) contributes
significantly in the high-frequency limit $\Delta/\Omega\to 0$)
\begin{eqnarray}\label{tunosc}
\langle \sigma_z(t)\rangle_{\Omega}=\cos(\Delta_{tun}t),
\end{eqnarray}
where $\Delta_{tun}=\Delta |J_{0}(A/\Omega)|$ is the renormalised
tunnelling frequency. When the amplitude $A$ of the high-frequency driving
is chosen to obey $J_{0}(A/\Omega)=0$, the tunnelling dynamics is  brought
(within this high frequency approximation)
to a complete standstill. This constitutes the celebrated phenomenon of
coherent destruction of tunnelling \cite{GrossHang,GrossHang2} which attracted much
attention and generated many applications over recent years,
see e.g. in \cite{GrifHangRev,KohlerRev}
and references therein.

\subsection{Driving-induced tunnelling oscillations (DITO)}

\begin{figure}
\centerline{\epsfxsize=6.5cm
\epsfbox{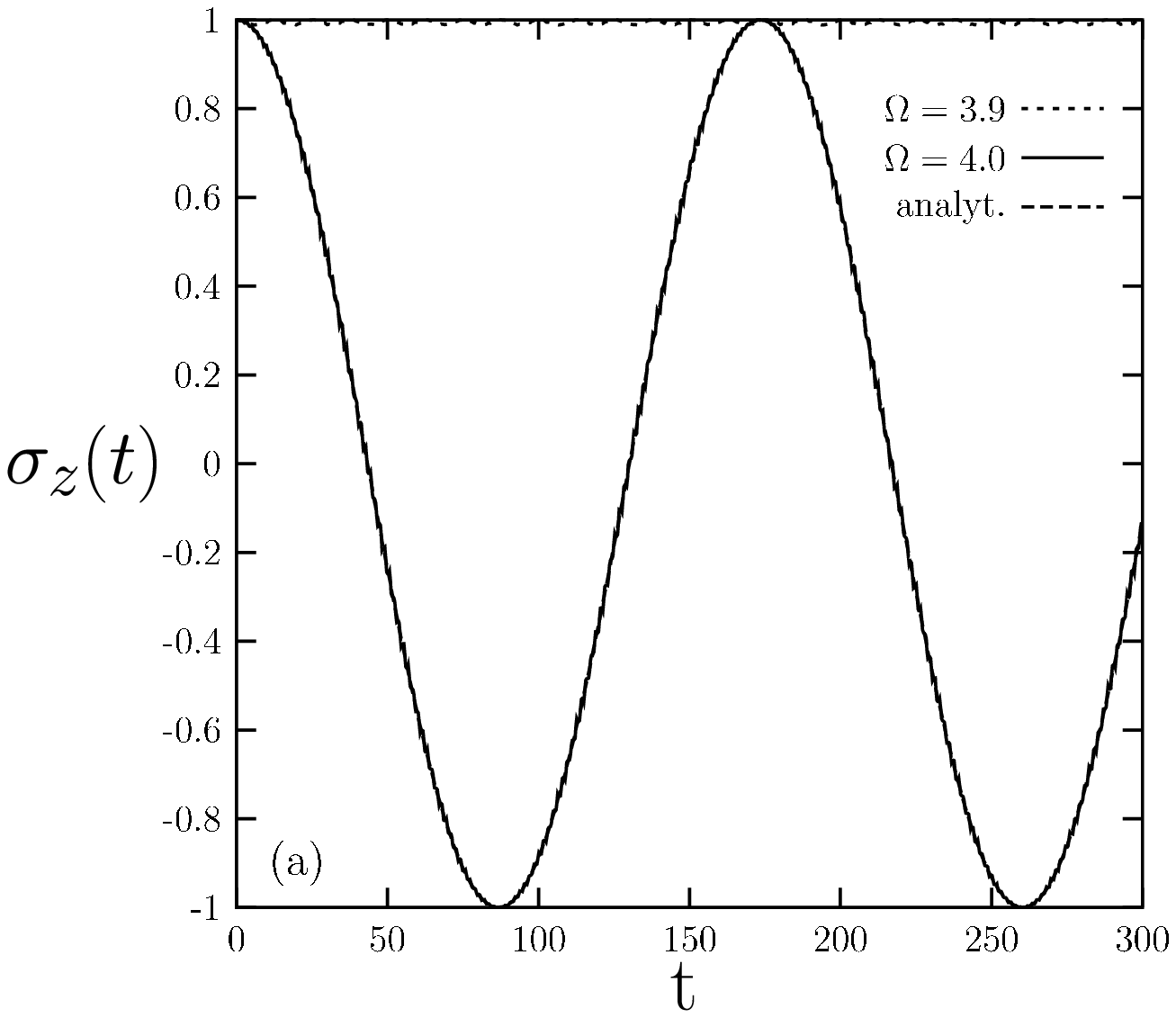}
\epsfxsize=6cm
\epsfbox{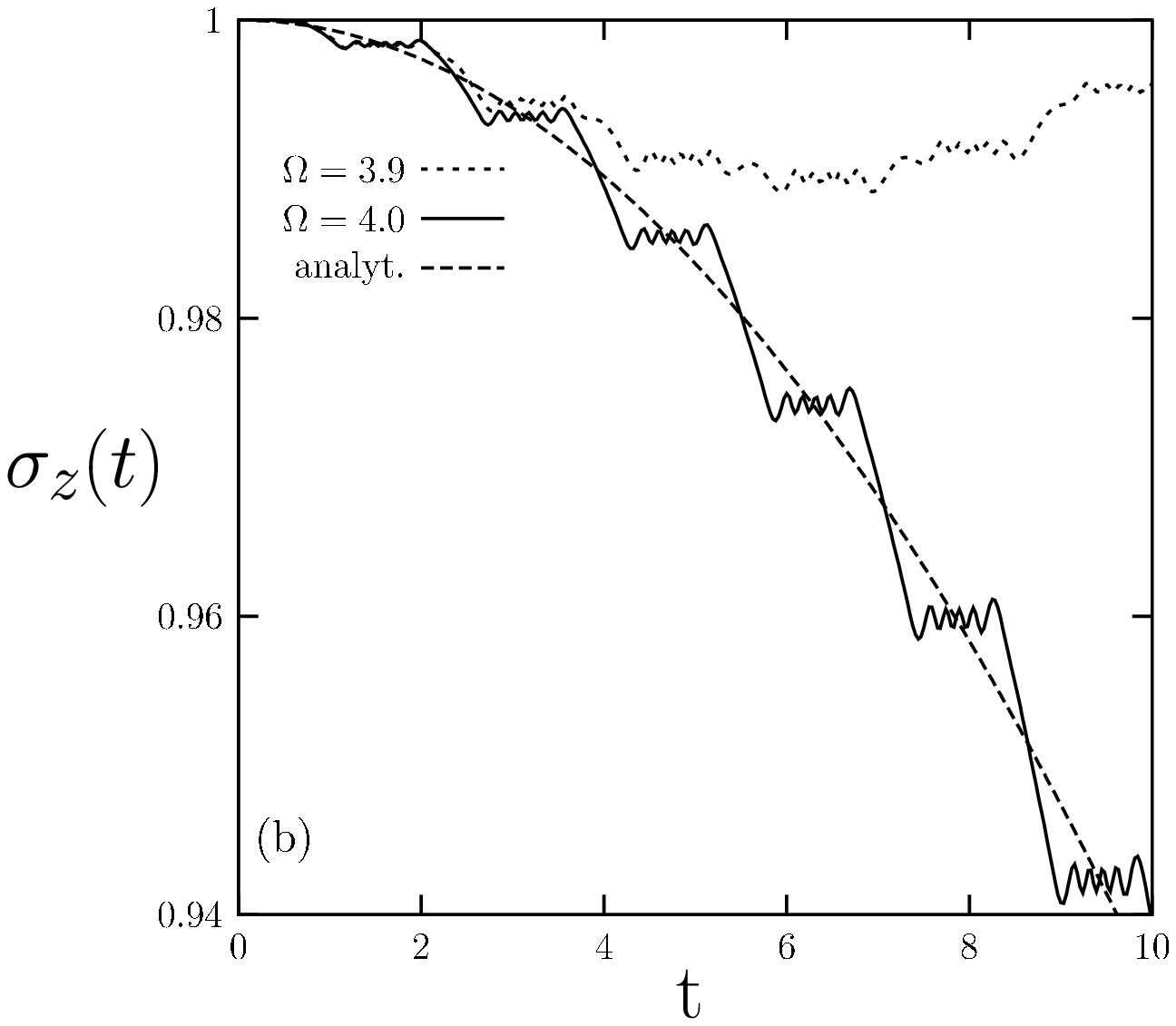}}
\caption{Driving-induced tunnelling oscillations. }
\label{DITOfig}
\end{figure}

Let us consider now the case of a large energy bias
$\epsilon_0\gg \Delta$.
In the absence of driving, the particle remains essentially localised
on the site ``1'', $\sigma_z(t) \approx 1$,
as can be deduced from the well-known exact solution
$\sigma_z(t)=[\epsilon_0^2+
\Delta^2\cos(\sqrt{\epsilon_0^2+\Delta^2}t)]/
[\epsilon_0^2+\Delta^2]$.  When, however, a high-frequency driving
$\Omega\gg \Delta$
is applied such that the resonance condition $n\Omega=
\sqrt{\epsilon_0^2+\Delta^2}\approx
\epsilon_0$
is approximately fulfilled,  large amplitude tunnelling oscillations
in Eq. (\ref{tunosc}) can be induced with a tunnelling frequency
$\Delta_{tun}=\Delta |J_{n}(A/\Omega)|$.  This phenomenon of
driving-induced tunnelling oscillations (DITO) -- being opposite
to CDT has been revealed in
Refs. \cite{Hartmann98,PRE2000,NakamuraPashkinPRL}.
It has recently
 been verified and observed  experimentally \cite{NakamuraPashkinPRL}.
This DITO-phenomenon is illustrated
in Fig. \ref{DITOfig} for an ``exotic'' 5-photon ($n=5$) resonance case
where the precise numerical solution of driven TLS dynamics using Eq.
(\ref{Bloch2})
and the approximation in Eq.
(\ref{tunosc}) with $\Delta_{tun}=\Delta |J_{n}(A/\Omega)|$
are plotted for the following set of parameters: $\Delta=0.1$,
$\epsilon_0=20$, $A=24$, $\Omega=3.9$ or $\Omega=4.0$.
For $\Omega=3.9$ the dynamics is almost localised
exhibiting small-amplitude
oscillation --
cf. dotted line near $\sigma_z(t)=1$ which is barely visible in
Fig. \ref{DITOfig}, a
but becomes clearly seen in Fig. \ref{DITOfig}, b due to a better
resolution on a different scale.
A relatively small change of the periodic field
frequency chosen to match
the resonance condition $5\Omega=\epsilon_0$ induces
large-amplitude tunnelling oscillations which are nicely described on
the long-time scale by the approximation in Eq. (\ref{tunosc})
with $\Delta_{tun}=\Delta|J_5(A/\Omega)|$. It
cannot be distinguished from the precise numerical solution
in Fig. \ref{DITOfig}, a. The frequency of these oscillations
is controlled by both the bare tunnelling frequency $\Delta$ and the
field amplitude $A$. It is worth
mentioning that DITO seems to be close in spirit to the famous Rabi
oscillations \cite{Rabi} (an interpretation given
in Ref. \cite{NakamuraPashkinPRL}), but are in  fact by no means
identical with
those. Rabi oscillations correspond usually to a particular case of
a small-amplitude, $A\ll \Omega$, resonant
driving, $\Omega=\sqrt{\Delta^2+\epsilon_0^2}$. For $\epsilon_0=0$,
the corresponding problem is equivalent
(in a rotated quasi-spin basis,
$\hat \sigma_x\to \hat \sigma_{z'}$,
$\hat \sigma_z\to \hat \sigma_{x'}$, $\hat \sigma_y\to \hat
\sigma_{y'}$, and with different initial conditions)
to the resonant dipole excitation
of a two-state atom with eigen-frequency $\omega_0=\Delta$.
In such a case, the frequency of Rabi oscillations
$\omega_R$ is determined approximately by the driving {\it amplitude}, i.e.
$\omega_R\approx A$ \cite{GrifHangRev,HanggiBookChap}.
This presents the most remarkable,
characteristic feature of Rabi oscillations.
The DITO-frequency  presents rather a driving-renormalised tunnelling frequency
as in the case of CDT.
The coarse-grained character of the result in Eq. (\ref{tunosc})
is illustrated in Fig. \ref{DITOfig}, b on a short-time scale
in comparison with the precise
numerical solution of the driven dynamics. This latter one
exhibits step-like
transitions with a number of oscillations on each step.
The number of oscillations corresponds to the number of
emitted (absorbed) photons. With the increase of $n$, the ``steps''
become longer and sharper.
In order to make a further ``step'' in the transfer
of population the two level system awaits for the next portion of $n$
photons to be emitted, or absorbed to match the resonance
condition $n\Omega=\epsilon_0$ (a quasiclassical
interpretation of the numerically observed step feature).

\section[Dissipative quantum dynamics in time-dependent fields]
{\label{dissip}Dissipative quantum
dynamics in strong \\ time-dependent fields}

\subsection{General formalism}

Without loss of generality  we consider a $N$-level, driven quantum system  characterised
by a time-dependent Hamilton operator $H_S(t)$ and which interacts $V_{SB}$ with
a thermal bath characterised by a Hamilton operator $H_B$. The
system-bath interaction is assumed here to be generally also   time-dependent. It is
characterised by the Hamilton operator $V_{SB}(t)$ which depends both
on the (relevant) variables of the system of interest and on the thermal bath variables.
The total Hamiltonian $H(t)$ thus reads
\begin{eqnarray} \label{Htot}
H(t)= H_S(t)+ V_{SB}(t) + H_B\;.
\end{eqnarray}
The dynamics of the density operator $\rho(t)$
of the total system is then governed by the corresponding
Liouville-von-Neumann equation, cf. Eq. (\ref{Liouville}).
Furthermore, the reduced density operator of interest is obtained by performing
a partial trace of $\rho(t)$ over the bath variables, i.e.
$\rho_S(t)={\rm Tr}_B \rho (t)$.  The
average $<A>$ of any operator $A$ which depends on the variables
of the system of interest
can  be calculated as the corresponding trace over the system
variables, i.e.,
 $<A>= {\rm Tr}_S (\rho_S(t)A)$. The  reduced density
 operator $\rho_S(t)$ , which also depends on the initial preparation scheme, thus
contains all the necessary information required to describe the
time-evolution
of the system of interest.  The main task consists in obtaining a tractable closed equation of motion
for $\rho_S(t)$. This can be achieved by applying to $\rho(t)$
a properly chosen projection operator $\Pi$, which projects the whole
dynamics onto  the subspace of the considered quantum system, thereby accounting indirectly for the
"irrelevant" bath variables, i.e.
$\rho_S(t)=\Pi \rho(t)$.
A  proper choice for the projection operator with the
idempotent property, $\Pi^2=\Pi$, is $\Pi:=\rho_B{\rm Tr}_B$
\cite{Zwanzig,Nakajima, ArgyresKelley},
where $\rho_B=\exp(-\beta H_B)/Z_B$ is the equilibrium
density operator of the bath; $Z_B={\rm Tr}_B\exp(-\beta H_B)$ is the
corresponding partition sum, and $\beta=1/(k_BT)$ denotes the inverse
temperature.
Then, $\rho(t)$ can identically be split as
$\rho(t)\equiv \rho_B \otimes\rho_S(t)+
\eta(t)$, where $\eta(t)=Q\rho(t)$ represents a cross-correlation term.
Here, $Q:=1-\Pi$ is the complementary
projection operator with the properties $Q\Pi=\Pi Q=0, Q^2=Q$.
By applying $\Pi$ and $Q$ to the Liouville-von-Neumann equation for $\rho(t)$,
 two coupled linear operator equations for $\rho_S(t)$ and $\eta(t)$
can be obtained, respectively, which in turn  yield a single closed equation for $\rho_S(t)$
after having eliminated the part $\eta(t)$. The  formally {\it exact} equation for
the reduced density operator, thus reads
\begin{eqnarray}\label{GMEexact}
\dot \rho_S(t)=-iL_S(t)\rho_S(t)-\int_0^t
\Gamma(t,t')\rho_S(t')dt' + I_0(t),
\end{eqnarray}
where
\begin{eqnarray}\label{exact_kernel}
 \Gamma(t,t')={\rm Tr}_B[ L_{SB}(t)  S_{S+B}(t,t')Q
L_{SB}(t')\rho_B]
\end{eqnarray}
denotes the memory kernel. In Eq. (\ref{exact_kernel}),
\begin{eqnarray}\label{SSB}
 S_{S+B}(t,t')=\mathcal{
T}\exp\{-i\int_{t'}^t[L_S(\tau)+L_B+QL_{SB}(\tau)]d\tau\}
\end{eqnarray}
is a Liouvillian  propagator. Furthermore,
$L_{S}(t)(\cdot)=[\tilde H_{S}(t),(\cdot)]/\hbar$,
$L_{B}(\cdot)=[H_{B},(\cdot)]/\hbar$,
$L_{SB}(t)(\cdot)=[\tilde V_{SB}(t),(\cdot)]/\hbar$ are the corresponding
Liouville operators,
where $\tilde H_S(t):=H_S(t)+<~V_{SB}(t)~>_B$ is the
renormalised Hamiltonian of the dynamical system and
$\tilde V_{SB}(t):=V_{SB}(t)-<~V_{SB}(t)>_B$ is the correspondingly
re-defined system-bath coupling.
\footnote{This issue deserves to be commented on in further detail: The generalised
quantum thermal forces acting on the system from  the bath
should be  on average unbiased. This implies that
the {\it thermal} average $<...>_B:={\rm Tr}_B(\rho_B...)$ of
a properly defined system-bath coupling,
$<\tilde V_{SB}(t)>_B:={\rm Tr}_B (\rho_B \tilde V_{SB}(t))$, should  be
zero,  i.e. $<\tilde V_{SB}(t)>_B=0$.
For this reason,
the systematic, {\it mean-field}  like contribution $<V_{SB}(t)>_B$
of the thermal ``force''  should be separated from the
very beginning and be included
in $\tilde H_S(t)$ without change of the Hamiltonian of the
{\it total} system. Obviously, this can {\it always} be achieved.
This formal
renormalisation is {\it always} assumed in the following (with
``tilde''
omitted when applicable).}
Moreover, $I_0(t)$ in Eq. (\ref{GMEexact})
\begin{eqnarray}\label{initial}
I_0(t)=-i {\rm Tr_B}\left ( L_{SB}(t)S_{S+B}(t,0)\mu(0)\right)
\end{eqnarray}
constitutes the initial correlation term, sometimes also termed "initial value term".

Note that the  GME (\ref{GMEexact})-(\ref{initial}) is still exact in
the subspace of the quantum
system for a quantum evolution started at $t_0=0$, i.e. no
approximations have been invoked so far
\cite{GrabertBook,Haake,FonsecaTalknerHanggi}.
Generally, a reduced quantum evolution contains some dependence on
the initial conditions
$\mu(0)$ in the irrelevant subspace.
We note, however, that for a factorised (uncorrelated)
initial preparation $\rho(0)=\rho_B\otimes \rho_S(0)$ ($\mu(0)=0$)
this initial correlation term vanishes identically, i.e. $I_0(t)=0$.
This standard  class of initial preparations will be  assumed in the following.

\subsubsection{Weak-coupling approximation}

In the second order approximation with respect to the system-bath
coupling $V_{SB}(t)$ (the so termed weak-coupling limit) one sets $L_{SB}(t)\to 0$
in $S_{S+B}(t,t')$, Eq. (\ref{SSB}). Moreover, let us assume a factorising
form of the system-bath coupling
$V_{SB}(t)=\frac{1}{2}
\sum_{\alpha} \kappa_{\alpha}(t)\hat \gamma_{\alpha} \hat
\xi_{\alpha} + h.c.$
where $\hat \gamma_{\alpha}$ denote some system operators,
$\hat \xi_{\alpha}$ are the bath operators, and $\kappa_{\alpha}(t)$
are the coupling strength functions. The complete set $\hat \gamma_{\alpha}$
is assumed to be closed under  commutation relations:
$[\hat \gamma_{\alpha},\hat \gamma_{\beta}]=\sum_{\delta}
\epsilon_{\alpha \beta \delta}\hat \gamma_{\delta}$ with
$\epsilon_{\alpha \beta \delta}$ being some structural constants
defining a corresponding Lie algebra
with generators $\hat \gamma_{\alpha}$. The Hamiltonian
$H_S(t)$ is represented as a linear superposition
\[
H_S(t)=\frac{1}{2}\sum_{\alpha}b_{\alpha}(t)
\hat \gamma_{\alpha}+h.c.
\]
in the corresponding algebra.
For N-level quantum systems the following set of operators
is  conveniently  used. It is given (here $\alpha:=(n,m)$)
by the set of operators
$\hat \gamma_{nm}:=|n\rangle \langle m|$ (they correspond
to the elements of corresponding Liouville space). Here
the the ket-vectors $|n\rangle$ provide an
orthonormal vector
basis with the scalar product $\langle n|m\rangle=\delta_{nm}$
in the corresponding
Hilbert space of the considered N-level quantum system.
The representation of the system Hamiltonian in this discrete
 basis reads
\begin{eqnarray}\label{HD}
H_S(t)=\sum_{nm} H_{nm}(t) \hat \gamma_{nm},
\end{eqnarray}
with  $H_{nm}(t)=H_{mn}^{*}(t)$.
It is evident that any quantum
system with a discrete number of states
can be represented in this way. The system-bath coupling
can be chosen in the form
\begin{eqnarray}\label{coupling}
V_{SB}(t)=\sum_{nm} \kappa_{nm}(t) \hat \gamma_{nm} \hat \xi_{nm},
\end{eqnarray}
with $\kappa_{mn}(t)=\kappa_{nm}^*(t)$ and
$\hat \xi_{mn}=\hat \xi_{nm}^{\dagger}$.
Moreover, the dissipative operator kernel in Eq. (\ref{GMEexact})
 reads in the
given approximation:
\begin{eqnarray}\label{kernel}
 \Gamma (t,t')(\cdot) = \sum_{n,n',m,m'}
 &\kappa_{nn'}&(t)\kappa_{mm'} (t')
\Big \{
K _{nn'mm'}(t-t')
\Big[\hat\gamma_{nn'},S(t,t')\hat\gamma_{mm'}(\cdot)
\Big]  \nonumber \\
 & - &K_{n'nm'm}^*(t-t')\Big[\hat\gamma_{nn'},
 S(t,t')(\cdot)\hat\gamma_{mm'} \Big]
\Big\},
\end{eqnarray}
where
\begin{eqnarray}
S(t,t')=\mathcal{ T}\exp\{-i\int_{t'}^t L_S(\tau)d\tau\}
\end{eqnarray}
is the Liouville evolution
operator of the physical system under consideration. Note that it does  include the external, time-dependent field
influences exactly. Moreover,
\begin{eqnarray}\label{autocorr}
K_{nn'mm'}(t):=\frac{1}{\hbar^2}< \hat
\xi_{nn'}(t)\hat \xi_{mm'}>_B=K_{m'mn'n}^*(-t),
\end{eqnarray}
is the autocorrelation tensor  of the thermal force operators
$\hat \xi_{nn'}(t):=e^{iH_Bt/\hbar}\\
\hat \xi_{nn'}e^{-iH_Bt/\hbar}$.
An expression formally similar to Eq. (\ref{kernel})  has been obtained
first, for a particular case of a spin 1/2 system (and for a time-independent
system-bath coupling)
in Ref. \cite{ArgyresKelley}.  For the reduced density matrix,
$\rho_{nm}(t):=\langle n | \rho_S(t) |m\rangle$ the following
generalised master equation (GME)
follows:
\begin{eqnarray}\label{matrixGME}
\dot \rho_{nm}(t)=-i\sum_{n'm'}L_{nmn'm'}(t)\rho_{n'm'}(t)-
\sum_{n'm'}\int_0^t
\Gamma_{nmn'm'}(t,t')\rho_{n'm'}(t')dt',
\end{eqnarray}
where $L_{nmn'm'}(t)=\frac{1}{\hbar}[
H_{nn'}(t)\delta_{mm'}-H_{m'm}(t)\delta_{nn'}]$ is the Liouville
superoperator, written  in the supermatrix representation
and the corresponding memory kernels read
\begin{eqnarray}\label{tetradicKern}
\Gamma_{nmn'm'}(t,t')=\sum_{kk'}\Big\{& &\kappa_{nk'}(t)
\kappa_{kn'}(t') K_{nk'kn'}(t-t')
U_{k'k}(t,t')U_{mm'}^*(t,t')\\ \nonumber+ &&
\kappa_{k'm}(t) \kappa_{m'k}(t') K_{mk'km'}^*(t-t')
U_{nn'}(t,t')U_{k'k}^*(t,t')\\ \nonumber- &&
\kappa_{nk'}(t) \kappa_{m'k}(t')
K_{k'nkm'}^*(t-t')
U_{k'n'}(t,t')U_{mk}^*(t,t')\\- &&
\kappa_{k'm}(t) \kappa_{kn'}(t')
K_{k'mkn'}(t-t')
U_{nk}(t,t')U_{k'm'}^*(t,t')\nonumber
\Big\} ,\;
\end{eqnarray}
where $U_{mm'}(t,t'):=\langle m| \mathcal{ T}
\exp\{-\frac{i}{\hbar}\int_{t'}^t H_S(\tau)d\tau\} |m'\rangle$ is
the evolution operator of the considered quantum system in the Hilbert space.
This result presents the most general form
of weak-coupling GME in arbitrary external fields. Generalised master
equations of a
similar form have been derived before, by making use of different methods and in
different notations
in Refs. \cite{Petrov86,PRE94,PRE96}~\footnote{For a periodic driving, the field influence on the relaxation
kernel can alternatively be taken into account
applying the corresponding
Floquet basis for periodically driven quantum dynamics.
This has been done in Refs.
\cite{Dittrich93,Kohler}. Our approach is valid, however,
for arbitrary time-dependence.}.
The kernel (\ref{tetradicKern}) satisfies two important properties which
must be strictly obeyed; these are (i):
\[
\Gamma_{nmn'm'}(t,t')=\Gamma_{mnm'n'}^{*}(t,t')
 \]
 (imposed by
the requirement that $\rho_S(t)$ must be  Hermitian, $\rho_S(t)=
\rho^{\dagger}_S(t)$), and (ii):
\[
\sum_n \Gamma_{nn n'm'}(t,t')=0
\]
(conservation of probability,
${\rm Tr}_S\,
\rho_S(t)=1$ for all times).

This driven
GME (\ref{matrixGME}) presents our "working horse" that will be
used frequently for the discussion of various  applications detailed below.

\subsubsection[Markovian approximation: 
Generalised Redfield Eqs.]{Markovian approximation: \\
Generalised Redfield Equations}

The integro-differential equations
(\ref{matrixGME})-(\ref{tetradicKern})
are, notably, non-local in time, i.e., they
 describe  a so-termed  ``non-Markovian'' quantum dynamics.
This  non-locality in time
makes their practical use rather cumbersome. A corresponding Markovian
approximation, which renders a description that is local in time,  is therefore  of great use in practice,
if it can be justified on physical grounds.
There are several ways to obtain such a Markovian approximation.
The most popular one is to perform a back propagation, i.e.,
$\rho_S(t') = S^{-1}(t,t')\rho_S(t)+O(\kappa^2)$,
in the kernel of GME making use of the Liouville evolution operator $S(t,t')$
of the dynamical subsystem. The corresponding master equation for the
reduced density matrix, which constitutes the generalisation of
the well-known Redfield equations \cite{Redfield} to the case of driven, open quantum systems
reads:
\begin{eqnarray}    \label{Redfield}
\dot \rho_{nm}(t)=-i\sum_{n'm'}L_{nmn'm'}(t)\rho_{n'm'}(t)-\sum_{n'm'}
R_{nmn'm'}(t)\rho_{n'm'}(t),
\end{eqnarray}
with the explicit {\it time-dependent}  relaxation tensor
\begin{eqnarray}\label{RRedfield}
R_{nmn'm'}(t)
 =  \sum_{kk'}\nonumber \\
 \times  \int_0^t \Big\{\sum_l&[&
\kappa_{nl}(t)\kappa_{kk'}(t')K_{nlkk'}(t-t')
U_{lk}(t,t')U_{n'k'}^*(t,t')\delta_{mm'}\nonumber \\ \nonumber
+ &&\kappa_{lm}(t)\kappa_{kk'}(t') K_{mlk'k}^*(t-t')
U_{m'k}(t,t')U_{lk'}^*(t,t')\delta_{nn'}]\\- &&
\kappa_{nn'}(t)\kappa_{kk'}(t') K_{n'nk'k}^*(t-t')
U_{m'k}(t,t')U_{mk'}^*(t,t')\nonumber \\- &&\kappa_{m'm}(t)
\kappa_{kk'}(t')
 K_{m'mkk'}(t-t')
U_{nk}(t,t')U_{n'k'}^*(t,t')
\Big\}dt'\,\,.\nonumber \\
\end{eqnarray}
This relaxation tensor  satisfies two important relations, namely,
$R_{nmn'm'}(t) = R_{mnm'n'}^{*}(t)$ ($\rho_S(t)$ is
Hermitian), and
$\sum_n R_{nn n'm'}(t)=0$ (imposed by the conservation of probability).
Notable, the upper limit of integral in (\ref{RRedfield})
is here the actual evolution time $t$ (instead of $\infty$) -- this feature
softens already
the well-known problem  which relates to a possible
violation of positivity at initial time scales of the quantum evolution \cite{Alicki} for  Redfield
equations  for certain  initial
conditions\footnote{This problem can be
resolved by the so-called slippage of the initial conditions, see in
\cite{SuarSilbOppen,GnutzHaake,GaspardNagaoka}.
Moreover, within the weak-coupling
approximation the effect of dissipation should be consistently taken into account
to the second order of the system-bath coupling  only; i.e., in the solutions of Redfield equations
(rather than in the relaxation kernels only).
The dissipation-induced frequency shifts (i.e. the Lamb shifts at $T=0$)
should also be
very small (compared to the corresponding eigen-frequencies of
quantum evolution in the absence of dissipation). Otherwise,
the theory needs to be  renormalised.  Notwithstanding these
essential restrictions, the Redfield equations provide one of
the most widely used tool
to describe open quantum systems in many areas of physics and physical chemistry
\cite{GrifHangRev,KohlerRev,Slichter,MayBook,Blum,PollardFriesner,MaySchreiber,EgorovaDomcke}.
}.
Upon neglecting (setting to zero)  the influence of external time-dependent fields in
 the relaxation tensor, by using the basis of {\it eigen-states} of
{\it time-independent} $H_S$, and setting $t\to\infty$
in (\ref{RRedfield}), we recover the commonly known form of the Redfield relaxation tensor.

It must be stressed that the physical nature
of the thermal bath operators was up to now still  not specified.
Those can be either be of bosonic,  fermionic nature, or also describe a spin
bath \cite{Shao}.
The corresponding autocorrelation tensor (\ref{autocorr}) has to be calculated
for every particular microscopic model.  We next address within this methodology  several
physical applications.

\section[Application I: Quantum relaxation  in 
two-level systems]
{Application I: Quantum relaxation  in \\
driven, dissipative  two-level systems\label{TLS}}

Let us consider a  two-level quantum system with time-dependent
eigenenergy levels 
\footnote{These levels
can correspond, e.g., to spatially separated localisation sites of a
transferring (excess) electron in a protein \cite{PetrovBook}. If
such electronic states possess very different dipole moments (the
difference can reach $50$ D \cite{Lockhart}), an external
time-dependent electric field will modulate the energy difference in
time due to the Stark effect. Such an electric field dependence of
the electronic energy levels can  be very strong
\cite{Lockhart,Bixon}. A large modulation of the local electric
field can be induced, e.g., due attachment/detachment of an ATP
molecule/products of its hydrolysis.
 A substantial shift of the electronic energy levels can then be
induced  \cite{Kurnikov}.
In a simple setting, the corresponding modulation
of an energy level
can be modelled  by a two-state Markovian
process  \cite{AstumianScience}.
The chemical source
of driving force can also be substituted by a direct application
of a stochastic electric
field \cite{Tsong,AstumianScience}. This latter possibility
has been demonstrated experimentally
for some ion pumps \cite{Tsong}.},
\begin{eqnarray}
H_S(t)=[E_1^{(0)}+\tilde E_1(t)]|1\rangle\langle 1|+
[E_2^{(0)}+\tilde E_2(t)]|2\rangle\langle 2|,
\end{eqnarray}
which is coupled  to a bath of independent
harmonic oscillators, possessing the spectrum $\{\omega_{\lambda}\}$, i.e.,
\begin{eqnarray}
H_B=\sum_{\lambda}
\hbar\omega_{\lambda}(b^{\dagger}_{\lambda}b_{\lambda}+\frac{1}{2}),
\end{eqnarray}
where $b^{\dagger}_{\lambda}$ and $b_{\lambda}$ are the bosonic creation
and annihilation operators, respectively.
The  interaction
with the thermal bath
causes the relaxation transitions between the eigenstates of
the dynamical system (``longitudinal'' interaction). Such
transitions are
absent otherwise and thus require either the emission, or the absorption of bath phonons.
The interaction is chosen to be of a  bi-linear form, reading
\begin{eqnarray}
V_{SB}=
 \hat \xi (|1\rangle\langle 2|+|2\rangle\langle 1|)\, ,
\end{eqnarray}
with the influence of the bath being presented by a {\it random force operator},
$\hat \xi_{12}=\hat \xi_{21}^{\dag}=\hat \xi$,
\begin{eqnarray}\label{xi}
\hat \xi=\sum_{\lambda}\kappa_{\lambda}
(b^{\dagger}_{\lambda}+b_{\lambda})\,.
\end{eqnarray}
Here and  elsewhere below, the (real-valued)
coupling constants $\kappa$'s
are included
into the fluctuating force $\hat \xi$. From a phenomenological perspective,
the considered model represents an analogue of the model in Sec.
{\ref{TLSclassic}}, where a classical random force assuming two values
is replaced by a quantum operator force which possesses a Gaussian statistics.
Moreover, a possible time-dependence of the energy levels is assumed here.

The corresponding  correlation function of this Gaussian quantum stochastic force
$K(t):=K_{1221}(t) = <\hat\xi_{12}(t)\hat\xi_{12}^{\dag}(0)>$
is complex-valued. It  reads explicitly
\begin{eqnarray}
\label{Kfunc}
K(t)=\frac{1}{2\pi}
\int_0^{\infty}J(\omega)[\coth(\frac{\hbar\omega}{2k_BT})\cos(\omega t)-
i\sin(\omega t)]d\omega.
\end{eqnarray}
with the bath spectral density given by
\begin{eqnarray}
\label{Jnm}
J(\omega):=\frac{2\pi}{\hbar^2}\sum_{\lambda}\kappa_{\lambda}^2
\delta(\omega-\omega_{\lambda}).
\end{eqnarray}
Upon extending to negative frequencies  $\omega<0$, it is convenient to formally define
$J(-\omega):=-J(\omega)$.

The complex nature of this bath correlation function  is crucial for the establishment of
thermal equilibrium at the finite temperatures.
The application of GME (\ref{matrixGME}), (\ref{tetradicKern}) to
the present case yields a closed system
of generalised master equations for the level populations
populations $p_n(t):=\rho_{nn}(t), n=1,2$, reading,
\begin{eqnarray}
\label{balance}
\dot p_1(t)& = & -\int_{0}^{t} [
w_{12}(t,t')p_1(t')- w_{21}(t,t')p_2(t') ]dt',
\nonumber \\
 \dot p_2(t)& = & \,\,\,\,\,\int_{0}^{t} [ w_{12}(t,t')p_1(t')-
 w_{21}(t,t')p_2(t') ]dt',
\end{eqnarray}
with memory kernels \footnote{One can immediately
see that if $K(t)$ would be real-valued, then the forward and backward
rate kernels were always equal. This would mimic the situation of an
infinite temperature as it was elucidated in Sect. 2 for a classical stochastic bath.}
\begin{eqnarray}\label{w12}
w_{12}(t,t') & = &2{\rm Re} [ K(t-t') \exp(i\epsilon_0(t-t')+i\tilde
\zeta(t,t')],\nonumber \\
w_{21}(t,t') & = &2 {\rm Re} [ K(t-t') \exp(-i\epsilon_0(t-t')-i\tilde \zeta(t,t')],
\end{eqnarray}
where $\epsilon_0=(E_{1}^{(0)}-E_{2}^{(0)})/\hbar$ and
$\tilde \zeta(t,t')$ is a functional of time-dependent driving,
Eq. (\ref{phase}), with
$\tilde\epsilon(t)=[\tilde E_1(t)-\tilde E_2(t)]/\hbar$.
We shall assume that $\tilde \epsilon(t)$ fluctuates (either randomly,
or periodically in time) around a zero mean value. In order to  obtain the quantum
relaxation averaged over the fluctuations of $\tilde \epsilon(t)$ one needs
to perform a corresponding stochastic averaging of the GME (\ref{balance}).
For an arbitrary stochastic process $\tilde \epsilon(t)$, this task
cannot be carried out exactly any longer; consequently
 one must resort to some  approximation scheme(s).

\subsection{Decoupling approximation for fast  fluctuating energy levels}

If the characteristic time
scale for $\tilde \epsilon(t)$ fluctuations $\tau_{\epsilon}$ is
very small in comparison with the characteristic system relaxation time scale
$\tau_r$, i.e. $\tau_{\epsilon}\ll \tau_r$, then
one can use a decoupling approximation by averaging for
$\langle p_{1,2}(t)\rangle_{\epsilon}$; namely,
$\langle \exp(\pm i\tilde\zeta(t,t')p_{1,2}(t')\rangle_{\epsilon}\approx
\langle \exp(\pm i\tilde\zeta(t,t')\rangle_{\epsilon}
\langle p_{1,2}(t')\rangle_{\epsilon}$. For  fast fluctuations of
the energy levels
the relaxation dynamics then follows
$\langle p_{1,2}(t)\rangle_{\epsilon}$ with  fast, superimposed
small-amplitude fluctuations whose amplitude
diminishes when the ratio $\tau_{\epsilon}/\tau_r \ll 1$  becomes smaller.
A subsequent Markovian approximation for the averaged dynamics yields
a master equation description of the form:
\begin{eqnarray}\label{fastB}
 \langle \dot p_{1}(t)\rangle_{\epsilon}& =&
-\langle W_{12}(\epsilon_0)\rangle_{\epsilon}
\langle p_{1}(t)\rangle_{\epsilon}+
\langle W_{21}(\epsilon_0)\rangle_{\epsilon}
\langle p_{2}(t)\rangle_{\epsilon}, \nonumber \\
 \langle \dot p_{2}(t)\rangle_{\epsilon}& =&
 \langle W_{12}(\epsilon_0)\rangle_{\epsilon}\langle p_{1}(t)\rangle_{\epsilon}-
\langle W_{21}(\epsilon_0)\rangle_{\epsilon} \langle p_{2}(t)\rangle_{\epsilon}
\end{eqnarray}
with the averaged transition rates reading \cite{PRE94},
\begin{eqnarray}\label{rates}
\langle W_{12}(\epsilon_0)\rangle_{\epsilon} & = &\int_{-\infty}^{\infty} e^{\frac{\hbar\omega}{k_BT}}
n(\omega)J(\omega)I(\epsilon_0-\omega)d\omega,\nonumber \\
\langle W_{21}(\epsilon_0)\rangle_{\epsilon} & = &\int_{-\infty}^{\infty}
n(\omega)J(\omega)I(\epsilon_0-\omega)d\omega \; .
\end{eqnarray}
Here $n(\omega)=1/[\exp(\hbar\omega/(k_BT))-1]$ is the Bose function,
and $I(\omega)$ is the spectral line shape of a Kubo
oscillator $\dot X(t)=i\tilde \epsilon(t) X(t)$ (see in Sect. 2).
From Eqs. (\ref{fastB}), (\ref{rates}) one can immediately deduce that in the
absence of fluctuations, where $I(\omega)=\delta(\omega)$,
 the thermal
equilibrium, $p_1(\infty)=e^{-\hbar\epsilon_0/k_BT}p_2(\infty)$,
is attained independently of the specific model for
$J(\omega)$ for arbitrary temperatures $T$.  Moreover,
the thermal detailed balance condition, $p_2(\infty)W_{21}(\epsilon_0)=
p_1(\infty)W_{12}(\epsilon_0)$ is obeyed always with the
thermal bath temperature $T$.
In other words, the
temperature $T_{\sigma}$ of the considered TLS,
defined through Eq.
(\ref{Tsigma}) coincides with the temperature of the thermal bath $T$,
$T_{\sigma}=T$.
This is in a sharp contrast to the stochastic bath modelling in Sec.
\ref{TLSclassic},
where we found that $p_1(\infty)=p_2(\infty)$ and $T_{\sigma}=\infty$.
Furthermore, one can  see that the thermal equilibrium at the
bath temperature $T$ becomes
violated either by periodic, or by stochastic nonequilibrium
fluctuations. This also implies that,
generally, $T_{\sigma}\neq T$.
Put differently, either periodic, or stochastic (thermally nonequilibrium)
fields  drive the system
{\it out of the thermal equilibrium} with the thermal bath.
This fact lies at the heart  for the emergence of
a diversity of interesting and often counter-intuitive nonequilibrium
effects which we shall address next.

\subsubsection{Control of quantum rates}

\begin{figure}

\centerline{\epsfxsize=6.5cm
\epsfbox{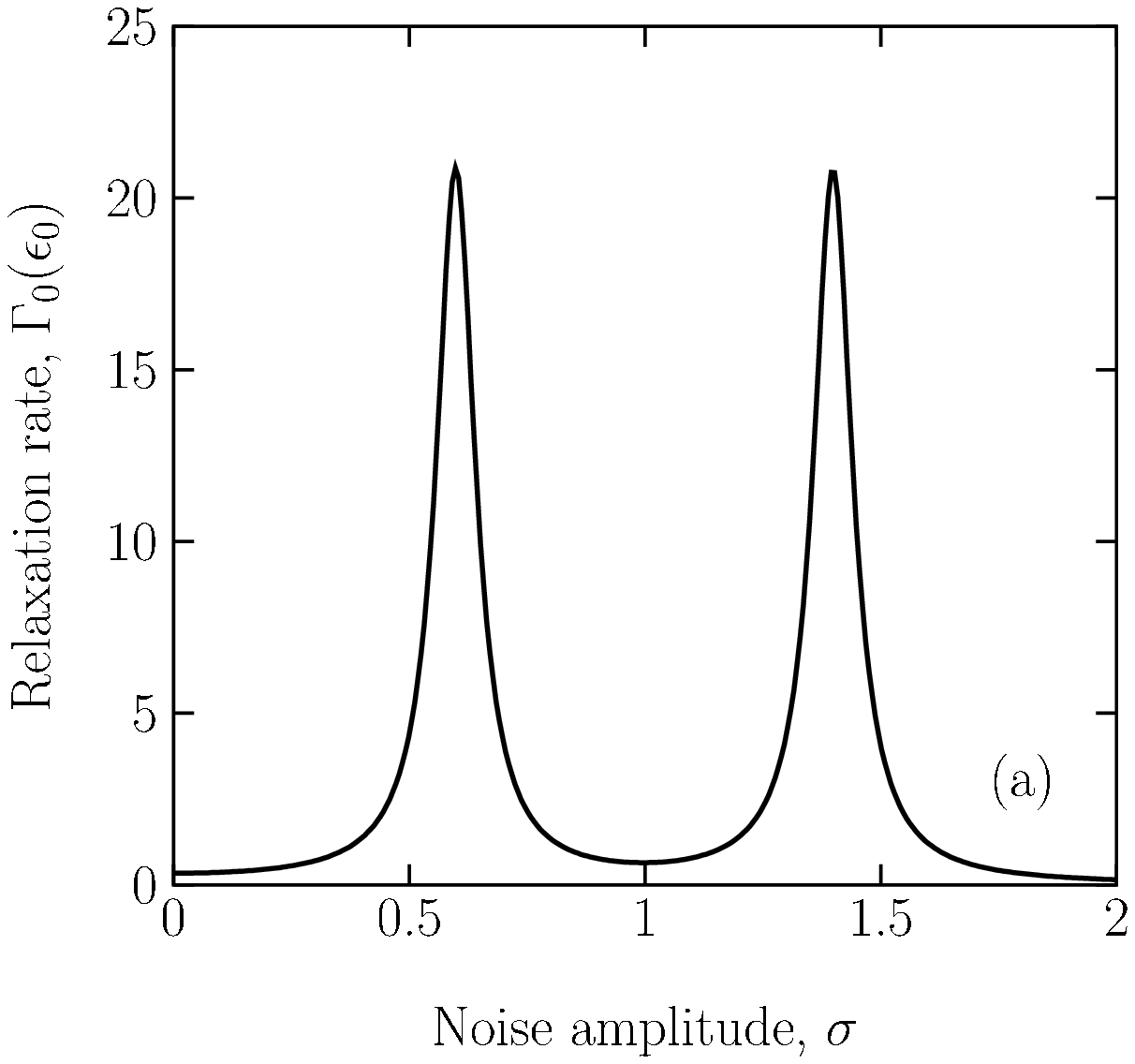}
\epsfxsize=6.5cm
\epsfbox{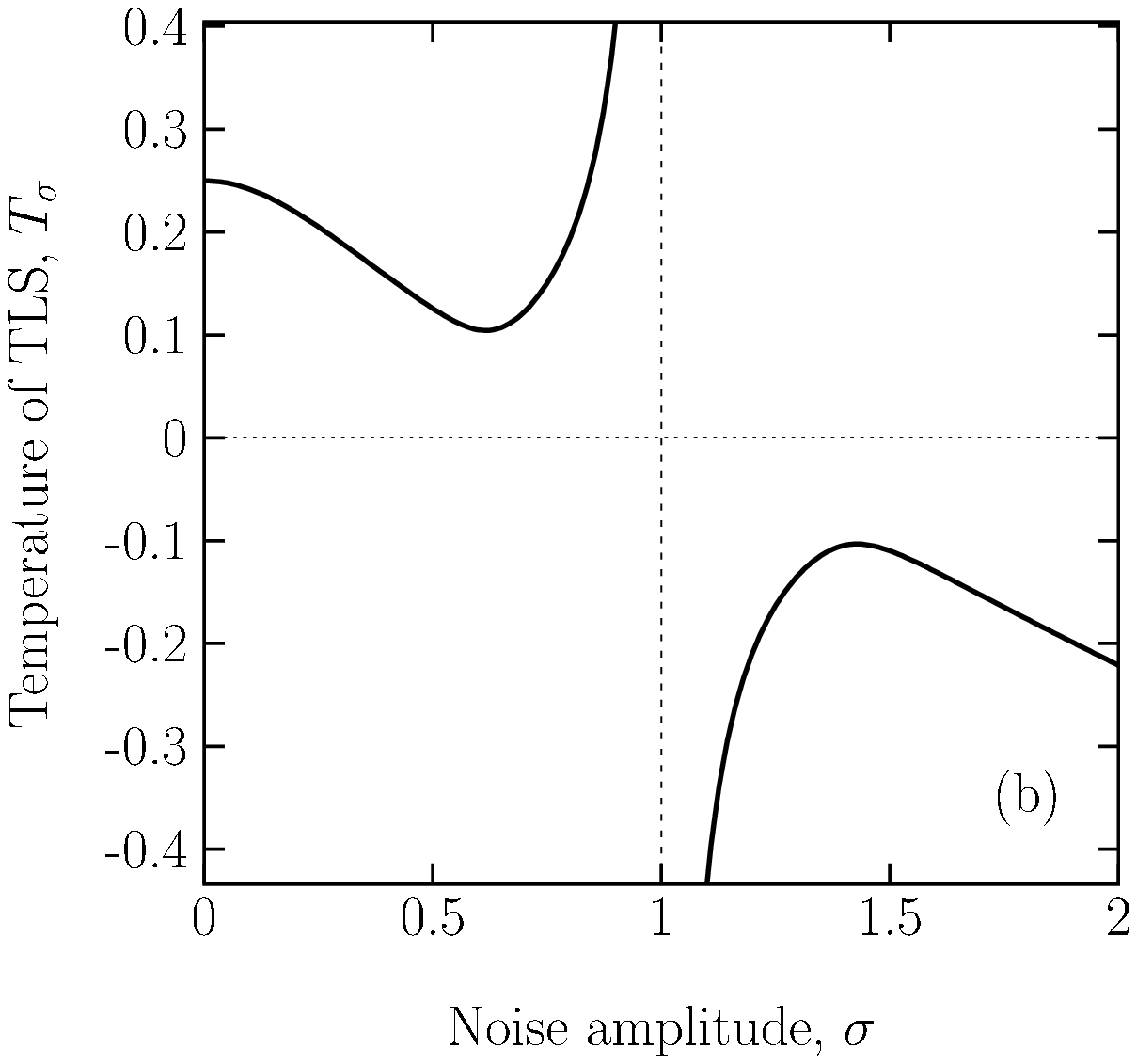}}
\caption{(a) The averaged relaxation rate of
TLS, $\Gamma_0(\epsilon_0)$, in  units of
$\kappa_0^2/(\hbar^2\Omega_0)$, is depicted versus the noise amplitude $\sigma$
(in units of $\Omega_0$) for the averaged energy bias
$\epsilon_0=0.4\Omega_0$ and the thermal bath temperature $T=0.25
\hbar\Omega_0/k_B$ and $\gamma=0.05\Omega_0$.
(b) The effective temperature of TLS, $T_{\sigma}$, in units of
$\hbar\Omega_0/k_B$ versus the noise amplitude $\sigma$ (in units of
$\Omega_0$) for the same set of parameters of TLS and the environmental
temperature $T$.
At $\sigma\approx 0.6\Omega_0$,
where the TLS is maximally cooled, the lower level is populated with
the probability being close to one. On the contrary, for $\sigma
\approx 1.4 \Omega_0$ the upper level becomes populated with the
probability being close to one; i.e.,  an almost complete inversion of populations
occurs. The model assumptions are well justified for the coupling
constant $\kappa_0\ll
0.05\hbar\Omega_0$ such that $\Gamma_0(\epsilon_0)\ll \gamma$.}
\label{Fig2}
\end{figure}

A first application is the
manipulation of the transition rates by many orders of magnitude by use of a
rapidly fluctuating, discrete
stochastic fields \cite{PetTes90,UFZ93,PRE94,PRE96}.
This scheme becomes  feasible when
the spectral
density of the bath $J(\omega)$ is sharply peaked around some vibrational
frequencies. The effect can be demonstrated for a quantum Brownian
oscillator model
of the bath: It corresponds to a single quantum vibrational
mode $\Omega_0$
which acquires a frictional
broadening width $\gamma$ due to a bi-linear coupling to other environmental
vibrational modes\footnote{
A fast  (on the time scale  $\tau_r$ of system relaxation) 
equilibration of this single mode with other
vibrational  modes is assumed. This imposes an important restriction
$\tau_r^{-1}:=\langle W_{12}(\epsilon_0)\rangle_{\epsilon}+
\langle W_{21}(\epsilon_0)\rangle_{\epsilon}\ll \gamma$
which can always be justified by a proper tuning of the coupling constant
$\kappa_0$. Furthermore, the
broadening of vibrational spectral lines in molecular systems
$\gamma$ exceeds typically
$\gamma>5$ cm$^{-1}$ (in spectroscopic units) which corresponds
$\gamma>10^{12}$ Hz in  units of the frequency. The considered
relaxation transitions must consequently occur more slowly, e.g., an electron
transfer can occur on a msec time scale \cite{PetrovBook}.}.
The corresponding spectral density
assumes the form \cite{Garg}:
\begin{eqnarray}\label{BrownOsc}
J(\omega)=\frac{8\kappa_0^2}{\hbar^2}
\frac{\gamma\Omega_0\omega}{(\omega^2-\Omega_0^2)^2+4\gamma^2\omega^2}.
\end{eqnarray}
Let us start out by considering first a control scenario of  quantum relaxation by use of
a symmetric dichotomous Markovian noise
$\tilde \epsilon(t)=\pm \sigma$ with $I(\omega)$
given by (\ref{kuboline}), where
$\epsilon_{1,2}=\pm \sigma$. For the
case\footnote{This case
presents a realistic situation for
experimental realizations of molecular systems
since a significant stochastic perturbation with an
energy exceeding one $k_BT$,
$\hbar\sigma\sim 25\;{\rm meV}$ (at room temperatures),  
corresponds in the units of
frequency $\sigma\sim 4\times 10^{13}$ 1/s.
The frequency $\nu$ of large
amplitude bistable conformational fluctuations of molecular
groups is typically much smaller.} $ \nu \ll \sigma$,
this spectral line shape consists of two sharply shaped peaks, located at
$\omega=\pm \sigma$ and possessing the width $\nu$. For
$\nu\ll \gamma$, which is typically the case,
this latter broadening can be neglected. Then,
$I(\omega)\approx \frac{1}{2}[\delta(\omega-\sigma)+
\delta(\omega+\sigma)]$, and the averaged rates simplify to read
\begin{eqnarray}
\label{rates:aver}
\langle W_{12}(\epsilon_0)\rangle_{\epsilon} & \approx &
\frac{1}{2}[W_{12}(\epsilon_0+\sigma)+W_{12}(\epsilon_0-\sigma)]
\nonumber \\ & = &
\frac{4\kappa_0^2\gamma\Omega_0}{\hbar^2}\Big(
\frac{(\epsilon_0-\sigma) e^{\frac{\hbar(\epsilon_0-\sigma)}{k_BT}}
n(\epsilon_0-\sigma)}
{[(\epsilon_0-\sigma)^2-\Omega_0^2]^2+4\gamma^2(\epsilon_0-\sigma)^2}
\nonumber \\
 & + &
\frac{(\epsilon_0+\sigma) e^{\frac{\hbar(\epsilon_0+\sigma)}{k_BT}}
n(\epsilon_0+\sigma)}
{[(\epsilon_0+\sigma)^2-\Omega_0^2]^2+4\gamma^2(\epsilon_0+\sigma)^2}
\Big )\\
\langle W_{21}(\epsilon_0)\rangle_{\epsilon} & \approx &
\frac{1}{2}[W_{21}(\epsilon_0+\sigma)+W_{21}(\epsilon_0-\sigma)]
\nonumber \\ & = &
\frac{4\kappa_0^2\gamma\Omega_0}{\hbar^2}\Big(
\frac{(\epsilon_0-\sigma)
n(\epsilon_0-\sigma)}
{[(\epsilon_0-\sigma)^2-\Omega_0^2]^2+4\gamma^2(\epsilon_0-\sigma)^2}
\nonumber \\
   & + &  \frac{(\epsilon_0+\sigma)n(\epsilon_0+\sigma)}
{[(\epsilon_0+\sigma)^2-\Omega_0^2]^2+4\gamma^2(\epsilon_0+\sigma)^2}
\Big )\,.\nonumber
\end{eqnarray}
From Eq. (\ref{rates:aver}) it follows
that if the quantum transition
frequency $\epsilon_0$ matches the vibrational frequency of
the medium $\Omega_0$ an increase of energy fluctuation size $\sigma$
(induced by
local electric field fluctuations in the medium)
from zero to some finite value
$\sigma\gg\gamma$  can drastically reduce the relaxation rate
$\Gamma_0(\epsilon_0)=\langle W_{12}(\epsilon_0)\rangle_{\epsilon}+
\langle W_{21}(\epsilon_0)\rangle_{\epsilon}$; even a practical blockade of
  relaxation transitions can occur \cite{PetTes90,UFZ93,PRE94,PRE96}.
On the contrary, in the case of a frequency mismatch between $\epsilon_0$
and $\Omega_0$ one in turn can dramatically enhance the rate of relaxation
transitions by tuning
the noise amplitude $\sigma$ appropriately \cite{PRE96},
see in Fig. \ref{Fig2}, a.

\subsubsection{Stochastic cooling and inversion of level populations}

A second effect relates to the blockage of the rate for backward
transitions $\langle W_{21}(\epsilon_0)\rangle_{\epsilon}$
 relative to
the forward rate $\langle W_{12}(\epsilon_0)\rangle_{\epsilon}$.
This can cause a  {\it stochastic cooling} of the TLS, where the temperature
$T_{\sigma}$  becomes {\it smaller} than the
temperature of the environment, i.e., $T_{\sigma}<T$. This
interesting phenomenon is demonstrated with Fig. \ref{Fig2} (b).
Similar in spirit, although different in the
physical mechanism is the laser
cooling (of the nuclear degrees of freedom) as it has been studied both, theoretically and experimentally for
polyatomic molecules \cite{Pollak}.

Moreover, for $\sigma>\Omega_0$
a noise-induced inversion of steady state averaged populations takes
place, i.e.,
for a sufficiently small positive energy
bias $\epsilon_0$ the higher energy level
becomes more populated. This constitutes the third very
important effect under discussion, see
Fig. \ref{Fig2}, b, where this pumping effect is
interpreted in terms of a {\it negative}
temperature $T_{\sigma}$.
 In other words, the considered nonequilibrium noise of a sufficiently large
amplitude is capable to pump quantum particles from the lower energy
level to the higher one.
This provides a possible archetype for quantum molecular pumps
driven by nonequilibrium noise.

This inversion of population can be accompanied by cooling.
Namely, the ensemble of TLSs becomes first
effectively cooled and only then heated up (through, formally,
$T_{\sigma}=\infty$ to $T_{\sigma}=-\infty$) until the inversion of population
occurs -- cf. in Fig. \ref{Fig2}, b.
For this pumping mechanism to work, an {\it inverted transport regime}
\cite{Marcus} is necessary; i.e., a
regime where the static, unfluctuating 
forward rate becomes smaller with the increasing energy bias
after reaching a maximum at $\epsilon_{max}$. In the present model,
this maximum is located
in the neighbourhood of $\Omega_0$.
More precisely, $\epsilon_{max}$
corresponds to the maximum in the difference between the
forward and the backward rates,
rather
than to the maximum of the forward rate alone.
The inversion
happens for $\sigma>\epsilon_{max}$ and a sufficiently
small energy bias $\epsilon_0$. A similar mechanism has been proposed
in Ref. \cite{PRE97} within a spin-boson
modelling of electron tunnelling in proteins, see below in Sec. \ref{secSB},
driven by nonequilibrium conformational fluctuations, e.g.,
utilising energy of
ATP hydrolysis.

The underlying mechanism
seems quite general. Indeed, the inversion
of populations occurs whenever the difference of averaged rates
$\langle \Delta W(\epsilon_0)\rangle_{\epsilon}:=
\langle W_{12}(\epsilon_0)\rangle_{\epsilon}-
\langle W_{21}(\epsilon_0)\rangle_{\epsilon}$
becomes negative, $\langle\Delta W(\epsilon_0)\rangle_{\epsilon}<0$,
for a positive bias $\epsilon_0>0$. In the discussed limiting
case, $\langle\Delta W(\epsilon_0)\rangle_{\epsilon}\approx
\frac{1}{2}[\Delta W(\epsilon_0+\sigma)+
\Delta W(\epsilon_0-\sigma)]$
with $\Delta W(-\epsilon)=-\Delta W(\epsilon)$. Therefore, when
$\sigma$ exceeds $\epsilon_{max}$, where $\Delta W(\epsilon)$
achieves a maximum and
$\frac{d}{d\epsilon}\Delta W(\epsilon)<0$ for $\sigma>\epsilon_{max}$,
the averaged  difference of forward and backward rates becomes negative
$\langle\Delta W(\epsilon_0)\rangle_{\epsilon}<0$,
for a positive energy bias $\epsilon_0>0$, i.e. an inversion of populations
takes place. In application to the quantum transport in a spatially
extended system, a similar effect results in the noise-induced absolute
negative mobility \cite{PLA98}, see below in Sec. \ref{sec7}. The existence of the
static current-voltage characteristics with a negative
{\it differential} conductivity part is important for the latter
phenomenon to occur.

\subsubsection{Emergence of an effective energy bias}

The fourth important effect,
the onset of which can be seen
already in the discussed archetype model, is rooted in a possible asymmetry
of the unbiased on average fluctuations. Namely, let us consider the
symmetric quantum system, $\epsilon_0=0$, with
asymmetric dichotomous fluctuations of the energy levels with zero mean,
see in Sec. \ref{KuboOsc}. Since in this case, the averaged propagator
of the corresponding Kubo-oscillator is {\it complex-valued},
${\rm Im}\; \langle
\exp[i\tilde \zeta(t,t')]\rangle_{\epsilon}\neq 0$, it can be readily seen from
Eq. (\ref{w12}) after invoking the decoupling approximation that
$\langle w_{12}(t,t')\rangle_{\epsilon}-\langle w_{21}(t,t')
\rangle_{\epsilon}\neq 0$ even if $\epsilon_0=0$. This means that
an effective asymmetry emerges. Moreover, the above difference is
proportional also to ${\rm Im}\; K(t-t')\neq 0$. If the autocorrelation
function  of the thermal bath, $K(t)$, were real (like for a stochastic
bath), then {\it no}
asymmetry between the forward and backward rates could emerge
in principle. Therefore, the discussed asymmetry
does emerge due to a subtle interplay of the equilibrium {\it quantum}
fluctuations of the thermal bath and nonequilibrium classical fluctuations
of the energy levels, both of which are unbiased on average.
Here is rooted the origin of quantum dissipative rectifiers put forward
in Ref. \cite{PRL98,EPL98}. The very same effect can
also be deduced from Eq. (\ref{rates}),
since the corresponding spectral line $I(\omega)$,
cf. Eq. (\ref{kuboline}), is {\it asymmetric}. Yet,
 ultimate insight is achieved in the
slow-modulation limit of the Kubo oscillator, $K_{\epsilon}:=
\sigma/\nu \gg 1$, like in Eq. (\ref{rates:aver}) where
the mean forward and backward rates are the static rates
$W_{12}(\epsilon)$ and $W_{21}(\epsilon)$ averaged over
the energy bias distribution,
$p(\epsilon_{1,2})=\langle \tau_{1,2}\rangle/(\langle \tau_1\rangle
+\langle \tau_2\rangle)$, correspondingly, i.e.,
$\langle W_{12}(0)\rangle_{\epsilon}=\sum_{j=1,2}p(\epsilon_{j})
W_{12}(\epsilon_j)$ and
$\langle W_{21}(0)\rangle_{\epsilon}=\sum_{j=1,2}p(\epsilon_{j})
W_{21}(\epsilon_j)$.

For several applications of  quantum transport in spatially
extended systems our  driving-induced breaking of symmetry
leads to a rectification current in  tight-binding Brownian
rectifiers \cite{PRL98,EPL98}, see in Sec. \ref{sec7}.

\subsection{Quantum relaxation
in strong periodic fields}

The considered strong nonequilibrium effects are present as well in the case
of a fast periodic driving,
$\tilde \epsilon(t)=A\cos(\Omega t+\varphi_0)$, with
a static phase $\varphi_0$ which is uniformly
distributed between $0$ and $2\pi$.
Then, the corresponding
spectral line shape form $I(\omega)$ is
$I(\omega)=\sum_{n=-\infty}^{\infty} J_{n}^2(A/\Omega)
\delta(\omega-n\Omega)$, where $J_{n}(z)$ is the Bessel
function of the first kind.

The rate expressions (\ref{rates:aver}) take on the form
\begin{eqnarray}\label{ratesPeriod2}
\langle W_{12}(\epsilon_0)\rangle_{\epsilon} & = &
\sum_{n=-\infty}^{\infty}J_n^2\Big ( \frac{A}{\Omega}\Big)
e^{\frac{\hbar[\epsilon_0-n\Omega]}{k_BT}}
n(\epsilon_0-n\Omega)J(\epsilon_0-n\Omega),\nonumber \\
\langle W_{21}(\epsilon_0)\rangle_{\epsilon} & = &
\sum_{n=-\infty}^{\infty}J_n^2\Big ( \frac{A}{\Omega}\Big)
n(\epsilon_0-n\Omega)J(\epsilon_0-n\Omega)\;.
\end{eqnarray}
Such an expansion of the transition
rates as a sum over different
emission (absorption) channels with  $n$
emitted (absorbed) photons with the corresponding probabilities
$p_n=J_{n}^2(A/\Omega)$ is similar to one
used by Tien and Gordon in a different context
\cite{TienGordon,GrifHangRev}.
For the averaged relaxation rate the above expression yields the
same result as in \cite{Petrov86} where the principal
possibility to regulate
quantum relaxation processes
in condensed molecular systems by strong periodic external
fields has been indicated.
Moreover, the inversion of populations by periodic driving
takes also place for the above model $J(\omega)$ and some properly
adjusted parameters of the periodic driving. For a periodically driven
spin-boson model (see below in Sec. 6) and a strong system-bath
coupling, this latter effect has been theoretically
predicted and described in Refs. \cite{CPL96,DakhCoal95}
(see also \cite{GrifHangRev} for a review and further references).
For the case of a weak system-bath coupling, the inversion of populations
in the spin-boson model has been shown in
Ref. \cite{PRE2000}.

\subsection{Approximation of time-dependent rates\label{time-rate}}

If the external field varies sufficiently slow on the characteristic
time-scale,  $\tau_{d}$, describing the decay of the  kernels in Eq. (\ref{balance}),
 an adiabatic approximation of time-dependent rates that follow
the time-variation of the energy levels
can be invoked; i.e.,
\begin{eqnarray}
 \dot p_{1}(t) & =&
- W_{12}(\epsilon(t)) p_{1}(t) +
W_{21}(\epsilon(t)) p_{2}(t), \nonumber \\
 \dot p_{2}(t) & =&
W_{12}(\epsilon(t)) p_{1}(t) -
W_{21}(\epsilon(t)) p_{2}(t),
\end{eqnarray}
where $W_{21}(\epsilon)=n(\epsilon)J(\epsilon)$
and $W_{12}(\epsilon)=\exp(\hbar\epsilon/k_BT)
W_{21}(\epsilon)$ are the static rates. For  discrete state noise,
this approximation holds whenever $\langle \tau_j\rangle \gg \tau_d$.
Then, the corresponding rates describe a discrete state
stochastic process and the
averaging method of Sec. \ref{rand} can be applied.
In this limiting case, the corresponding
Laplace-transformed averaged populations can be given in
exact analytical form.  In the considered case this corresponds
to the averaging of a Kubo oscillator with an imaginary frequency. The
corresponding averaged solution can be analytically
inverted into the time domain in the case of two-state Markovian fluctuations.
This solution is generally bi-exponential. Two limiting cases can be distinguished which
can be classified by the Kubo number $K_{W}$ of the {\it rate}
fluctuations, i.e.
by the product of the variance
of the rate fluctuations multiplied with the corresponding
autocorrelation time.
In a {\it slow modulation} limit (in terms of the {\it rate} fluctuations),
$K_W\gg 1$ and
the (ensemble) averaged relaxation is approximately described
by a quasi-static averaging of the time-dependent solutions
with the static,
 ``frozen'' energy bias randomly distributed.
 It assumes a bi-exponential form, but can be
multi-exponential and anomalously slow in a more general
case of multi-state
fluctuations.  The opposite limit of
{\it fast modulation} (in terms of the {\it rate} fluctuations, $K_W\ll 1$)
corresponds to the averaged rate description which is detailed above and
which invokes
the decoupling approximation and in addition can possibly  involve
a  {\it slow modulation} limit
in terms of the {\it energy} level fluctuations, $K_{\epsilon}\gg 1$. The resulting averaged
relaxation process remains approximately single-exponential.

In view of the presence
of many different time scales the underlying physics is
nontrivial. Therefore, it is useful to be able to resort to a case study
where the stochastic averaging can be performed {\it exactly}.
Such an archetypal  investigation has been put forward in Ref. \cite{Goychuk95}
and has been applied to the stochastic spin-boson model in Refs.
\cite{PRE95,PRE97}.

\subsection{Exact averaging for dichotomous Markovian fluctuations
\label{exact}}

By use of the conservation of probabilities, the system of integro-differential
equations (\ref{balance}) can be reduced to a single equation for
the {\it difference of populations} $\sigma_z(t)=p_1(t)-p_2(t)$. It reads:
\begin{eqnarray}
\label{GMEsigma}
\dot \sigma_z(t)=-\int_{0}^{t}f(t,t')\sigma_z(t')dt' -
\int_0^{t}g(t,t')dt'
\end{eqnarray}
with the integral kernels
\begin{eqnarray}
\label{fg}
f(t,t') & = & f_0(t,t')\cos[\epsilon_0(t-t')+\tilde\zeta(t,t')],
\nonumber\\
g(t,t')& = & g_0(t,t')\sin[\epsilon_0(t-t')+\tilde\zeta(t,t')]\;,
\end{eqnarray}
where
\begin{eqnarray}
\label{fg0}
f_0(t,t') & = &f_0(t-t')= 4\;{\rm Re} [K(t-t')],
\nonumber\\
g_0(t,t') & = & g_0(t-t')=-4\;{\rm Im} [K(t-t')]\;.
\end{eqnarray}
The kernel $f(t,t')$ in Eq. (\ref{GMEsigma}) denotes a stochastic 
functional of
the driving field $\tilde \epsilon(t)$ on the time interval
$[t',t]$ (with times later than $t'$) whereas
$\sigma_z(t')$ is a functional of the dichotomous Markovian process (DMP)
for the times prior to $t'$.
The task of stochastic averaging of the
product of such functionals,
$\langle f(t,t')\sigma_z(t')\rangle$,  can become very difficult, see
\cite{Hanggi78}. However, in the case
$\tilde \epsilon(t)=\sigma \alpha(t)$, where $\alpha(t)=\pm 1$ is
symmetric DMP
with unit variance and autocorrelation time $\tau_c=1/\nu$
this task can be solved exactly
by referring to a theorem by Bourret, Frisch and Pouquet \cite{BourFrischPouq}
(for a different proof of this remarkable exact result,
see also Ref. \cite{Klyatskin}). It reads: $\langle f(t,t'+\tau)
\alpha(t'+\tau)\alpha(t')\sigma_z(t')\rangle=\langle f(t,t'+\tau)\rangle
\langle \alpha(t'+\tau)\alpha(t')\rangle \langle
\sigma_z(t')\rangle + \langle f(t,t'+\tau)\alpha(t'+\tau)\rangle
\langle \alpha(t')
\sigma_z(t')\rangle$ for $\tau\geq 0$. By passing to the limit $\tau\to 0$
and using the characteristic property of the DMP, namely that $\alpha^2(t)=1$ (without averaging),
the above relation yields an important corollary \cite{Goychuk95}:
\begin{eqnarray}
\label{Bourret}
\langle f(t,t')\sigma_z(t')\rangle=\langle f(t,t')\rangle
\langle
\sigma_z(t')\rangle + \langle f(t,t')\alpha(t')\rangle
\langle \alpha(t')
\sigma_z(t')\rangle\;.
\end{eqnarray}
This result is beyond the decoupling approximation, given by the first term in the sum.
The result for the cross-correlation function
$\langle \alpha(t) \sigma_z(t)\rangle$ is more intricate.
The equation of motion for this cross-correlation function can be
obtained due to a theorem by
 Shapiro-Loginov \cite{ShapLog,ShapLog1}, reading,
\begin{eqnarray}
\label{Shap}
\frac{d}{dt}\langle \alpha(t)\sigma_z(t)\rangle =
-\nu \langle \alpha(t)\sigma_z(t)\rangle +
\Big \langle \alpha(t)\frac{d\sigma_z(t)}{dt}\Big\rangle\;.
\end{eqnarray}
Use of this relation in turn generates an integro-differential equation for
$\langle \alpha(t)\sigma_z(t)\rangle$, where the problem
of decoupling of $\langle \alpha(t) f(t,t')\sigma_z(t')\rangle$
emerges. It can be solved in the
same way as in Eq. (\ref{Bourret}), namely
\begin{eqnarray}
\label{Bourret2}
\langle \alpha(t) f(t,t')\sigma_z(t')\rangle &= &\langle\alpha(t)
f(t,t')\rangle
\langle
\sigma_z(t')\rangle \nonumber \\ & + & \langle \alpha(t) f(t,t')\alpha(t')\rangle
\langle \alpha(t')
\sigma_z(t')\rangle\;.
\end{eqnarray}
All these averaged functionals, like $\langle f(t,t')\rangle$,
$\langle \alpha(t)f(t,t')\rangle$,
$\langle f(t,t')\alpha(t')\rangle$,\\
$\langle \alpha(t)f(t,t')\alpha(t')\rangle$ can be expressed
in terms of the averaged propagator of the corresponding Kubo
oscillator $S^{(0)}(t-t')
=\langle \exp[i\sigma\int_{t'}^{t}\alpha(\tau)
d\tau] \rangle$ given in Eq. (\ref{propKubo})
with $\chi=\nu/2$ (zero asymmetry),
and its derivatives $S^{(n)}(t):=\frac{1}{\sigma^n}
\frac{d^n}{dt^n}S^{(0)}(t)$
\cite{Goychuk95,PRE95}.
Applying the general results in Eqs. (\ref{Bourret}), Eq. (\ref{Shap}) and
 (\ref{Bourret2}) to Eq. (\ref{GMEsigma})
yields a closed system of two integro-differential 
equations \cite{Goychuk95,PRE95}:
\begin{eqnarray}
\label{GMEsigmaAv}
\frac{d}{dt}\langle \sigma_z(t)\rangle  =
& - & \int_{0}^{t}\Big (  S^{(0)}(t-t') f_0(t-t')\cos[\epsilon_0(t-t')]
\langle \sigma_z(t')\rangle \nonumber\\
 & - & S^{(1)}(t-t')
f_0(t-t') \sin[\epsilon_0(t-t')]\langle\alpha(t') \sigma_z(t')\rangle
\nonumber\\
& + & S^{(0)}(t-t') g_0(t-t')\sin[\epsilon_0(t-t')]
\Big)dt', \\
\frac{d}{dt}\langle \alpha(t)\sigma_z(t)\rangle  = & - &\nu
\langle \alpha(t)\sigma_z(t)\rangle \nonumber \\
& + & \int_{0}^{t}\Big ( S^{(2)}(t-t') f_0(t-t')\cos[\epsilon_0(t-t')]
\langle \alpha(t')\sigma_z(t')\rangle \nonumber\\
 & + & S^{(1)}(t-t')
f_0(t-t') \sin[\epsilon_0(t-t')]\langle \sigma_z(t')\rangle
\nonumber\\
& + & S^{(1)}(t-t')
g_0(t-t') \cos[\epsilon_0(t-t')]
\Big)dt'\;. \nonumber
\end{eqnarray}
A subsequent Markovian approximation for Eq. (\ref{GMEsigmaAv}) then yields
\cite{Goychuk95}:
\begin{eqnarray}
\label{sigmaMark}
\frac{d}{dt}\langle \sigma_z(t)\rangle & = & -\Gamma_0
\langle \sigma_z(t)\rangle- \Gamma_1
\langle \alpha(t)\sigma_z(t)\rangle - r_0, \nonumber \\
\frac{d}{dt}\langle \alpha(t)\sigma_z(t)\rangle   & = &
-\Gamma_1
\langle \sigma_z(t)\rangle- (\nu+\Gamma_2)
\langle \alpha(t)\sigma_z(t)\rangle - r_1
\end{eqnarray}
with
\begin{eqnarray}
\Gamma_k=\int_{-\infty}^{\infty}\coth\left( \frac{\hbar\omega}{2k_BT}\right)
J(\omega)
I_k(\epsilon_0-\omega) d\omega, \nonumber \\
r_k=\int_{-\infty}^{\infty}
J(\omega)
I_k(\epsilon_0-\omega) d\omega\;,
\end{eqnarray}
where $I_k(\omega)=(-\omega/\sigma)^k I(\omega)$ and
$I(\omega)$ is given in Eq. (\ref{kuboline}) with
$\epsilon_{1,2}=\pm \sigma$. It can be shown that
all known limiting cases are reproduced from this remarkable result.

For the case of weakly coloured noise $K_{\epsilon}\ll 1$ (i.e. the
fast modulation limit of the Kubo oscillator), the spectral line
$I(\omega)$ becomes a Lorentzian with the width $D=\sigma^2/\nu$.
The same result holds true in the white noise limit
$\sigma\to\infty, \nu\to\infty$, with $D=const$ and $K_{\epsilon}\to
0$. In these limits
 $\Gamma_1$ is negligible small, $\Gamma_1\approx 0$,
 and the relaxation is described by
the averaged rate $\Gamma_0$. Precisely the same result re-emerges
also for white Gaussian noise $\tilde\epsilon(t)$
with the noise intensity $D$.
The spectral line $I(\omega)$
becomes narrower when $\nu$ increases, -- this constitutes the celebrated
motional narrowing limit of
NMR \cite{Kubo54,Anderson53} --, and approaches
zero when $\nu\to \infty$ (with $\sigma$ kept constant).
Such infinitely fast fluctuations have no
influence on the considered rate process; the field-free description  is thus
reproduced with the thermal equilibrium being restored.

In the slow modulation limit of the Kubo-oscillator ($K_{\epsilon}\gg1$),
$I(\omega)\approx I_2(\omega)\approx
\frac{1}{2}[\delta(\omega+\sigma)+
\delta(\omega-\sigma)]$
and $I_1(\omega)\approx
\frac{1}{2}[\delta(\omega+\sigma)-
\delta(\omega-\sigma)]$, with the corresponding line widths neglected.
In this case,
the approximation of time-dependent fluctuating
rates following adiabatically to
 the energy levels fluctuations becomes justified. The relaxation is generally
bi-exponential with the two rates given by
\begin{eqnarray}
\lambda_{1,2}=\frac{\nu}{2}+\Gamma_0\pm \frac{1}{2}
\sqrt{(\Gamma_{+}-\Gamma_{-})^2+\nu^2}\;.
\end{eqnarray}
Here, $\Gamma_{\pm}=\coth[\hbar(\epsilon_0\pm\sigma)/2k_BT]J(\epsilon_0\pm
\sigma)$ is the relaxation rate in the quasi-static limit
and $\Gamma_0=(\Gamma_{+}+\Gamma_{-})/2$. Furthermore, if
$\nu\gg \Gamma_0$, we have $\lambda_1\approx \nu$
(note that the corresponding
exponent $\exp(-\lambda_1 t)$ contributes, however, with a very small
weight), and $\lambda_2\approx \Gamma_0$ (with a weight which
approximately equals one): This in turn implies that
 the relaxation is practically single exponential with 
rate $\Gamma_0$ and corresponds to the fast modulation limit in terms
of the fluctuating rates.

\section[Application II: Driven electron transfer and spin-boson model]
{\label{secSB}Application II: Driven electron transfer \\ 
 within a spin-boson description }

Let us proceed  with an application of our general theory to the celebrated
{\it driven} spin-boson model \cite{GrifHangRev}.
This model is of special importance since it describes a large variety of physical phenomena
\cite{LeggettReview,WeissBook}, such as relevant aspects of
 electron transfer in molecular systems.

\subsection{Curve-crossing problems with dissipation}
\begin{figure}[h!]
\centerline{
 \epsfxsize=8cm
\epsfbox{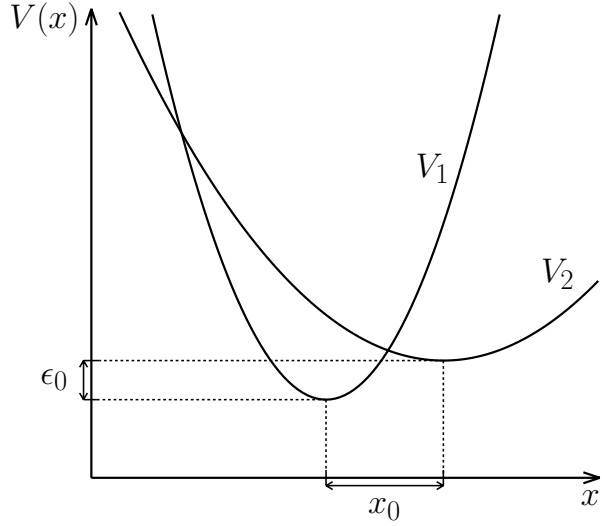}}
  \caption{Sketch of the diabatic electronic curves. Note that two crossing points occur
   in the presence of different curvatures \cite{CPL2003}.}
  \label{parabola}
  \end{figure}

The simplest case of a
two-state donor acceptor electron transfer reaction can be considered as
a curve-crossing problem within the description of two
diabatic electronic states $|1\rangle$ and $|2\rangle$
with electronic energies
$V_{1}(x)$ and $V_{2}(x)$ that depend on a nuclear reaction coordinate
 $x$ \cite{advance,MayBook,FrauenfelderWolynes,HTB,HJSP}
(cf. Fig. \ref{parabola}). Namely, after separating nuclear and
electronic degrees
of freedom within the Born-Oppenheimer approximation, the electron tunnelling
process coupled to the nuclear dynamics (modelled by the reaction
coordinate $x$) can be described by the following
Hamiltonian
\begin{eqnarray}\label{H1}
 H_{tun}(x,p,t) & = &\left [\frac{\hat p^2}{2M} + V_1(x,t)
\right ] |1\rangle \langle 1| +
\left [\frac{\hat p^2}{2M} +V_2(x,t)\right ]|2\rangle \langle 2|
\nonumber \\
& +& \frac{1}{2}\hbar\Delta(t) \Big (|1\rangle \langle 2| +
|2\rangle \langle 1| \Big ).
\end{eqnarray}
The time-dependent electronic curves in Eq. (\ref{H1})
\begin{eqnarray}
V_{1,2}(x,t)=\frac{1}{2}M\Omega_{1,2}^2(x\pm x_{0}/2)^2
\pm\hbar
\epsilon_0/2 - d_{1,2}\mathcal{ E}(t),
\end{eqnarray}
can generally possess different curvatures in the parabolic
approximation with minima energetically separated by $\hbar\epsilon_0$
and separated by a distance $x_0$ (the tunnelling distance). Moreover,
such electronic states generally possess electric dipole moments
$d_{1,2}$ (their coordinate dependence is neglected) and thus the discussed
energy levels will generally become dependent either on the stochastic
microscopic fields of the environment, or on an externally applied electric field
$\mathcal{ E}(t)$.
The corresponding time-dependence can likewise reflect also
some nonequilibrium conformational dynamics.
Moreover, the coupling or  tunnelling matrix element $\Delta(t)$ can also
parametrically depend on a nonequilibrium reaction
coordinate which generally introduces an explicit stochastic time-dependence.
The reaction coordinate $x$ is coupled to the rest of vibrational
degrees of freedom of the environment. This introduces a dissipation mechanism into the tunnelling
problem which can be modelled by a bilinear coupling of $x$ to the thermal bath
of harmonic oscillators \cite{CaldeiraLeggett,Garg},
\begin{eqnarray}\label{H-B}
 H_{\rm BI} =
\frac{1}{2}\sum_{i}\Big\{\frac{\hat  p_i^2}{m_i}+m_i\omega_i^2\Big [
x_i-\frac{c_i}{m_i\omega_i^2}x
\Big]^2
\Big \}.
\end{eqnarray}
It is worthy to point out that the frequencies $\Omega_{1,2}$ of the
oscillator $x$ can depend on the electronic state.
In other words, the relevant vibration modes can become either softer, or more
rigid depending on the electronic state.
In the following we neglect this possible effect and
assume that  $\Omega_1=\Omega_2=\Omega_0$,
but note the studies in Refs. \cite{Tang,LeeMedStuch,CPL2003} for the more general situation.
Moreover, one assumes that the reaction coordinate relaxes
rapidly (with respect to the time scale of electron transfer)
into  thermal equilibrium with the bath of oscillators.
Then, a canonical transformation from the ``reaction coordinate + N bath
oscillators´´ to ``N+1 new bath oscillators'' brings the original
problem into the spin-boson form, i.e.,
\cite{Garg}\label{sb}
\begin{eqnarray}
 H(t)  = \frac{1}{2}\hbar\epsilon(t)
\hat \sigma_z  +
\frac{1}{2}\hbar\Delta (t) \hat \sigma_x +  \frac{1}{2}x_0
\hat \sigma_z\sum_{\lambda} \tilde c_{\lambda}\tilde x_{\lambda} \nonumber \\ +
\frac{1}{2}\sum_{\lambda}\Big\{\frac{\tilde p_\lambda^2}{\tilde
m_\lambda}+\tilde
m_\lambda\tilde \omega_\lambda^2 \tilde x_\lambda^2
\Big \},
\end{eqnarray}
where $\epsilon(t)=\epsilon_0-(d_1-d_2)\mathcal{ E}(t)/\hbar$.
The coupling between the
quasi-spin and boson bath is
characterised by a so-called spectral density \footnote{This definition is
related to the one given
in Eq. (\ref{Jnm}) by: $J(\omega)=2x_0^2\tilde J(\omega)/\hbar$
with $\kappa_{\lambda}=x_0\tilde c_{\lambda}\sqrt{\hbar/(2\tilde m_{\lambda}
\tilde\omega_{\lambda})}$, $\tilde x_{\lambda}=
\sqrt{\hbar/(2\tilde m_{\lambda}
\tilde\omega_{\lambda})}(b^{\dagger}_{\lambda}+b_{\lambda})$.}
$\tilde
J(\omega)=(\pi/2)\sum_i(\tilde c_\lambda^2/
\tilde m_\lambda\tilde \omega_\lambda)\delta(\omega~-~\tilde
\omega_\lambda)$~\cite{LeggettReview}.
Moreover, we assume for the low frequency behaviour an Ohmic-like coupling between the reaction coordinate
and the environmental vibrational modes (which corresponds in the classical limit to
a viscous frictional force $F=-\zeta \dot x$ acting on the reaction
coordinate $x$) which in turn yields the effective spectral density
 $\tilde J(\omega)=\zeta\omega\frac{\Omega_0^4}
{(\omega^2-\Omega_0^2)^2+4\omega^2\gamma^2};\; \gamma=\zeta/2M $.
This scheme corresponds to the model of a damped Brownian harmonic oscillator
used in Eq. (\ref{BrownOsc}) with
$\kappa_0=\sqrt{(\hbar\Omega_0)\lambda}$, where
 $\lambda=Mx_0^2\Omega_0^2/2$
is the reorganisation energy.
The coupling
strength $\zeta$ can  be related to the dimensionless (Kondo) parameter
$\alpha=\frac{\zeta
x_0^2}{2\pi\hbar}=\frac{2}{\pi}\frac{\lambda}{\hbar\Omega_0}\frac{\gamma}{\Omega_0}$.
The use of the representation of bosonic operators in Eq. (\ref{sb})
then yields
 \begin{eqnarray}
\label{spinboson}
H(t)=\frac{1}{2}\hbar\epsilon(t)\hat \sigma_z+
\frac{1}{2}\hbar\Delta(t)\hat \sigma_x+
\frac{1}{2}\hat \sigma_z \eta(t)
\sum_{\lambda}\kappa_{\lambda}
(b^{\dagger}_{\lambda}+b_{\lambda}) \nonumber \\ +\sum_{\lambda}
\hbar\omega_{\lambda}(b^{\dagger}_{\lambda}b_{\lambda}+\frac{1}{2})
\end{eqnarray}
(the ``tilde'' over $\omega_\lambda$ is  omitted here).
 To address formally the most
general case we assume  in addition that the system-bath coupling can be
modulated in time as well, i.e., $\kappa_{\lambda}\to \kappa_{\lambda}\eta(t)$
with some prescribed time-dependent function $\eta(t)$.

\subsection{Weak system-bath coupling}

Let us consider first the case of a weak system-bath coupling.
The corresponding generalised master equations are obtained
by applying
Eqs. (\ref{matrixGME}), (\ref{tetradicKern})
(in the representation
of $\hat\gamma_{nm}$) to the considered spin-boson model.
This yields after  some cumbersome calculations
by using  the quasi-spin basis the following GMEs:
\begin{eqnarray} \label{GMEBloch}
\dot \sigma_x(t) &=& -\epsilon(t) \sigma_y(t) -\int_{0}^{t}
\Gamma_{xx}(t,t')\sigma_x(t')dt' - \int_{0}^{t}
\Gamma_{xy}(t,t')\sigma_y(t')dt' -A_x(t), \nonumber \\
\dot \sigma_y(t) &=& \epsilon(t) \sigma_x(t) -\Delta(t) \sigma_z(t)-
\int_{0}^{t}
\Gamma_{yx}(t,t')\sigma_x(t')dt' \nonumber \\ & - & \int_{0}^{t}
\Gamma_{yy}(t,t')\sigma_y(t')dt' - A_y(t), \\
\dot \sigma_z(t) &=& \Delta(t) \sigma_y(t), \nonumber
\end{eqnarray}
with the kernels reading
\begin{eqnarray}\label{BlochKern}
\Gamma_{xx}(t,t') & = & \eta(t)\eta(t')
{\rm Re}[K(t-t')] {\rm Re}
[ U_{11}^2(t,t') +
U_{12}^2(t,t')], \nonumber \\
\Gamma_{yy}(t,t') & = & \eta(t)\eta(t')
{\rm Re}[K(t-t')] {\rm Re}
[ U_{11}^2(t,t') -
U_{12}^2(t,t')], \nonumber \\
\Gamma_{xy}(t,t') & = & \eta(t)\eta(t')
{\rm Re}[K(t-t')]{\rm Im}
[ U_{11}^2(t,t') -
U_{12}^2(t,t')], \nonumber \\
\Gamma_{yx}(t,t') & = -& \eta(t)\eta(t')
{\rm Re}[K(t-t')]{\rm Im}
[ U_{11}^2(t,t') +
U_{12}^2(t,t')],
\end{eqnarray}
and the inhomogeneous terms given by
\begin{eqnarray}
A_{x}(t)&  = &2\int_{0}^{t}\eta(t)\eta(t'){\rm Im}[K(t-t')]{\rm
Im[U_{11}(t,t')U_{12}(t,t')]}dt', \nonumber\\
A_{y}(t)& = &2\int_{0}^{t}\eta(t)\eta(t'){\rm Im}[K(t-t')]{\rm
Re[U_{11}(t,t')U_{12}(t,t')]}dt'\; .
\end{eqnarray}
The  evolution operator of the driven TLS in the absence of coupling, which defines the Hamiltonian $H_D(t)$ of the
driven, nondissipative dynamics, is denoted by
\[ U_{nm}(t,t')=
\langle n |\mathcal{ T}
\exp\left [-\frac{i}{\hbar}\int_{t'}^{t}H_D(\tau)d\tau \right]
|m\rangle. \]

This propagator enters the
 above memory kernels; it
can be  found numerically
from the solution of the corresponding Schr\"odinger equation for
an arbitrary time-dependence. Moreover, in the case of a periodic
driving,  an expansion into Floquet modes
is conveniently applied, see in
\cite{HanggiBookChap}, and further references therein.
Other methods, e.g., the use of a Magnus expansion \cite{Magnus}
are also possible. Due to the unitary  quantum evolution
in the absence of dissipation we have $U_{22}(t,t')=U_{11}^*(t,t')$
and $U_{21}(t,t')=-U_{12}^*(t,t')$ with $\det[U_{nm}(t,t')]=1$
for arbitrary time-dependence of $\epsilon(t)$.

Time-nonlocality of the generalised master
equations in Eq. (\ref{GMEBloch}) makes them difficult to study
from a numerical viewpoint. To work with a memoryless,
Markovian description presents, therefore, a pivotal
advantage. If  the dissipation is very weak, this  description suffices
to capture the main influences of dissipation on the driven quantum
dynamics, i.e.,  the emergence of an exponential relaxation (and decoherence)
described by some small rate constants and corresponding,
dissipation-induced frequency shifts,
i.e. the Lamb shifts occurring even at $T=0$.
Both the relaxation rates and the frequency shifts
are proportional, in the lowest order, to
$\kappa_{\lambda}^2$.
Applying  Eqs. (\ref{Redfield}),
(\ref{RRedfield})  to the considered dynamics
yields the following driven
Bloch-Redfield equations:
\begin{eqnarray} \label{BlochRed}
\dot \sigma_x(t) &=& -\epsilon(t) \sigma_y(t) -
R_{xx}(t)\sigma_x(t)-R_{xz}(t)\sigma_z(t)-A_x(t), \nonumber \\
\dot \sigma_y(t) &=& \epsilon(t) \sigma_x(t) -\Delta(t) \sigma_z(t)-
R_{yy}(t)\sigma_y(t)
- R_{yz}(t)\sigma_z(t)-A_y(t), \\
\dot \sigma_z(t) &=& \Delta(t) \sigma_y(t) \nonumber
\end{eqnarray}
with the relaxation matrix elements reading
\begin{eqnarray}\label{BlochRelax}
R_{xx}(t) & = & R_{yy}(t)=\int_{0}^{t}\eta(t)\eta(t')
{\rm Re}[K(t-t')]
[ |U_{11}(t,t')|^2-
| U_{12}(t,t')|^2]dt', \nonumber \\
R_{xz}(t) & = & 2\int_{0}^{t}\eta(t)\eta(t'){\rm Re}[K(t-t')]{\rm
Re[U_{11}(t,t') U_{12}(t,t')]}dt',\nonumber \\
R_{yz}(t) & = & -2\int_{0}^{t}\eta(t)\eta(t'){\rm Re}[K(t-t')]{\rm
Im[ U_{11}(t,t') U_{12}(t,t')]}dt'\;.
\end{eqnarray}

In the common case of a time-independent tunnelling matrix element, i.e.  $\Delta(t)=const$ and a time-independent
system-bath coupling, i.e.
$\eta(t)=1$ (what is assumed in the following), this result reduces to the driven
Bloch-Redfield equations derived in Ref. \cite{PRE2000}.
Note the different signs of $\Delta$ and $\epsilon$ used throughout
this work and in Ref. \cite{PRE2000}, as well as some other cited
references.

\begin{figure}
\centerline{
 \epsfxsize=6.5cm
\epsfbox{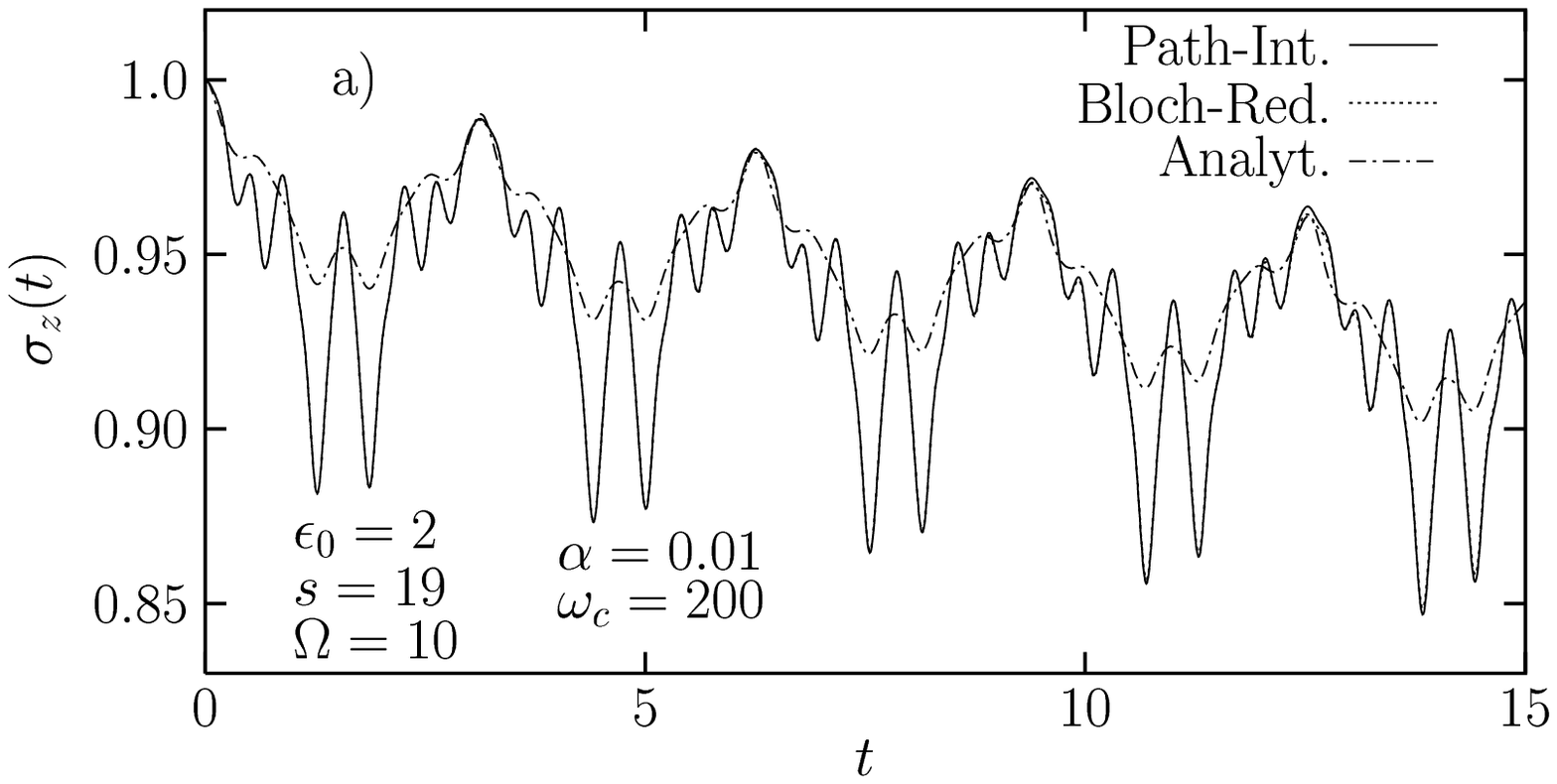}
\epsfxsize=6.5cm
\epsfbox{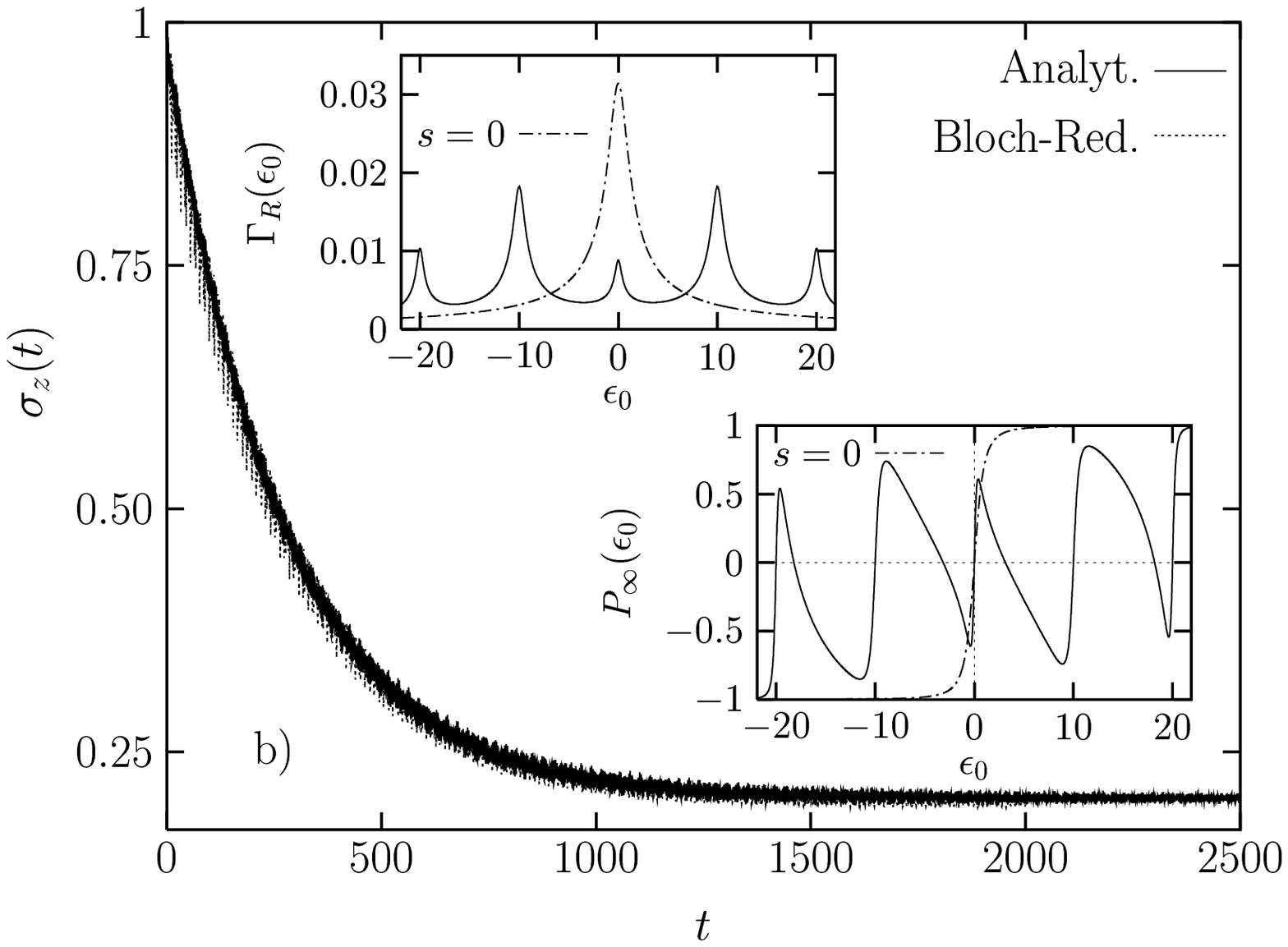}}
\caption{(a) Numerical comparison of the driven Bloch-Redfield
equations of Ref. \cite{PRE2000} (dotted line) and the
path-integral GME
of Ref. \cite{GWW} (full line) for an oscillatory high-frequency driving
$\Omega\gg\omega_0$ (data taken from Ref. \cite{PRE2000}).
Both depicted numerical solutions practically coincide within
line width. The dashed-dotted line depicts a quasi-analytical
solution (for details see in \cite{PRE2000}) of the driven path-integral GME.
It captures well the main features of the driven dynamics,
lacking only some finer details.
Time and frequencies are measured in
units of $\Delta^{-1}$ and $\Delta$, correspondingly. The used parameter sets are depicted
 in the figure.
(b) Corresponding asymptotic long-time dynamics: the numerical
solution of the driven Bloch-Redfield equations (dotted line) is compared
with the quasi-analytical solution of driven path-integral GME
(full line). Both solutions agree well within the width of the
small-amplitude, driving induced oscillations.
The two insets depict the analytical results for the rate of
averaged relaxation $\Gamma_R(\epsilon_0)$ and the difference of
asymptotic populations $P_{\infty}(\epsilon_0):=-\lim_{t\to\infty}
\sigma_z(t)$, respectively.
The rate of incoherent relaxation
 $\Gamma_R$ exhibits  characteristic resonance
peaks, being located at multiple integers of the driving frequency $\Omega$.
These peaks are shifted replicas of the dc-driven rate with different weights,
i.e. the case with no oscillatory forcing acting
(i.e. a vanishing driving amplitude $s =0$). Thus,
a suitable chosen static field $\epsilon_0$ can enhance or suppress
the decay of populations.
The asymptotic population difference $P_\infty$ exhibits a
non-monotonic  dependence versus the asymmetry $\epsilon_0$ when combined
with a high frequency driving field. For appropriate values of bias $\epsilon_0$ , a
 population inversion takes place ($P_\infty <0$ when $\epsilon_0>0$,
 and {\it vice versa}). }
\label{Fig3}
\end{figure}

For constant bias $\epsilon(t)=\epsilon_0$, and constant tunnelling
coupling
$\Delta(t)=\Delta$,
\begin{eqnarray}
U_{11}(t,t') & = &\cos[\omega_0(t-t')/2]-i\frac{\epsilon_0}{\omega_0}
\sin[\omega_0(t-t')/2],\nonumber \\
U_{12}(t,t') &=& -i\frac{\Delta}{\omega_0}
\sin[\omega_0(t-t')/2]\;,
\end{eqnarray}
where $\omega_0=\sqrt{\epsilon_0^2+\Delta^2}$.
Then,  the equations reduce to the
non-driven Bloch-Redfield
equations of
Ref. \cite{Aslangul}.  Some different
weak coupling master equations for the driven spin-boson model have been
derived in Ref. \cite{GWW} using the path integral approach. The
equation for $\sigma_z(t)$ (not shown here) has the form of a
{\it closed} integro-differential equation of  rather involved form.
In the limit of vanishing dissipation
it reduces to Eq. (\ref{eq:exact2}) derived within a projection operator
formalism.

The {\it numerical} equivalence of the our driven Bloch-Redfield
equations and the
weak-coupling integro-differential equation of Ref. \cite{GWW}
has been demonstrated in Ref. \cite{PRE2000}, both by comparison
of the numerical solutions of both equations for the
initial-to-intermediate
part of the relaxation time scale and by comparison of the
numerical solution of
the Bloch-Redfield equation and an approximate analytical solution
of the weak-coupling  GME of path-integral approach on the {\it whole}
relaxation time-scale. This numerical comparison
has been performed for periodically driven case,
$\epsilon(t)=-\epsilon_0-s\cos(\Omega t)$, for the Ohmic bath with
exponential cutoff, $J(\omega)=4\pi\alpha\omega e^{-\omega/\omega_c}$,
where $\alpha$ is the dimensionless coupling strength (Kondo
parameter) which has to be sufficiently small \footnote{
An important restriction is:
$\alpha(\Delta/\omega_0)^2\ln(\omega_c/\omega_0)\ll 1$
for $\omega_c\gg\omega_0$.
It stems from the requirement of the smallness of
the frequency Lamb shift, $\omega_0\to\omega_r$,
at $T=0$. This restriction is most crucial for
$\epsilon_0=0$, where $\omega_0=\Delta$ and $\omega_r=
\Delta_r\approx\Delta [1-\alpha\ln (\tilde \omega_c/\Delta)]\approx
\Delta\exp[-\alpha\ln (\tilde\omega_c/\Delta)]\approx\Delta
(\frac{\Delta}{\tilde\omega_c})^{\alpha/(1-\alpha)}$ (for $\alpha\ll 1$,
to the linear order in $\alpha\ln(\tilde\omega_c/\Delta)$).
Thus,
this frequency shift is consistent  with the
renormalisation in section 6.3.
For a large asymmetry $\epsilon_0\gg \Delta$, the validity range of
Bloch-Redfield equations in
$\alpha$ becomes broader.}.

Both approaches agree quite well,  cf. in Fig. \ref{Fig3}.
 The presented approach, however, is technically more convenient. The results possess
a broad range of applications; for example,
it allows one to study a mechanism of  suppression
of quantum decoherence by strong periodic fields for a
two-level atom dynamics in
an optical cavity  \cite{Thorwartetall}.
The investigation of similar mechanisms is also of prime importance for the investigation of the
quantum decoherence in various quantum information processing applications \cite{FonsecaKohlerHanggi}.

\subsection{Beyond weak coupling theory: Strong system-bath coupling}

Thus far we concentrated on the case of  weak-coupling to quantum
thermal heat bath, or the regime of weak dissipation, respectively.
The analytic theory is,  however, not restricted to the case of weak
dissipation only. In fact, by use of a combination with the method
of canonical (unitary) transformations one can  study the opposite
limit of strong dissipation and weak tunnelling. To do so, let us
consider the spin-boson problem in Eq. (\ref{spinboson}) in the case
of a strong coupling between the quasi-spin and the bath degrees of
freedom. As a primary effect, the bath oscillators will become
shifted due to this coupling to new  positions which  depend on the
spin state. If the tunnelling coupling $\Delta$ were {\it absent},
then the small
polaron  unitary 
transformation
\cite{Frohlich,Holstein,Davydov,LangFirsov,PetrovBook,SilbeyHarris,
LeggettReview,Wurger}
\begin{eqnarray}
\hat U=\exp[\frac{1}{2}\hat \sigma_z\hat R], \;\;\;
\hat R=\sum_\lambda
\frac{\kappa_\lambda}{\hbar\omega_\lambda} (B_\lambda^{\dagger}-B_\lambda)
\end{eqnarray}
 to the new basis of displaced bath oscillators
 $B^{\dagger}_\lambda=U^{\dagger}b^{\dagger}_\lambda
 U=b_\lambda^{\dagger}+\frac{\kappa_\lambda}{2\hbar\omega_\lambda}\hat\sigma_z$,
$B_\lambda=U^{\dagger}b_\lambda
 U=b_\lambda+\frac{\kappa_\lambda}{2\hbar\omega_\lambda}\hat\sigma_z$
and boson-dressed spin states, $|\tilde n\rangle:=\hat U^{\dagger}
 |n\rangle$
would in fact  diagonalise
the Hamiltonian,
solving thereby the problem
of finding the eigenstates of the {\it total} system {\it exactly}.
For this reason, the corresponding canonically
transformed
basis  of phonon-dressed quasi-spin states (polaronic states)
and displaced bath oscillators is well suited
for an approximate treatment in the case of
weak intersite tunnelling and strong system-bath coupling.
In this new polaronic basis the Hamiltonian reads,
\begin{eqnarray}\label{pbasis}
 H(t) &  = & \frac{1}{2}\hbar\epsilon(t) \Big [|\tilde 1\rangle
\langle \tilde 1|-|\tilde 2\rangle
\langle \tilde 2| \Big ] \nonumber \\  &+ &  \frac{1}{2}\hbar\Delta(t)
 \Big ( <e^{\hat R}>_B|\tilde 1\rangle
\langle \tilde 2| + <e^{-\hat R}>_B|\tilde 2\rangle
\langle \tilde 1|\Big) \nonumber \\
& + & \frac{1}{2}\hbar\Delta(t)\Big ([e^{\hat R}- <e^{\hat R}>_B]|\tilde 1\rangle
\langle \tilde 2| + [e^{-\hat R}-<e^{-\hat R}>_B]|\tilde 2\rangle
\langle \tilde 1| \Big )\nonumber \\
 & + & \frac{1}{2}\sum_{\lambda}\hbar\omega_\lambda
  (B^{\dagger}_\lambda B_\lambda + 1/2 )
  -\lambda \hat I/4
\end{eqnarray}
where
\begin{eqnarray}
\lambda=\frac{\hbar}{2\pi}\int_0^\infty
\frac{J(\omega)}{\omega}d\omega
\end{eqnarray}
is the reorganisation energy.

Since $<\exp[\pm \hat R ]>_B=\exp[<\hat R^2>_B/2]=\exp[-D]$, where
$D=\frac{1}{4\pi}\int_0^
\infty[J(\omega)\coth(\beta\hbar\omega)/\omega^2]d\omega$
and $\beta=1/(k_BT)$,
the effective tunnelling coupling, $\Delta_r:=\Delta\exp(-D)$,
between the polaronic states is  exponentially suppressed
by the Debye-Waller factor \cite{Holstein,Davydov,PetrovBook}. For
the  relevant case of Ohmic coupling, $J(\omega)=4\pi\alpha\omega
\exp(-\omega/\omega_c)$,
$D\to \infty$ and $\Delta_r\to 0$ due to the infrared divergence of
the corresponding integral. One can attempt to remove this
divergence by using instead
of $\kappa_\lambda$ in the polaron transformation some variational
parameters to be determined from the requirement of a minimum of the (free) energy
of the whole system \cite{SilbeyHarris}. An approximate solution of the
corresponding variational problem by
using the Peierls-Bogolyubov-Feynman upper  bound for the
free energy \cite{Peierls,Callen}
leads \cite{SilbeyHarris} to a self-consistent equation
for $\Delta_r$ which at $T=0$  and for the symmetric case
$\epsilon(t)=0$ reads,
\begin{eqnarray}\label{Deltar}
\Delta_r=\Delta\exp\Big [-\frac{1}{4\pi}\int_0^
{\infty}\frac{J(\omega)}{(\omega+\Delta_r)^2}d\omega\Big].
\end{eqnarray}
Numerically, it can be solved by iterations. An approximate analytical
 solution is also available in the limiting case
$\omega_c\gg \Delta$ for $\alpha<1$. It
yields the celebrated renormalised tunnelling matrix element,i.e.,
$\Delta_r=\Delta\Big (\frac{\Delta}{\tilde\omega_c}\Big)
^\frac{\alpha}{1-\alpha}$ \cite{Chakravarty,BrayMoore,
LeggettReview,SilbeyHarris,WeissBook}
with $\tilde \omega_c=C\omega_c$, where $C$
is some constant which depends on the precise form of cutoff function in
$J(\omega)$. In this case, the use of the variationally optimised
polaron basis allows one to obtain an effective
Bloch-Redfield description which interpolates
 well between  weak and strong dissipation, see for
 the undriven case the study in Ref. \cite{SilbeyHarris}. The
corresponding generalisation of this approach onto the
driven case for an intermediate coupling strength $\alpha<1$ remains yet to be done.
Within our approach this generalisation
is rather straightforward.

We proceed further with the case of a strong coupling, i.e. $\alpha\geq
1$, where $\Delta_r$ does iterate to zero for any fixed value of
$\omega_c$. This fact indicates the famous dissipation-induced localisation  transition
\cite{BrayMoore,Chakravarty,LeggettReview}. In this case, the discussed
divergence is not removable; it is real. The polaronic states are
strictly localised in this case. This is also  the feature that causes
 the localisation phase transition in the dissipative tight-binding
model \cite{Schmid,Guinea,WeissBook}. The third line in Eq. (\ref{pbasis})
presents a (small) time-dependent interaction between the dressed
system and the bath
which can be handled in perturbation theory in the lowest order of tunnelling
coupling $\Delta$.
Applying the GME (\ref{matrixGME})
to the considered
case of an Ohmic bath yields \cite{JCP95,PRE95}
a GME in the form of Eqs. (\ref{GMEsigma}), (\ref{fg})
wherein $f_0(t,t')$ and $g_0(t,t')$ assume, however,  a distinct different form; namely,
\begin{eqnarray}\label{f0}
f_0(t,t') & = &\Delta(t)\Delta(t')\exp[-{\rm Re}\, Q(t-t')]\cos[{\rm Im}\, Q(t-t')],
\nonumber \\
g_0(t,t') & = &\Delta(t)\Delta(t')\exp[-{\rm Re}\, Q(t-t')]\sin[{\rm Im}\, Q(t-t')],
\end{eqnarray}
where
\begin{eqnarray} \label{Q}
Q(t)=\int_{0}^{t}dt_1\int_{0}^{t_1}K(t_2)dt_2+i\lambda t/\hbar
\end{eqnarray}
denotes the doubly-integrated autocorrelation function
of the bath, $K(t)$, in Eq. (\ref{Kfunc}).
For $\Delta(t)=const$ the same generalised master equation was derived
in Ref. \cite{Dakh94} using a different approach. It has been
derived also in Refs. \cite{GrifSasWeis} using the path-integral method
within the so-called noninteracting blip approximation (NIBA).
In the case $\Delta(t)=const$ and $\epsilon(t)=const$, it reduces
to the NIBA master equation of Refs. \cite{Aslangul85,Dekker,MorCukTij}.

Notably, the driven NIBA
master equation is valid
for $\alpha\geq 1$ at $T=0$ and $\epsilon_0=0$ (and sufficiently
small $\Delta\ll\lambda/\hbar=2\alpha\omega_c$).
It can also be
used, however, for $\alpha<1$ for an asymmetric case, $\epsilon_0\neq 0$,
and/or for $T>0$,
where the dynamics (in the absence of driving) is incoherent
and where $\Delta_r=0$. The parameter domain, where this latter condition is
fulfilled, is defined from the
solution of a (more complicated than Eq. (\ref{Deltar}))
self-consistent equation for $\Delta_r$
which generally depends on the static bias $\epsilon_0$,
temperature $T$, cutoff $\omega_c$,
It can be solved  only numerically:
in particular, for $\epsilon_0\neq 0$ and $T=0$, the renormalised
tunnelling coupling vanishes,
$\Delta_r=0$, already for $\alpha>1/2$. Moreover, even for
zero energy bias,
$\epsilon_0=0$, the renormalised tunnelling
coupling vanishes at a
sufficiently high temperature,
$\pi\alpha k_BT>\hbar\Delta$ \cite{SilbeyHarris}.
Even more, for $\Delta_r \neq 0$, the incoherent tunnelling
regime holds obviously when $k_BT\gg \hbar\Delta_r$.
Surprisingly, however, for the symmetric situation, $\epsilon_0=0$,
the NIBA  master
equation turns out to be a very
good approximation even for arbitrarily small $\alpha$ and $T$
(including coherent dynamics)
in the so-called
scaling limit $\omega_c\gg \Delta$ with $\Delta_r$ fixed.
This remarkable fact is rationalised within the
path-integral approach \cite{WeissBook}. Some understanding
can be obtained by observing that in the limit of vanishing dissipation
$\alpha\to 0$ the NIBA master equation is exact and it reduces to the one in
Eq. (\ref{eq:exact2})
, for the initial condition being $\sigma_z(0)=\pm 1$.
This, however, amounts to a singular
limit which must be handled with care.

\subsubsection{Fast fluctuating energy levels}

Let us assume for the following an  incoherent quantum dynamics with a
time-independent tunnelling matrix element $\Delta(t)=const$.
In the case of fast stationary fluctuating energy levels the procedure of
Sec. \ref{TLS} leads (after Markovian approximation)
to an averaged dynamics in Eq. (\ref{fastB})
with the time-averaged transition rates given by
\begin{eqnarray}\label{GenGoldenRule}
\langle W_{12}(\epsilon_0)\rangle_{\epsilon} &=
&\frac{1}{2}\Delta^2 {\rm Re}\int_0^{\infty} e^{i\epsilon_0t-Q(t)}\langle
S(t)\rangle_{\epsilon}dt \\ \langle
W_{21}(\epsilon_0)\rangle_{\epsilon}& = & \frac{1}{2}\Delta^2 {\rm
Re}\int_0^{\infty} e^{-i\epsilon_0t-Q(t)}\langle
S^*(t)\rangle_{\epsilon}dt,
\end{eqnarray}
where
\begin{eqnarray}\label{Phi}
\langle S(t)\rangle_{\epsilon}:=\langle S(t+t_0,t_0)\rangle_{\epsilon} =
\langle e^{i\int_{t_0}^{t+t_0}\tilde\epsilon(t')dt'}\rangle_{\epsilon}
\end{eqnarray}
is the averaged propagator of the corresponding Kubo oscillator which does not
depend anymore on the initial time $t_0$, or the initial phase of driving.
These averaged rates can be also given in the equivalent spectral
representation form,
like in Eq. (\ref{rates}),
\begin{eqnarray}\label{ratesFC}
\langle W_{12}(\epsilon_0)\rangle_{\epsilon} & = &\frac{\pi}{2}
\Delta^2\int_{-\infty}^{\infty}
 FC(\omega)I(\epsilon_0-\omega)d\omega,\nonumber \\
\langle W_{21}(\epsilon_0)\rangle_{\epsilon} & = &\frac{\pi}{2}\Delta^2
\int_{-\infty}^{\infty}
e^{-\frac{\hbar\omega}{k_BT}} FC(\omega)I(\epsilon_0-\omega)d\omega,
\end{eqnarray}
where
\begin{eqnarray}\label{FC}
FC(\omega)=\frac{1}{2\pi}\int_{-\infty}^{\infty} \exp[i\omega t-Q(t)] dt
\end{eqnarray}
is the Franck-Condon factor \footnote{i.e. the  thermally weighted
overlap of the wave functions of displaced quantum oscillators.}
which describes spectral line
shape due to multi-phonon transitions
\cite{HuangRyss,KuboToyozawa,advance},
and $I(\omega)$ denotes the spectral line shape
of the  Kubo oscillator, $\dot X(t)=i\tilde\epsilon(t) X(t)$.
The result in Eq. (\ref{GenGoldenRule}) is in essence the Golden Rule
result generalised here to fast fluctuating nonequilibrium fields. This
fact underlines the generality and importance of the
{\it nonequilibrium} Golden Rule result
which is very useful in many applications. Many profound nonequilibrium
effects described in this work can be rationalised within its framework.
The structure of this result
has a clear physical interpretation. Namely, $FC(\omega)$ in (\ref{FC}) is nothing but
the spectral line shape of a {\it quantum} Kubo oscillator with
the frequency modulated by the quantum Gaussian force
$\hat \xi(t)$ in Eq. (\ref{xi}) (in the corresponding
Heisenberg representation)
which has the complex-valued equilibrium autocorrelation
function in Eq. (\ref{Kfunc}). Due to the Gaussian character of
the quantum random force, this spectral line shape in Eq. (\ref{FC}) is expressed merely
in terms of the doubly-integrated autocorrelation function $K(t)$
and the reorganisation energy term in Eq. (\ref{Q}).
Due to the equilibrium character of quantum fluctuations, $FC(\omega)$ possesses
a symmetry property, $FC(-\omega)=e^{-\beta\hbar\omega}
FC(\omega)$, which is enforced by the thermal detailed balance condition. It holds
independently of the form of the
bath spectral density $J(\omega)$ \cite{LeggettReview}.
 Thus, the thermal equilibrium for
{\it localised} energy levels \footnote{Reminder: we consider the case
$\Delta_r=0$, or $k_BT\gg \hbar\Delta_r$.},
$p_1(\infty)=e^{-\hbar\epsilon_0/k_BT}p_2(\infty)$, holds always in the
absence of {\it nonequilibrium} fluctuations of the energy levels.
Furthermore, by splitting  $\hat \xi$ into a sum
of two arbitrary statistically
independent components (two subsets of quantum bath oscillators),
$\hat \xi=\hat \xi_1+\hat\xi_2$  one can show that $FC(\omega)$
can exactly be represented as a frequency convolution
of the corresponding (partial) Franck-Condon factors
$FC_1(\omega)$ and $FC_2(\omega)$ \cite{advance,WeissBook},
namely,
\begin{eqnarray}\label{convolution}
FC(\epsilon)=\int_{-\infty}^{\infty}
FC_1(\omega)FC_2(\epsilon-\omega)d\omega.
\end{eqnarray}
Such frequency convolution can be generalised to an arbitrary number of
partitions.
The {\it nonequilibrium} Golden Rule in Eq. (\ref{ratesFC})
presents an additional frequency convolution with the spectral line shape $I(\omega)$
of the
{\it nonequilibrium} Kubo oscillator
 which
corresponds to a generally non-Gaussian and
nonequilibrium stochastic force.  $I(\omega)$  does now no longer possess  the
above symmetry imposed by thermal detailed balance. Thus,
the violation of the thermal detailed balance condition by
the nonequilibrium
fluctuations lead generally to intriguing nonequilibrium
effects described in Sec. \ref{TLS}, and below.
It is important
to notice that the localised states can be stabilised by strong,
fast oscillating periodic fields \cite{Makri} and
the Golden Rule description is generally improved for such fields \cite{JCP2000}.
This latter fact can be readily understood
from the representation of the (quantum) stochastic force as a sum
of statistically independent components. Namely,
if $\Delta_r=0$, due the interaction with a subset of oscillators,
the addition of an interaction with further oscillators cannot
enhance $\Delta_r$. It will work always in the direction to
make the effective tunnelling coupling smaller (when $\Delta_r\neq 0$), thus
improving
the perturbation theory in $\Delta$. Replacing equilibrium oscillators
with a fast fluctuating field does not change this trend.

\subsubsection[Exact averaging over dichotomous fluctuations]
{Exact averaging over dichotomous fluctuations \\ of 
the energy levels}

An exact averaging of the NIBA master equation of the driven
spin-boson model in the dichotomous Markovian
field is possible by analogy with the consideration pursued
in Sec. \ref{exact}.
The result is formally the same as in
Eq. (\ref{GMEsigmaAv}) with $f_0(t-t')$ and $g_0(t-t')$
given but in Eq. (\ref{f0}) (with $\Delta(t)$ = const)
\cite{PRE95}. An interesting feature is that for $\epsilon_0=0$
the equations for the average $\langle \sigma_z(t)\rangle$
and the correlator $\langle \alpha(t)\sigma_z(t)\rangle$
are decoupled. Moreover, in the dissipation-free  case, $Q(t)=0$,
the solution of equation for  $\langle \sigma_z(t)\rangle$
with the initial condition $\langle\sigma_z(0)\rangle=1$ yields
the same result as
in Eq. (\ref{sigmaX}) with the following substitutions implemented, i.e.,
$\langle\sigma_z\rangle\to\langle\sigma_x\rangle$,
$\sigma\to\Delta$, $\Delta\to\epsilon_0$.
This  finding provides a rather nontrivial
cross-check  of the validity of
different methods of stochastic averaging.

\subsubsection{Electron transfer in fast oscillating periodic fields}

Let us next focus on the case with  strong and fast periodic driving fields $\tilde
\epsilon(t)=A\cos(\Omega t+\varphi_0)$, yielding
\begin{eqnarray}\label{ratesPeriod}
\langle W_{12}(\epsilon_0)\rangle_{\epsilon} & = & \frac{\pi}{2}
\Delta^2 \sum_{n=-\infty}^{\infty}J_n^2\Big ( \frac{A}{\Omega}\Big)
FC(\epsilon_0-n\Omega),\nonumber \\
\langle W_{21}(\epsilon_0)\rangle_{\epsilon} & = & \frac{\pi}{2}
\Delta^2
\sum_{n=-\infty}^{\infty}J_n^2\Big ( \frac{A}{\Omega}\Big)
e^{-\frac{\hbar[\epsilon_0-n\Omega]}{k_BT}}
FC(\epsilon_0-n\Omega).
\end{eqnarray}
This result  of the Golden Rule type
for the nonadiabatic
electron transfer (ET) rates in strong periodic fields has been  derived
 in \cite{CPL96} and independently in \cite{DakhCoal95}. In particular,
the quasi-static (Gaussian) approximation for $FC(\omega)$ for
$k_BT\gg\hbar\omega_c$ with
$K(t)$ replaced by $K(0)\approx 2k_BT\lambda/\hbar^2$ in Eq. (\ref{Q})  leads  in absence of driving, i.e.  A=0,
independent of the detailed structure
of $J(\omega)$, to the celebrated Marcus-Dogonadze-Levich rate
expression \cite{Marcus,LevichDogon}
for the ET rates with
\begin{eqnarray}\label{Marcus}
FC(\omega)=\frac{\hbar}
{\sqrt{4\pi\lambda k_BT}}\exp\Big (-\frac{(\hbar\omega-\lambda)^2}
{4\lambda k_BT}\Big) \;.
\end{eqnarray}
This approximation
is suitable for a
thermal bath with a low frequency cut-off and in the high-temperature limit, e.g., for polar solvents.
This presents a semiclassical limit for the Franck-Condon factor.
If some high-frequency (quantum) vibrational mode $\omega_0$
couples to
ET with the coupling constant $\kappa_0$
in addition to the low-frequency vibrations, being of relevance
for ET in molecular aggregates,
then a different model for $FC(\omega)$ is more appropriate, namely
\cite{Jortner,advance},
\begin{eqnarray}\label{JortnerBixon}
FC(\omega)=\frac{\hbar}{\sqrt{4\pi\lambda k_BT}}
e^{- D_0}
\sum_{p=-\infty}^{\infty}I_{|p|}(x)e^{
-\frac{p\hbar\omega_0}{2k_BT}} \nonumber \\ \times
\exp\Big (-\frac{(\hbar\omega-\lambda+p\hbar\omega_0)^2}
{4\lambda k_BT}\Big)
\end{eqnarray}
where $S=(\frac{\kappa_0}{\hbar\omega_0})^2$,
$D_0=S\coth(\frac{\hbar\omega_0}{2k_BT})$,
$x=S/\sinh(\frac{\hbar\omega_0}{2k_BT})$,
and $I_p(x)$ is the modified Bessel function.
The periodic driving may induce an  inversion of ET
transfer direction and modulate ET transfer rates by orders
of magnitude. This has been theoretically predicted in
\cite{DakhCoal95,CPL96} for both of the above-mentioned models of
$FC(\omega)$. \footnote{The use of an improved perturbation theory in $\Delta$  in the case
of fast fluctuating fields does {\it not}
imply that the Golden Rule rates cannot be enhanced
by such {\it nonequilibrium}
fields.
A large enhancement of the forward (backward) rate can
occur, e.g., when
the absorption of $n$ photons
helps to overcome the corresponding forward (backward)
activation barrier of the thermally-assisted incoherent tunnelling.
For example,
for the generalised Marcus rates a  condition is
$\epsilon_0\mp\lambda/\hbar\pm n\hbar\Omega=0$,
with the field amplitude $A$ chosen such that the
probability
$J_n^2(A/\Omega)$ of the corresponding reaction channel
is maximised. }

\subsubsection{Dichotomously fluctuating tunnelling barrier}

Another relevant situation involves the case of fluctuating
tunnelling matrix
element $\Delta(t)$
and  constant energy bias $\epsilon(t)=\epsilon_0=const$.
In the superexchange picture of
ET this corresponds to a physical situation where the stochastic dynamics
of the bridge states, which mediate the electron transfer between
the donor and acceptor molecules, introduces an explicit, stochastic time-dependence
into $\Delta(t)$.
Generically, this corresponds to a fluctuating tunnelling barrier.
In the case of  dichotomous Markovian fluctuations $\Delta(t)=\Delta_0+
\Delta \alpha(t)$, the stochastic averaging of the NIBA master equation
can  be done exactly \cite{JCP95}. Towards this goal one makes use of the
Shapiro-Loginov theorem (\ref{Shap}) and the following
exact decoupling property \cite{BourFrischPouq,HanggiProceedings,Klyatskin}:
\begin{eqnarray}
\langle \alpha(t)\alpha(t')\sigma_z(t')\rangle=
\langle \alpha(t)\alpha(t')\rangle \langle\sigma_z(t')\rangle\;.
\end{eqnarray}
Applying these two theorems and
using the DMP property, $\alpha^2(t)=1$, the averaging of the GME
yields the following exact
results \cite{JCP95}:
\begin{eqnarray}
\label{GMEsigmaAv2}
\frac{d}{dt}\langle \sigma_z(t)\rangle  =
& - & \int_{0}^{t}\Big (  [\Delta_0^2
+ \Delta^2 e^{-\nu(t-t')}]  f(t-t')\langle \sigma_z(t')\rangle
\nonumber\\
 & + & \Delta_0\Delta[1+e^{-\nu(t-t')}]
f(t-t')\langle\alpha(t') \sigma_z(t')\rangle
\nonumber\\
& + &  [\Delta_0^2
+ \Delta^2 e^{-\nu(t-t')}] g(t-t')
\Big)dt', \\
\frac{d}{dt}\langle \alpha(t)\sigma_z(t)\rangle  = & - &\nu
\langle \alpha(t)\sigma_z(t)\rangle
 -  \int_{0}^{t}\Big (  [\Delta^2
+ \Delta_0^2 e^{-\nu(t-t')}]  f(t-t')\langle \alpha(t')
\sigma_z(t')\rangle
\nonumber\\
 & + & \Delta_0\Delta[1+e^{-\nu(t-t')}]
\{f(t-t')\langle \sigma_z(t')\rangle + g(t-t')\}
\Big)dt'
\;, \nonumber
\end{eqnarray}
where
\begin{eqnarray}\label{fDelta}
f(t) & = &\exp[-{\rm Re}\, Q(t)]\cos[{\rm Im}\, Q(t)]\cos[\epsilon_0 t],
\nonumber \\
g(t) & = &\exp[-{\rm Re}\, Q(t)]\sin[{\rm Im}\, Q(t)]\sin[\epsilon_0 t].
\end{eqnarray}
For the case of vanishing dissipation, $Q(t)=0$, and for $\Delta_0=0$,
the solution of this integro-differential equation
for $\langle \sigma_z(t)\rangle$ for the initial condition
$\langle\sigma_z(0)\rangle=1$ yields  the same result as in Eq.
(\ref{sigmaZ}). This agreement
provides an additional test for the mutual consistency of different
methods of stochastic averaging used here.

 Furthermore, in the absence of dissipation the
 rate of incoherent relaxation exhibits a resonance-like feature as a function of
 the frequency $\nu$ of the barrier fluctuations. Namely, a resonance
 occurs when $\nu$
 matches the transition frequency $\epsilon_0$, i.e.
 $\nu=\epsilon_0$ (see Eq. (\ref{resonance}) in Sec. \ref{TLSclassic}).
 This presents
 a physical stochastic resonance, which should not to be identified with a well-known
 phenomenon of noise-assisted Stochastic Resonance
 \cite{SRReview}. It occurs when a stochastic frequency of the driving
 matches an eigenfrequency of a quantum transition. In the presence
 of dissipation, this resonance feature is maintained, but becomes modified.
 Namely, the resonance can occur at $\nu=|\epsilon_0\pm \lambda/\hbar|$,
 rather than at $\nu=\epsilon_0$
 \cite{JCP95}. This resonance is responsible for the interesting phenomenon
 of a stochastic acceleration of  dissipative quantum tunnelling which
 is predicted by the theory \cite{JCP95}: For the case that
 $\Delta:=\Delta_0$, when $\Delta(t)$ fluctuates between zero and
 $2\Delta_0$ the rate of incoherent transfer can exceed that for the static
 tunnelling barrier with tunnelling coupling strength
 $\Delta(t)=2\Delta_0=$const. At the first sight,
 this effect seems
 paradoxical; it must be remembered, however, that the considered noise
 is {\it nonequilibrium} and it is capable of pumping energy into the system
 enhancing thereby the rate of incoherent quantum
 tunnelling. Unfortunately, for the parameters typical for molecular ET
 the experimental conditions for this effect to occur can barely be met experimentally
 because the required frequency $\nu$ is too high. Nevertheless, this
 fact does not invalidate the principal possibility of the discussed
 effect for some other physical
 systems in view of the generality of the model set-up.

 In contrast, when $\Delta(t)$
 fluctuates very slowly on the time-scale of decay of
 kernels $f(t)$
 and $g(t)$, which  corresponds roughly to the inverse of the width of
 corresponding Franck-Condon
 factor $FC(\omega)$,
 then  our theory predicts --  after use of the Markovian
 approximation  -- the known results which corresponds to the
 approximation of a dichotomously fluctuating rate
 \cite{GehlMarchChand}, see also discussion in Sec. \ref{time-rate}.
 The corresponding problem of such fluctuating rates is known under the label
  of {\it dynamical disorder}
 and can be met in quite different areas of physics and chemistry
 \cite{ZwanzigR}.
 Depending on the relation between the stochastic frequency $\nu$ and
 the values of transfer rates corresponding to the
 ``frozen'' instant realizations of $\Delta(t)$, the transfer kinetics
 can exhibit different regimes of a (i) quasi-static disorder,  (ii) an averaged
 rate description, and (iii) a gated regime \cite{JCP95}.
 In the latter case, the mean transfer time becomes locked
 to the autocorrelation time of the fluctuations \cite{McCammon,PRE97}.

The influence of strong laser fields on the electron transfer with
nonequilibrium dynamical
disorder \cite{JCP97}, or driven by nonequilibrium conformational
fluctuations  \cite{PRE97}
has been studied within the obtained NIBA master equation
approach in Refs. \cite{JCP97,PRE97}. In particular, it has been
shown there, that a strong periodic field can induce a turnover between the
nonadiabatic regime of electron transfer and a gated regime. Moreover,
the direction of electron transfer in the gated regime can be inverted,
whereas the mean transfer time remains chiefly
controlled by the nonequilibrium stochastic fluctuations
and it is not influenced by periodic field \cite{JCP97}.
These theoretical predictions discussed here
are still awaiting  their experimental
realization. The  area of  chemically gated, or
chemically driven
electron transfer \cite{Davidson}, that is
 the electron transfer controlled by nonequilibrium
fluctuations due to spontaneous release of energy by breaking
some energy-rich
chemical bonds (e.g., due to the ATP hydrolysis), is currently still in its infancy
 \cite{Davidson}.

\section[Driven transport in dissipative tight-binding models]
{\label{sec7}Quantum transport in dissipative tight-\\ 
              binding  models subjected to strong \\
              external fields}

A salient application of our theoretical efforts relates  to
charge and particle transfer in spatially extended molecular structures.
It can be described within a model similar to the
Holstein model of a molecular crystal
\cite{Holstein,MikkelsenRatner}. Namely, one considers a molecular chain
using the assumption that only one energy, namely a lowest, unoccupied molecular orbital(LUMO)-level
describes the  electron transfer (or highest occupied, HOMO-level in case of a hole transport)
per molecule, or molecular group.
These energy levels are coupled to the local intramolecular
vibrations
which are thermalised. The transferring particle is  delocalised due to a
tunnelling coupling between the nearest neighbours. The intersite
coupling between
the intramolecular vibrations is however neglected (like in the Einstein
model of optical phonons),
i.e. the electron (or hole) energy levels in neighbouring molecules (or molecular
groups) are assumed to
fluctuate  independently.
In other words, one assumes uncorrelated identical
thermal baths formed by vibrational degrees of freedom of
each molecule in a molecular chain. Such a model is close in spirit to one
used for exciton transfer within a stochastic Langevin  description \cite{ReinekerBook}.
In the approximations employed below, this model becomes equivalent to the
model of a Quantum Brownian Motion within a single band,
tight-binding description. In an external electric field $\mathcal{ E}(t)$,
the latter one reads  \cite{GrifHangRev,KohlerRev}:
\begin{eqnarray}\label{QBM}
H_{\rm TB}(t)=
-\frac{\hbar\Delta}{2}\sum_{n=-\infty}^{\infty}(|n\rangle\langle
n+1|+|n+1\rangle\langle n|)-e \mathcal{ E}(t) \hat x \; +
H_{\rm BI},\\
 H_{\rm BI}=\frac{1}{2}\sum_{i}\Big [\frac{\hat
p_i^2}{m_i}+m_i\omega_i^2 \Big (\hat
q_i-\frac{c_i}{m_i\omega_i^2}\hat x\Big ) ^2 \Big ], \nonumber
\end{eqnarray}
where $\hat x=a\sum_{n}^{}
n|n\rangle\langle n|$ is the operator of the coordinate (within
the single band description).
The model in Eq. (\ref{QBM}) can  be derived from a
different perspective than the Holstein model, namely, by starting out from
a model of Quantum Brownian Motion in a periodic potential \cite{WeissBook,
GrifHangRev,ReimannRev} and by restricting
the corresponding consideration to the lowest band for the tunnelling particle
in the deep quantum regime. We  consider this model in the limit of a strong
coupling by applying the small polaron transformation which now reads
$\hat U=\exp[-i\hat x\hat P/\hbar]$, $\hat P=\sum_i
c_i\hat p_i/(m_i\omega_i^2)$. In the polaron basis, the Hamiltonian
reads
\begin{eqnarray} \label{TBpolaron}
H_{\rm TB}(t) =
-\frac{\hbar\Delta_r}{2}\sum_{n=-\infty}^{\infty}(|\tilde n
\rangle\langle
\tilde n+1|+|\tilde n+1\rangle\langle \tilde n|)-e \mathcal{ E}(t) \hat x
\nonumber \;\\
  -
\sum_{n=-\infty}^{\infty}(
\hat\xi|\tilde n\rangle\langle \tilde n+1| +h.c.)+
\frac{1}{2}\sum_{i}\Big [\frac{\hat p_i^2}{m_i}+m_i\omega_i^2
\tilde Q_i^2 \Big ],
\end{eqnarray}
where $\Delta_r=\Delta \langle e^{-ia\hat P/\hbar}\rangle_B=
\Delta e^{-a^2\langle \hat P^2\rangle_B/2\hbar^2}$ is the renormalised
tunnelling coupling (polaron band width), $\tilde Q_i:= \hat U q_i\hat
U^{-1}=\hat q_i-\frac{c_i}{m_i\omega_i^2}\hat x$ are displaced
bath oscillators and $\hat\xi=
\frac{\hbar}{2}[\Delta e^{-ia\hat P/\hbar}-\Delta_r]$ is the quantum
random force operator in the polaron basis which is considered further as
a small perturbation.
Note that $\hat x$
is not changed. Assuming a strong Ohmic dissipation with $\alpha\geq 1$
yields $\Delta_r=0$ at $T=0 K$ and for $\mathcal{ E}(t)=0$.
This indicates the celebrated localisation phase
transition \cite{Schmid,Guinea}, which alternatively can also be interpreted as a polaron
band collapse. In the presence of a constant electric
field and/or for $T>0$ this
localisation transition occurs for smaller values of $\alpha$.
Given our case of strong coupling, the transport occurs predominantly via
incoherent tunnelling hops between the nearest sites of localisation.
As a side remark, we note that also in the
dissipation-free case the Bloch
band  can collapse in presence of strong periodic fields \cite{Holthaus},
known as
the effect  of dynamical localisation \cite{KenkreDunlap}.
Use of Eq. (\ref{matrixGME}) for the case in
(\ref{TBpolaron}) with $\Delta_r=0$ yields for the diagonal elements
of the reduced
density matrix a set of coupled generalised master equations
\begin{eqnarray}\label{NIBAband}
\dot
\rho_{nn}(t) &=&\int_{0}^{t}\{W^{(+)}(t,\tau)\rho_{n-1n-1}(\tau)
+W^{(-)}(t,\tau)\rho_{n+1n+1}(\tau)
\nonumber \\
 &-& [W^{(+)}(t,\tau)+W^{(-)}(t,\tau)]\rho_{nn}(\tau) \}d\tau
\end{eqnarray}
with memory kernels
\begin{eqnarray}
W^{(\pm)}(t,\tau)=\frac{1}{2}\Delta^2e^{-{\rm Re}\;Q(t-\tau)}
\cos[{\rm Im} \;Q(t-\tau)\mp
\frac{ea}{\hbar}\int_{\tau}^{t}\mathcal{ E}(t^{\prime})dt^{\prime}  ].
\end{eqnarray}
The very same equations are obtained in the NIBA approximation
of the path-integral approach \cite{HartGrifHang}.
The Holstein like model which has been discussed at the beginning
of this Section
yields in  similar
approximations
the same set of GME's (with a trivial
renormalisation of the coupling constant in the identical bath
spectral densities $J_n(\omega)=J(\omega)$) \cite{PLA98}.
The stationary electrical current carried
by one particle reads
$j=e\lim_{t\to\infty}\frac{d}{dt}\langle x(t)\rangle$, where
$\langle x(t)\rangle=a\sum_n n\rho_{nn}(t)$ denotes the mean particle
position in the considered infinite chain. It
obeys (this result follows immediately from Eq. (\ref{NIBAband}))
\begin{eqnarray}\label{intermediate}
\frac{d}{dt} \langle x(t) \rangle = a\int_{0}^{t}
[W^{+}(t,\tau) - W^{-}(t,\tau)]d\tau.
\end{eqnarray}
This current in (\ref{intermediate}) still needs to be averaged of the
stochastic field realizations.
\footnote{In the case of periodic driving, this additional
averaging is obsolete by defining the current in a self-averaged
manner as
$j=e\lim_{t\to \infty}\frac{\langle x(t)\rangle}{t}$}
This objective is again reduced to the averaging of an effective Kubo oscillator
which can be done exactly for many different models of stochastic driving.
We decompose the electric field $\mathcal{ E}(t)$ into the sum of the
mean, or constant field $\mathcal{E} _0$ and a fluctuating, unbiased component
$\tilde{ \mathcal{E}} (t)$, i.e.
$\mathcal{ E}(t)=\mathcal{ E} _0+\tilde{ \mathcal{ E}} (t)$.
 The resulting
expression for the averaged current $j(\mathcal{ E}_0)$ can be put into
 two equivalent forms.
First, it can be written in terms of  a time integral \cite{PRL98,EPL98},
\begin{eqnarray}\label{first}
j(\mathcal{ E}_0)=ea\Delta^2\int_{0}^{\infty} \exp[-{\rm Re}\; Q(\tau)]
\sin[{\rm Im}\; Q(\tau)] {{\rm Im} }[e^{iea \mathcal{ E}_0\tau/\hbar}
\langle S(\tau) \rangle]d\tau,
\end{eqnarray}
where $\langle S(\tau) \rangle$ is given in Eq. (\ref{Phi}) with
$\tilde \epsilon(t)=ea\tilde{ \mathcal{ E}}(t)/\hbar$ and $Q(t)$ in
Eq. (\ref{Q}).
Alternatively, the current expression can be given as a
frequency convolution in a spectral representation form, i.e.,
\begin{eqnarray}\label{second}
j(\mathcal{ E}_0)=\int_{-\infty}^{\infty}
j_{dc}(\omega) I(ea\mathcal{ E}_0/\hbar-\omega)d\omega,
\end{eqnarray}
where
\begin{eqnarray}
j_{dc}(\omega)=
\frac{\pi}{2}ea\Delta^2(1-e^{-\hbar\beta\omega}) FC (\omega) \;.
\end{eqnarray}

$I(\omega)$ denotes
 the spectral line shape corresponding to $\langle S(\tau) \rangle$.
The dc-current obeys the symmetry property
$j_{dc}(-\omega)=-j_{dc}(\omega)$ which is imposed by the thermal detailed
balance symmetry,
$FC(-\omega)=e^{-\hbar\beta\omega} FC(\omega)$ with
$FC(\omega)$ in (\ref{FC}). It is
important to note that the averaged current in (\ref{second})
does not obey such a symmetry requirement.

\subsection{Noise-induced absolute negative mobility}

As a first application of the above results we consider the
phenomenon of ANM, or absolute negative
mobility, where the transferring particles move around {\it zero
bias} in  {\it opposite direction} to the average applied force.
 This effect has been anticipated for
semiconductors in strong periodic fields almost thirty years ago
using a Boltzmann equation approach \cite{Pavlovich,Bass}. The first
experimental realization was obtained
 in 1995
for semiconductor superlattices  \cite{Keay}. The
corresponding experimental
results were seemingly
consistent \cite{Keay} with a mechanism of incoherent sequential tunnelling
like one just described.
The occurrence of the ANM phenomenon
for a sinusoidal driving
within the considered dissipative
tight-binding model has been demonstrated  in Ref. \cite{HartGrifHang}.

The question we addressed in Ref.\cite{PLA98} within a Holstein like
model was whether an external stochastic field can also induce ANM.
The occurrence of such noise-induced ANM has been shown for
dichotomous Markovian fields. ANM presents a multi-state analogy of
the effect of inversion of populations in TLS described in Sec.
\ref{TLS}. A simple criterion for ANM to occur can be found within
the quasi-static approximation for the spectral line shape
$I(\omega)$. For a symmetric dichotomous field $\tilde{ \mathcal{
E}}(t)=(\hbar\sigma/ea)\alpha(t)$ with the inverse autocorrelation
time $\nu$, this quasi-static approximation holds whenever
$\sigma\gg \nu$, being almost always the case in the relevant regime
of parameters even if the field fluctuations are fast on the
time-scale of the charge transfer. Then, $I(\omega)\approx
\frac{1}{2}[\delta(\omega-\sigma)+ \delta(\omega-\sigma)]$ and
$j(\mathcal{ E}_0)=\frac{1}{2} [j_{dc}(ea\mathcal{ E}_0/\hbar
-\sigma)+j_{dc}(ea\mathcal{ E}_0/\hbar +\sigma)]$.

Given  the symmetry property, $j_{dc}(-\sigma)=-j_{dc}(\sigma)$, one can
conclude that the phenomenon of ANM will occur in {\it any} such system
with the static current-voltage characteristics
 $j_{dc}(\sigma)$ assuming a maximum at some
$\sigma_{max}$ which is complemented by a corresponding regime of
{\it differential}  negative conductance occurring for
$\sigma>\sigma_{max}$. Then, $j(\mathcal{ E}_0)<0$ for a
sufficiently small static force,  $e\mathcal{ E}_0>0$, whenever
$\sigma>\sigma_{max}$ \cite{PLA98}, i.e. whenever  the charge
transfer is driven into the regime of negative differential
conductance by some appropriately chosen alternating, two-state
stochastic fields. This mechanism is quite general and robust. It
does not depend on the details of the dissipation mechanism. In
particular, for the Gaussian $FC(\omega)$ in Eq. (\ref{Marcus}), we
obtain
\begin{eqnarray}
j_{dc}(ea\mathcal{ E}_0/\hbar)=
\frac{\pi}{2}\frac{ea\Delta^2\hbar}{\sqrt{\pi\lambda k_BT}}
\exp\Big [-\frac{\lambda^2+(ea\mathcal{ E}_0)^2}{4\lambda k_BT} \Big]
\sinh \Big[\frac{ea\mathcal{ E}_0}{2k_BT} \Big].
\end{eqnarray}
This corresponds to a (nonadiabatic) small polaron conductance
\cite{Holstein,Emin}
with the differential mobility, 
$\mu(\mathcal{ E}_0)=\frac{d v(\mathcal{ E}_0)}{d\mathcal{ E}_0}$, 
obeying in the linear response regime
\begin{eqnarray}
\mu(0)=\sqrt{\frac{\pi}{2W_p}} \frac{ea^2 V^2}{\hbar(k_BT)^{3/2}}
e^{-W_p/2k_BT} \;,
\end{eqnarray}
where
$W_p=\lambda/2$ is the polaron binding energy and $V=\hbar\Delta/2$.
For this nonadiabatic small polaron model the regime of negative differential mobility
occurs for $\mathcal{ E}_0>\mathcal{ E}_{max}$ with $\mathcal{ E}_{max}$ defined
implicitly by the equation $ea \mathcal{ E}_{max}=
2W_p\coth(\frac{ea\mathcal{ E}_{max}}{2k_BT})$. Quasi-one-dimensional systems
exhibiting this small polaron conductance (in the nonadiabatic ET regime with
respect to $\Delta$) can be considered along with
the semiconductor superlattices as possible candidates to
exhibit the phenomenon of noise-induced ANM  experimentally. A finite photo-induced
small polaron mobility
of the hole type is found, for example, in columnar liquid crystals
\cite{Kreozis,Shiyan,Bredas}.
We then estimate the value of $\mathcal{ E}_{max}$ for
these systems with the lattice period of about $a=0.35$ nm to be in the
range of $5\cdot 10^6$ V/cm, which is rather large.
 For superlattices with a larger period $a$,
$\mathcal{ E}_{max}$ can be much less \cite{Keay}. Basically, this crucial
quantity is determined by two factors: (i) the width of $FC(\omega)$
due to multi-phonon transitions (it depends on the precise mechanism of
dissipation and should be made as small as possible)
and (ii) the lattice period $a$ (it should be engineered as large as possible).
These criteria can serve as a useful guides in identifying
the  appropriate experimental materials.

\subsection{Dissipative quantum  rectifiers}

Yet another intriguing application is provided by the fluctuation-induced quantum transport
in the {\it absence} of a mean electric field, $\mathcal{ E}_0=0$. Similar
nonequilibrium phenomena are known under the notion of Brownian motors,
or Brownian ratchets \cite{Magnasco93,Prost94,BartHangKis94,AstumBier94,Doering94,Bartussek1996,HanggiBartussek,
AstumianHanggi,HanggiReimann,ReimannRev,HanggiMarchesoniNori}. 
The first case of a
quantum ratchet in a periodic
spatially asymmetric
(ratchet)-potential was studied theoretically in Ref. \cite{ReimGrifHang97}
within a semi-classical approach and
for an
adiabatically varying driving field.
In Ref. \cite{PRL98,EPL98,JPCB2001},
we put forward
periodic dissipative
nonadiabatic quantum rectifiers \cite{popular} operating
in the absence of spatial asymmetry. The current is produced by
a nonlinear transport mechanism due to an interplay between
 equilibrium quantum fluctuations and an unbiased,
 but asymmetric nonequilibrium
external noise \cite{PRL98}. Likewise, an asymmetric periodic
driving of the harmonic mixing type
can be used instead of the nonequilibrium noise \cite{EPL98,JPCB2001}.
Our rectifier behaves genuinely
quantum mechanically and corresponds to the case
of a strong dissipation
when the transport
mechanism is incoherent
and the transport proceeds by
incoherent tunnelling hopping as outlined above.
The origin of the resulting current can be traced to  Eq. (\ref{first})
and Eq. (\ref{second}). Namely, $j(0)\neq 0$, when
$\langle S(\tau) \rangle$ assumes a complex values, i.e.,
${\rm Im} \langle S(\tau) \rangle \neq 0$. This corresponds to a
complementary criterion which follows from Eq. (\ref{second}).
Namely, $j(0)\neq 0$, when the corresponding spectral line $I(\omega)$
is asymmetric, $I(-\omega)\neq I(\omega)$.

In particular, this
is the case of asymmetric dichotomous field of zero mean, cf. Sec.
\ref{KuboOsc} and Eq. (\ref{kuboline}),
which takes on the (frequency scaled) two discrete
values $ea\tilde{ \mathcal{ E}}_{1,2}/\hbar=\epsilon_{1,2}=
\mp \sigma e^{\mp b/2}$,
where $b$ characterises the field asymmetry and $\sigma$ is
the (scaled)  rms of field fluctuations. The emergence of a finite current in this case
 can readily be seen in the quasi-static approximation of
$I(\omega)$ for $\sigma\gg \nu$, $I(\omega)\approx p_1\delta(\omega-
\sigma e^{-b/2})+p_2\delta(\omega+\sigma e^{b/2})$ with $p_{1,2}=
|\epsilon_{2,1}|/(|\epsilon_1|+\epsilon_2)$. In this adiabatic
(with respect to driving) approximation,
\begin{eqnarray}\label{jquasistatic}
j(0)=p_2 j_{dc}(\sigma e^{b/2})-p_1 j_{dc}(\sigma e^{-b/2}).
\end{eqnarray}
In the semiclassical high-temperature approximation
for $FC(\omega)$ in Eq. (\ref{Marcus}) (this corresponds to
 noise-driven small
polaron transport), one can see that the rectification current appears as a
nonlinear response to the external, unbiased on average driving. Namely,
to the lowest order, the current is proportional to $\langle \mathcal{ E}^3(t)\rangle$,
$j(0)\propto \langle \mathcal{ E}^3(t)\rangle\approx
b\sigma^3$ ($b\ll 1$), with a nontrivial prefactor. Moreover, the
current flows into the direction of $\langle e^3\mathcal{ E}^3(t)\rangle$,
which is the direction of the larger force realization,
if the applied random force is sufficiently small. With an increase
of the noise rms $\sigma$ the current can however {\it reverse} its direction. In the
considered approximations and for a small driving asymmetry $b\ll 1$
this occurs when $\sigma$ exceeds some maximum associated with
$FC(\omega)$. Thus, this change of the current
direction from the expected to physically
counter-intuitive direction is closely
related to the mechanism of noise-induced absolute negative mobility,
as it has been outlined in the previous subsection.

\begin{figure}
\centerline{
 \epsfxsize=8cm
\epsfbox{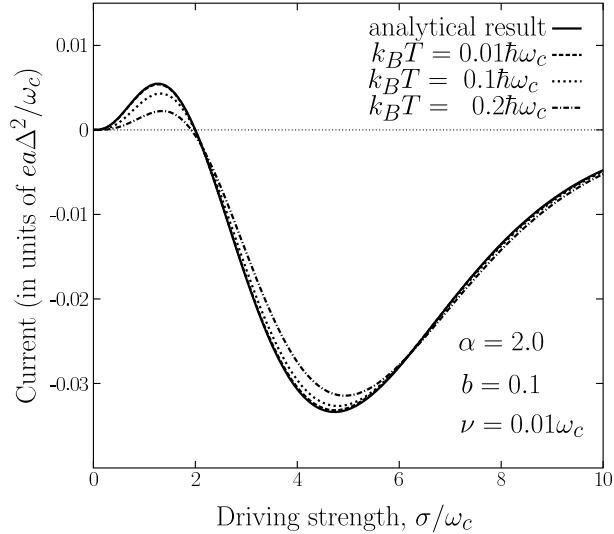}}
\caption{Noise induced rectification of current for an asymmetric
dichotomous driving field
{\it vs.} rms of field fluctuations
at different temperatures. The set of used parameters is
 indicated in the figure.}
\label{scriptFig4}
\end{figure}

Moreover, the current can flow in the physically counter-intuitively
direction also for small applied forces when the coupling strength
$\alpha$ is sufficiently small. For $T=0$ a very insightful
approximate analytical expression can be obtained in the adiabatic
limit for driving and in the lowest order of the asymmetry parameter
$b$. Namely, assuming an Ohmic friction mechanism  for the thermal
bath with an exponential cutoff $\omega_c$, $FC(\omega)$ can be
exactly evaluated at $T=0$ to yield
$FC(\omega)=\frac{1}{\omega_c\Gamma(2\alpha)} (
\frac{\omega}{\omega_c})^{2\alpha-1} \exp[-\omega/\omega_c]
\Theta(\omega)$, where $\Theta(\omega)$ is the Heaviside step
function \cite{WeissBook}. For $b\ll 1$ in (\ref{jquasistatic}),
this then yields the averaged, zero bias current value, reading for
$\alpha>1/2$:
\begin{eqnarray}\label{current-analyt}
j(0)\approx b \frac{\pi}{2}\frac{ea\Delta^2}{\omega_c\Gamma(2\alpha)}
\Big (\frac{\sigma}{\omega_c}\Big)^{2\alpha-1}\Big(\alpha-1-
\frac{\sigma}{2\omega_c}\Big)\exp(-\sigma/\omega_c) \;.
\end{eqnarray}

The  result in Eq. (\ref{current-analyt}) predicts that for
$\alpha\leq 1$ the current flows into the physically
counter-intuitive direction. Furthermore, for $\alpha>1$, the
rectification current flows first in the expected, natural
direction, but it changes subsequently its direction   for
$\sigma>\sigma_*=2(\alpha-1)\omega_c$. Moreover, the absolute value
of current has two maxima at $\sigma_{max}=(2\alpha-1\pm
\sqrt{2\alpha-1})\omega_c$ for $\alpha>1$ and one maximum at
$\sigma_{max}= (2\alpha-1 + \sqrt{2\alpha-1})\omega_c$ for
$1/2<\alpha \leq1$. Furthermore, the current diminishes for large
$\sigma$. In the low-temperature limit, all these features are in
the remarkable agreement with the numerical evaluation of Eq.
(\ref{first}) in Ref. \cite{PRL98}. A related comparison is provided
in Fig. \ref{scriptFig4} for $\alpha=2$. For
$k_BT=0.01\hbar\omega_c$, the agreement is indeed excellent,  except
for very small values $\hbar\sigma\ll k_BT$. On the scale of
$\sigma$ variation used in Fig. \ref{scriptFig4} the rectification
tunnel current seems be maximal at $T=0$ for most $\sigma$. From
this point of view, the rectification results from an interplay
between the zero-point quantum fluctuations and the nonequilibrium
noise, i.e. has a manifestly quantum origin.

For larger values of $\alpha$ than those
depicted  in Fig. \ref{scriptFig4}, the rectification
current can be enhanced for sufficiently small $\sigma/\omega_c$ 
by increasing temperature
and go through a maximum, exhibiting thereby the phenomenon of Quantum
Stochastic Resonance \cite{Lofst,GrifoniPRL, GrifoniPRE, GrifoniPRE2} in the nonlinear
current response \cite{PRL98}.

\subsection{Limit of vanishing dissipation}
In the limit of vanishing dissipation $Q(t)\to 0$,
the result in Eq. (\ref{first}) predicts that $j(0)=0$,
independently of the form and strength of driving. This prediction
should be considered, however, with care since the result in Eq. (\ref{first})
is not
valid for very small $\alpha$ and $T$, because we have assumed throughout
incoherent transport regime where either $\Delta_r=0$, or
the temperature is
sufficiently high, $k_BT\gg\hbar\Delta_r$.
Nevertheless, the dissipationless
single-band, infinite tight-binding model can be solved exactly
in arbitrary time-dependent fields
\cite{KenkreDunlap,ProninBandraukOvchin,PRL98,LNP2000,JPCB2001,Korsch}.
The corresponding exact
solution for the current then shows \cite{PRL98,LNP2000,JPCB2001} that
the {\it stationary} current
is forced to vanish identically by the stochastic fluctuations of driving.
Put differently, in
the absence of quantum dissipation such a rectified current can
exist at most as a transient phenomenon.
As a matter of fact, the {\it stationary} rectification
current within the single-band tight binding description is due to
a nonlinear interplay of quantum dissipation and external nonequilibrium forces.
Its origin presents a highly nonlinear and nonequilibrium {\it statistical effect}.
This result does not hold for more general situations. For the case of the full potential problem,
with its intrinsic interband transitions,
a finite, stationary current can
be generated even in the absence of dissipation; it results as a
{\it dynamical
effect} due to an interplay
of a nonlinear dynamics and the
breaking of some space-time symmetries
by the driving mechanism \cite{Flach,LNP2000,Schanz}.

It must be emphasised, however, that the full potential problem
has little relation
to the electron transport  in molecular chains which is our main
focus here.
This is so, because the
tight-binding description emerges for the electron (or hole)
transport processes in molecular systems
in a very different way, being not the result of
a truncation of a full potential problem to the
description within the lowest band only.

\subsection{Case of harmonic mixing drive}

Another instance of  quantum rectifiers in  presence of dissipation is realized with
 a harmonic mixing driving \cite{Seeger, MarchesoniHM,EPL98},
\begin{eqnarray}\label{gh-mix}
\mathcal{ E}(t)= E_1\cos(\Omega t)+E_2\cos(2\Omega t +\phi),
\end{eqnarray}
with the driving strengths $E_1,E_2$, angular frequency $\Omega$ and
a relative phase $\phi$, respectively.
This model seems more
promising and readily can be implemented with experimental realizations.

\begin{figure}
\centerline{\epsfxsize=8cm
\epsfbox{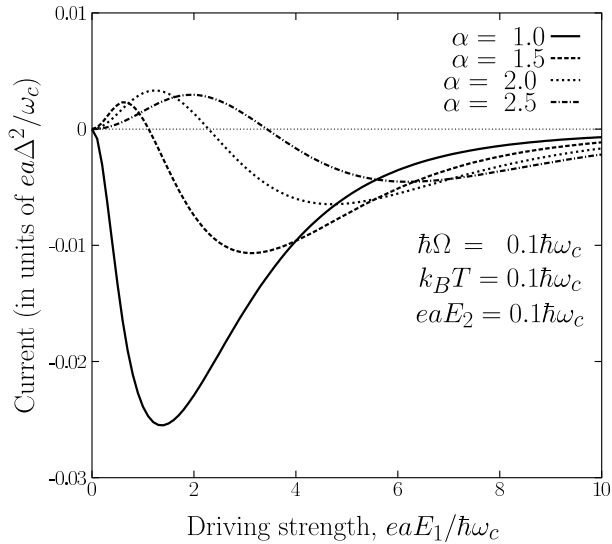}}
\caption{The rectified quantum  current induced by a harmonic mixing drive is depicted
{\it vs.} the strength of the first harmonic
at different values of the coupling strength $\alpha$.
The strength of the second
harmonic is held fixed.
The used parameters are given in the figure.}
\label{scriptFig5}
\end{figure}

The  corresponding expression for
$\langle S (\tau)\rangle$ reads \cite{EPL98}
\begin{eqnarray}\label{pr}
\langle S(\tau)\rangle
 = \sum_{k=-\infty}^{\infty}
J_{2k}\Big (2\xi_1\sin(\Omega\tau/2)\Big)
J_{k}\Big(\xi_2\sin(\Omega\tau)\Big)e^{-ik(\phi+\pi/2)},
\end{eqnarray}
where $\xi_{1,2}=ea E_{1,2}/(\hbar\Omega)$ and $J_n(z)$ are standard
Bessel functions. With its help the current in Eq. (\ref{first})
can be evaluated numerically for the Ohmic model with the exponential
cutoff, where the exact analytical expression for $Q(t)$ is
available \cite{WeissBook,GrifHangRev,PLA98}. Independently of other parameters
the current vanishes identically for $\phi=\pi/2,3\pi/2$, where
${\rm Im} \langle S(\tau)\rangle=0$ {\it exactly}. Otherwise,
the current can be different from zero. For sufficiently
high temperatures and weak fields applied,
$j(0)\propto \langle \mathcal{ E}^3(t)\rangle=\frac{3}{4}E_1^2 E_2\cos(\phi)$
with a nontrivial quantum prefactor. At $T=0$, the current
response is not analytical in the driving amplitude.
Unfortunately, in this case we do not find a simple
approximate analytical expression for the current
like the one in Eq. (\ref{current-analyt}). Some numerical calculations
\cite{EPL98}, see also in Fig. \ref{scriptFig5}, reveal a series of nontrivial
features as the occurrence of current inversion and the occurrence of
current maxima similar to the the case of stochastic dichotomous driving.
Moreover, in the case of harmonic mixing driving the direction of the
rectification current can be conveniently controlled by the phase $\phi$.
For a sufficiently large dissipation strength $\alpha$, the rectification
current response can also exhibit a Quantum Stochastic Resonance
feature \cite{SRReview,Lofst,GrifoniPRL, GrifoniPRE, GrifoniPRE2,GH_QSR,
GH_QSR2,Buchleit}, i.e.,
it exhibits a maximum versus the
temperature $T$. An experimental realization of the dissipative quantum
rectifiers in the studied incoherent tunnelling regime can be expected
for the semiconductor superlattices \cite{Wacker} and for a small polaron
like transport in molecular chains.

\section{Summary}

We have surveyed, extended and justified in great detail the results
of recent research which relates to the quantum dynamics with
fluctuating parameters. The nature of those fluctuations, that usually
stem from the influence of externally applied fields, or intrinsic
degrees of freedom, is either of stochastic or coherent deterministic
origin. Our general findings are applied to specific situations,
encompassing solid state applications, but mainly we did focus on the
theme of driven electron transfer in molecular aggregates.

A first major result is the derivation of
the generalised non-Markovian master equations and the generalised
Redfield equations for  quantum systems composed of a finite,  discrete number of states
which are subject to  the influence of external either stochastic,
or periodic fields. The resulting kinetic
equations allow one to study a
rich variety of different physical problems within a unified framework.
In simplest cases, the relevant part of the reduced
dynamics is  either described
by the Markovian  balance equation of the Pauli master equation type \cite{Pauli}
which is generalised to include the influence of explicit time-dependent,
external field manipulations into the quantum
transition rates; those becoming therefore functionals of the driving field,
or by its  generalisation which accounts as well for
the memory  effects in the corresponding dissipative kernels.

In the case of fast (on the time scale of the averaged
relaxation process) fluctuating, or oscillating fields
these quantum kinetic equations can be averaged.
The relaxation transitions can be described by the averaged quantum
transition rates of the Golden-Rule type. These averaged transition
rates, however, generally do not satisfy the detailed balance condition  at
the temperature of the thermal bath. This  violation of the thermal
detailed balance, being induced by the   {\it nonequilibrium} driving fields, in turn paves
a roadway for identifying several intriguing
nonequilibrium phenomena.

An important case is provided by symmetric dichotomous driving fields
for which our approximate theories and  considerations can be made rigorous, tested and
reaffirmed \cite{Goychuk95,PRE95,JCP95} because the corresponding
averaging can be performed exactly.

The problem of averaging a quantum dynamics in stochastic
fields modelled by non-Markovian processes of the continuous
time random walk type with a discrete number of states (with Markovian
processes emerging as a particular limiting case)
has been investigated  in the absence of dissipation in Sec. 2.
Using a classical stochastic path integral approach,
we obtained some general exact results on the averaging of quantum
propagator of the driven quantum system over the {\it stationary}
realizations of such non-Markovian jump processes.
In  particular, the exact result for the Laplace transform of the correspondingly averaged
quantum propagator has been obtained.
This novel result bears  the  potential
for future applications since it opens a way for a rigorous study of
an extreme case of $1/f^{\alpha}$ noise, implying
long-range  temporal correlations, where standard perturbation theory
is expected to fail. As a first important application, we obtained the
spectral line shape of the corresponding Kubo oscillator and the
Laplace-transformed averaged evolution of a spin 1/2- system that is  driven by a
symmetric alternating renewal process possessing an arbitrary distribution
of the residence times. This implies a very broad class of autocorrelation
functions including those which correspond to noise sources with
$1/f^\alpha$ power spectrum. This general result is shown to reproduce the known solution in
the Markovian limit, i.e. when the
the residence time intervals are exponentially distributed.

Starting out with section 4, we have investigated the combined influence of 
fluctuating parameters and
dissipation on the evolution of the corresponding quantum dynamics. 
In doing so, we considered both
the role of  fluctuations of the energy bias, and/or the effects of 
a fluctuating
intersite tunnelling matrix element. A generalised master equation was obtained
which corresponds to the known NIBA approximation obtained within the quantum
path-integral approach.
The obtained master equation was averaged exactly both over dichotomous
fluctuations of the energy bias  and dichotomous fluctuations
of the tunnelling coupling, i.e.  the case with a
fluctuating tunnelling barrier.
These results have been used  to study a rich repertoire of nonequilibrium phenomena for
electron transfer in condensed media with dynamical
disorder and possibly being  driven by stochastic or deterministic, coherent periodic laser fields.

In section 7 we have studied the quantum transport in extended
quantum systems within a tight-binding description, with the
dynamics being subjected to a strong system-bath coupling and weak
tunnelling, i.e. in the limit of an incoherent hopping regime.  A
general result for the quantum-noise assisted transport current,
being  averaged over the field fluctuations, has been derived. The
corresponding expression  is shown to be equivalent to the NIBA
approximation result of a corresponding quantum path integral
treatment.

Our theory for  dissipative systems with fluctuating parameters
predicts scores of interesting nonequilibrium phenomena  that are
the result of a stunning interplay between equilibrium quantum
fluctuations and  nonequilibrium perturbations. A few noteworthy
such effects are: (i) the suppression, or acceleration of  quantum
transition rates by many orders of magnitude; (ii) a noise-induced
enhancement of the thermally assisted quantum tunnelling; (iii) the
inversion of populations in the spin-boson  model; or (iv) a
noise-induced absolute negative mobility in quantum transport. We
further elaborated on the theme of dissipative quantum rectifiers.
Several of these novel predictions are presently being investigated
in a number of research groups, both theoretically and
experimentally. Our  research in particular also impacts such timely
activities like the investigation of the electronic transport in
infrared laser driven molecular wires \cite{HanggiGroup}. Here, the
fermionic thermal baths are provided by the electronic reservoirs in
the leads and the electron transport through the wire is mainly
coherent. This corresponds to the regime of a weak dissipation
within our approach, being opposite to the regime of incoherent
tunnelling. The role of the size, inter-electrode coupling effects,
etc. \cite{HRYREV} as well as the inelastic Coulomb repulsion
effects \cite{PetrovHanggi}, are also  important for molecular
wires. This brings about further complications that still need to be
investigated theoretically with greater detail. Experimental
progress is presently also forthcoming \cite{Reichert}: This particularly holds true
for quantum Brownian motors  and quantum rectifiers as witnessed
by the exemplary set of recent experimental studies
 \cite{Linke, Weiss,Majer,Grynberg,Lorke,Renzoni}. We share the confident belief that
this research topic will remain flourishing and, moreover, will
invigorate the readers in pursuing their own future research in this area.

\newpage
{\bf Acknowledgements \\} The authors greatly appreciate the
fruitful collaboration with E.G. Petrov, V. May, V. Teslenko, M.
Grifoni, L. Hartmann, M. Thorwart, J. Casado-Pascual and M. Morillo
on various topics covered in this review. Moreover, we acknowledge
stimulating and helpful discussions with P. Reineker, P. Reimann, P.
Talkner, G.-L. Ingold, A. Nitzan, E. Pollak, J. Klafter, 
S. Kohler, J. Lehmann,
M. Stra\ss, U. Kleinekath\"ofer, M. Thoss, F.K. Wilhelm, W. Domcke,
H.-J. Korsch, H. Grabert, J. Ankerhold, H. Linke, H. Schanz, S.
Flach, O. Evtushenko, Y. Zolotaryuk, and U. Weiss. Parts of this
work are based on the Habilitation thesis by one of us (I.G.). This
research has been supported by  the DFG through the collaborative
research centre, SFB-486, project A10 and the Volkswagen foundation,
via project no. I/80424.

\end{document}